\documentclass[12pt,a4paper]{article}

\pdfoutput=1
\usepackage{jheppub,psfrag,slashed,cancel,lscape,caption,array,graphicx,subcaption}
\usepackage[utf8]{inputenc}
\usepackage{amsmath}
\usepackage{nicefrac}  % for 1/2 
\usepackage{mathtools}
\usepackage{mathrsfs}
\usepackage{multirow}
\usepackage{booktabs} %nice tables 
\usepackage{chngcntr}
\usepackage{nicefrac}
\usepackage{cleveref}
\usepackage{orcidlink}
\usepackage{relsize}
\usepackage{tikz}
\usetikzlibrary{decorations.pathmorphing}
\usetikzlibrary{decorations.markings}
\usetikzlibrary{positioning, shapes, snakes, arrows}
\usetikzlibrary{patterns}

\tikzset{
    wl/.style={line width=1pt},
    graviton/.style={line width=.8pt, -latex,decorate, decoration={snake, segment length=4pt,amplitude=1.8pt, pre length=.15cm, post length=.25cm}},
    worldlineStatic/.style={dotted, line width=1pt},
	worldline/.style={gray, line width=1pt},
	worldlineBold/.style={black, line width=.6pt},
	zUndirected/.style={line width=1pt},
	zParticle/.style={line width=1pt,postaction={decorate},decoration={markings,mark=at position .6 with {\arrow[#1]{latex}}}},
	zParticle2/.style={line width=1pt,postaction={decorate},decoration={markings,mark=at position .7 with {\arrow[#1]{latex}}}},
	worldlineCut/.style={dotted,line width=1pt,postaction={decorate},decoration={markings,mark=at position .7 with {\arrow[#1]{latex}}}},
	worldlineCut2/.style={dotted,line width=1pt,postaction={decorate},decoration={markings,mark=at position .6 with {\arrow[#1]{latex}}}},
	zParticleF/.style={line width=1pt,postaction={decorate}},
	cscalar/.style={line width=1pt,postaction={decorate},decoration={markings,mark=at position .6 with {\arrow[#1]{latex}}}},
	cscalar2/.style={line width=1pt,postaction={decorate},decoration={markings,mark=at position .8 with {\arrow[#1]{latex}}}},
	photon/.style={line width =.8pt, decorate, decoration={snake, segment length=4pt, amplitude=1.8pt,  pre length=.1cm, post length=.1cm}},
	photonRed/.style={red, line width =.8pt, decorate, decoration={snake, segment length=4pt, amplitude=1.8pt,  pre length=.1cm, post length=.1cm}},
	cross/.style={cross out, line width =.8pt, draw=black, minimum size=2*(#1-\pgflinewidth), inner sep=0pt, outer sep=0pt},
cross/.default={4pt}
}

 \newcommand{\gustav}[1]{}
 \newcommand{\jan}[1]{}
 \newcommand{\kays}[1]{}
 \newcommand{\gus}[1]{}

\newcommand{\Z}{\mathcal{Z}}
\newcommand{\ihbar}{\sfrac{\i}{\hbar}}
\newcommand{\iO}{\i 0^{+}}

\DeclareFontFamily{OT1}{pzc}{} 
\DeclareFontShape{OT1}{pzc}{m}{it}{<-> s * [1.350] pzcmi7t}{}
\DeclareMathAlphabet{\mathpzc}{OT1}{pzc}{m}{it}

\def\cL{\mathcal{L}}
\def\cN{\mathcal{N}}
\def\cO{\mathcal{O}}

\def\cM{\mathcal{M}}

\def\eps{\epsilon}

\def\d{\mathrm{d}}

\def\pat{\partial}
\def\mn{{\mu\nu}}
\def\ab{{\alpha\beta}}

\def\i\math

\def\bH{\hat{b}}

\def\dd{\delta\!\!\!{}^-\!}

\def\d{\mathrm{d}}
\def\eps{\epsilon}

\renewcommand{\i}{\ensuremath{\mathrm{i}}}
\renewcommand{\d}{\ensuremath{\mathrm{d}}}

\newcommand{\s}[1]{\relax}

\newcommand{\zenodo}{\href{https://doi.org/10.5281/zenodo.17242015}{\tt zenodo.org} \cite{zenodo}}

\def\nn{\nonumber}

\def\eqn#1{eq.~\eqref{#1}}

\def\rcite#1{ref.~\cite{#1}}

\newcommand{\be}{\begin{equation}}
\newcommand{\ee}{\end{equation}}
\newcommand{\ba}{\begin{align}}
\newcommand{\ea}{\end{align}}
\ifx\genfrac\sdflkaj\else\fi
\newcommand{\sfrac}[2]{{\textstyle\frac{#1}{#2}}}

\def\centerarc[#1](#2)(#3:#4:#5){ \draw[#1] ($(#2)+({#5*cos(#3)},{#5*sin(#3)})$) arc (#3:#4:#5); }
\begin{document}

\begin{flushright}
\begingroup\footnotesize\ttfamily
	HU-EP-25/33 \\
    QMUL-PH-25-29
\endgroup
\end{flushright}

\vspace{15mm}

\begin{center}
{\LARGE\bfseries 
	Unitarity and the On-Shell Action of Worldline Quantum Field Theory
\par}

\vspace{15mm}

\begingroup\scshape\large 
	Kays Haddad\,\orcidlink{0000-0002-1182-2750},${}^{1}$ 
	Gustav Uhre Jakobsen\,\orcidlink{0000-0001-9743-0442},${}^{1,2}$
    Gustav~Mogull\,\orcidlink{0000-0003-3070-5717},${}^{1,2,3}$
	and Jan~Plefka\,\orcidlink{0000-0003-2883-7825}${}^{1}$
\endgroup
\vspace{3mm}
					
\textit{${}^{1}$Institut f\"ur Physik, Humboldt-Universit\"at zu Berlin, 
 10099 Berlin, Germany} \\[0.25cm]
\textit{${}^{2}$Max-Planck-Institut f\"ur Gravitationsphysik
(Albert-Einstein-Institut), \\ 14476 Potsdam, Germany } \\[0.25cm]
\textit{${}^{3}$Centre for Theoretical Physics, Department of Physics and Astronomy, \\ \!\!\!\!Queen Mary University of London, Mile End Road, London E1~4NS, United Kingdom}

\bigskip
  
\texttt{\small\{kays.haddad@physik.hu-berlin.de, gustav.uhre.jakobsen@hu-berlin.de, jan.plefka@hu-berlin.de, g.mogull@qmul.ac.uk\}}

\vspace{10mm}

\textbf{Abstract}\vspace{5mm}\par
\begin{minipage}{14.7cm}
We develop the on-shell action formalism within Worldline Quantum Field Theory (WQFT) to describe 
scattering of spinning compact bodies in General Relativity in the
post-Minkowskian (PM) expansion.
The real on-shell action is constructed from vacuum diagrams with causal (retarded) propagators from which scattering observables such as momentum impulse and spin kick 
follow via Poisson brackets of the initial scattering data. 
Furthermore, we explore the implications of unitarity at the level of the worldline and show how generalised unitarity techniques can be adapted to WQFT to efficiently
compute multi-loop contributions. 
Our work establishes a concrete link between WQFT and amplitude-based methods, elucidating how unitarity cuts ensure equivalence between the on-shell action derived from either approach.
Extending the state-of-the-art, we complete the full on-shell action --- including dissipative terms ---
at (formal) 3PM order and up to quartic spin interactions on both massive bodies.
\end{minipage}\par

\end{center}
\setcounter{page}{0}
\thispagestyle{empty}
\newpage

\tableofcontents

\section{Introduction}

The need to provide accurate predictions for the coming third generation of gravitational waveform detectors
\cite{LISA:2017pwj,Punturo:2010zz,Ballmer:2022uxx}
continues to inspire work on the gravitational two-body problem.
Given the absence of exact analytic solutions ---
unlike the Newtonian two-body problem, where Kepler's laws provide a complete result ---
we are compelled to use various approximations.
These can broadly be put into two categories: numeric and perturbative.
The first kind, known as numerical relativity (NR), involves solving Einstein's equations to a high
degree of numerical precision using supercomputers~\cite{Pretorius:2005gq,Campanelli:2005dd,Baker:2005vv,Scheel:2025jct}.
The second involves finding exact, analytic solutions order-by-order in a perturbative expansion parameter.
The three choices typically considered are: post-Minkowskian (PM)
\cite{Bjerrum-Bohr:2022blt,Kosower:2022yvp,Buonanno:2022pgc},
wherein one expands in the ratio of size and inverse separation between the two bodies;
post-Newtonian (PN)~\cite{Blanchet:2013haa,Porto:2016pyg,Levi:2018nxp},
involving also an expansion in the relative velocity,
considered small compared with the speed of light $v/c\ll1$;
and gravitational self-force (GSF)
\cite{Mino:1996nk,Poisson:2011nh,Barack:2018yvs,Gralla:2021qaf},
which expands in the mass ratio $m_1/m_2$.
Each captures information about the two-body problem in different physical regimes.
PN is ideal for quasi-circular bound orbits,
PM for scattering or highly elliptic orbits,
and GSF for extreme mass ratios.

By themselves, perturbative solutions to the two-body problem are generally not enough
to enable accurate gravitational waveform modelling.
In order to make predictions in non-perturbative regimes ---
comparable masses, strong fields or fast velocities ---
resummations are also required.
A good example is the effective-one-body (EOB)
\cite{Buonanno:1998gg,Buonanno:2000ef},
which deforms the motion of a probe mass moving around a Schwarzschild or Kerr metric.
Regardless of the precise resummation used,
a key requirement is that the resummed model reproduces the true two-body motion
when expanded in a perturbative regime.
This raises a natural question about how to best encode the perturbative two-body data.
While a two-body Hamiltonian offers one good option (at least for the conservative dynamics)
it suffers from gauge-dependency,
meaning that one may write down equivalent versions related by a canonical transformation.
Recent work on scattering in the PM expansion has shown that scattering observables,
such as the scattering angle for aligned-spin configurations, provide a good alternative
\cite{Antonelli:2019ytb,Khalil:2022ylj,Damour:2022ybd,Rettegno:2023ghr,Buonanno:2024vkx,Buonanno:2024byg,Damour:2025uka,Long:2025nmj,Clark:2025kvu}.
For mis-aligned spin configurations, where motion is not confined to a single scattering plane,
the best choice of observable is less obvious.\footnote{Although, recently ref.~\cite{Akpinar:2025tct} argued that aligned-spin scattering may encode more 
information than previously thought.}

Our ability to derive PM-expanded scattering observables for the classical two-body problem
has been revolutionised by Quantum Field Theory (QFT)-based methods.
An important first step was the non-spinning conservative scattering dynamics
at 3PM ($G^3$) order
\cite{Bern:2019nnu,Bern:2019crd,Cheung:2020gyp,DiVecchia:2020ymx,Kalin:2020fhe}.
This was extended to include radiation~\cite{Damour:2020tta,Herrmann:2021lqe,Herrmann:2021tct,DiVecchia:2021bdo,Heissenberg:2021tzo,Bjerrum-Bohr:2021din,Damgaard:2021ipf,Kalin:2022hph}
and subsequently pushed to 4PM order~\cite{Bern:2021dqo,Bern:2021yeh,Dlapa:2021vgp} (conservative)
\cite{Dlapa:2022lmu,Dlapa:2023hsl,Damgaard:2023ttc} (full dissipative).
In the spinning case, the 1-loop scattering angle has been derived up to quartic spins
\cite{Guevara:2018wpp,Chen:2021kxt},
and higher-order predictions have also been made~\cite{Bern:2022kto,Aoude:2022trd,Aoude:2022thd,Aoude:2023vdk,Bautista:2023szu,Bohnenblust:2024hkw}.
At two-loop order, analytic results also exist as high as quartic order in spin for one spinning body
\cite{FebresCordero:2022jts,Akpinar:2025bkt}.
In all cases, the use of generalised unitarity~\cite{Bern:1994zx,Bern:1994cg}
has enabled efficient derivations of the integrands involved.
The link between these unbound scattering observables with bound-orbit observables
has also been thoroughly explored
\cite{Kalin:2019rwq,Kalin:2019inp,Cho:2021arx,Saketh:2021sri},
and first steps taken to address the tail effect~\cite{Bini:2024tft,Dlapa:2024cje,Dlapa:2025biy}.

As part of this broad effort, the Worldline Quantum Field Theory (WQFT) formalism
has proven itself a crucial workhorse~\cite{Mogull:2020sak,Jakobsen:2023oow}.
Building on the closely related PM-based Worldline EFT~\cite{Kalin:2020mvi},
WQFT combines first- and second-quantised degrees of freedom,
i.e.~particles and fields, enabling a method that directly targets classical contributions from the outset.
So far, the main emphasis has been on scattering observables ---
including the momentum impulse, gravitational waveform and spin kick,
which have a well understood retarded causality prescription~\cite{Jakobsen:2022psy,Kalin:2022hph}.
Early results included two-loop scattering observables up to quadratic order in spin
\cite{Jakobsen:2022fcj,Jakobsen:2022zsx},
then three-loop observables up to linear order in spin~\cite{Jakobsen:2023ndj,Jakobsen:2023hig}
and with adiabatic tidal effects~\cite{Jakobsen:2022psy,Jakobsen:2023pvx}.
More recently, the non-spinning scattering dynamics have been pushed as far as 5PM
and first order in GSF~\cite{Driesse:2024xad,Driesse:2024feo}.
As a result, we are now tantalisingly close to a complete picture of the scattering dynamics at $G^5$ order,
once one also counts powers of $G$ within the spin vectors $S_i^\mu \sim Gm_i^{2}$.
The link between QFT and worldline methods has also been explored~\cite{Mogull:2020sak,Damgaard:2023vnx,Capatti:2024bid}.

In a crucial development by the current authors~\cite{Haddad:2024ebn},
the WQFT approach was recently extended to arbitrary powers in spins,
pushing past the limitation to quadratic spins previously imposed by the use of an
$\cN=2$ supersymmetric worldline action~\cite{Jakobsen:2021lvp,Jakobsen:2021zvh}.
This was done by introducing bosonic oscillators $\alpha_i^\mu(\tau)$ on the worldline,
through which the spin tensor is encoded as $S_i^{\mu\nu}=\s{2}2\i m_{i}\bar\alpha_i^{[\mu}\alpha_i^{\nu]}$.
These oscillators are inspired by bosonic string theory in the tensionless limit,
and the associated worldline action enjoys a bosonic version of the supersymmetry
which we refer to as a ``BUSY''.
The BUSY is generated using a set of Poisson brackets,
and serves to ensure conservation of the covariant spin-supplementary condition (SSC)
$S^{\mu\nu}\pi_{\nu}=0$.
In that paper we performed calculations of 1-loop scattering observables up to quartic 
order in spins.
Other recent developments involving WQFT include electromagnetic self-force~\cite{Jakobsen:2023tvm}, scalar QED \cite{Wang:2022ntx},
calculating complete scattering trajectories~\cite{Mogull:2025cfn},
superior methods of calculating in the probe limit~\cite{Hoogeveen:2025tew},
curved backgrounds \cite{Cheung:2023lnj,Cheung:2024byb,Bjerrum-Bohr:2025bqg},
 spinning waveforms \cite{Bohnenblust:2025gir}, higher-spin with
SUSY \cite{Bonocore:2024uxk,Bonocore:2025stf} and investigations on the double copy 
\cite{Shi:2021qsb,Comberiati:2022cpm}.

In this paper we demonstrate how, through the use of bosonic oscillators,
the WQFT can straightforwardly be used to compute the on-shell action
for scattered spinning massive bodies at arbitrary powers in spin.
The real on-shell action --- not to be confused with the eikonal phase, which is complex ---
serves as a generating function for scattering observables including the
momentum impulse $\Delta p_i^\mu$ and spin kick $\Delta S_i^\mu$.
Thus, it is ideal for matching to resummed waveform models including EOB.
The on-shell action is computed in the WQFT as a sum over vacuum diagrams,
with an appropriate causality prescription giving the right $\iO$ prescriptions on internal propagators.
Scattering observables are subsequently generated using 
the Poisson brackets of the background parameters taken at past infinity~\cite{Alessio:2025flu}.
From a QFT perspective, the on-shell action has been shown to arise
from the Magnus series~\cite{Kim:2024svw,Kim:2025hpn},
which yields an appropriate $\iO$ prescription.
In this paper we take a more straightforward approach:
working backwards from the scattering observables, 
where the retarded causality prescription is well known,
and using the brackets to learn the causality prescription for the on-shell action.

To fully demonstrate our grasp of the formalism,
in this paper we also derive the full on-shell action, including dissipative terms,
up to 3PM order and quartic order in spins --- a two-loop computation.
This builds on recent work using scattering amplitudes~\cite{FebresCordero:2022jts,Akpinar:2025bkt}
and extends these results to spins on \emph{both} of the two massive bodies.
To perform this calculation,
we present for the first time the applicability of on-shell methods such as generalised unitarity
to the computation of observables in WQFT.
While calculations at the precision considered in this paper do not involve
a large number of diagrams or large-multiplicity vertices,
unitarity enjoys a better scaling with increasing spin orders compared with the reduction of tensor integrals.
The results of this paper yield an important expansion of the tools at our
disposal for computing observables in WQFT.

The outline of our paper is as follows.
In \cref{sec:onShellAction} we introduce the on-shell action in WQFT,
and show how it generates scattering observables by way of the background brackets.
In \cref{sec:symmetries} we re-introduce the bosonic worldline action of \cite{Haddad:2024ebn},
and derive the constrained brackets that account for conserved quantities.
Then, in \cref{sec:diagObservables}, we show how observables are generated at a diagrammatic level
by the brackets, which in turn leads to our $\iO$ prescription for the on-shell action.
A convenient grouping of the diagrams involved in the on-shell action is leveraged to import generalised unitarity to WQFT in \cref{sec:Unitarity}, which facilitates the connections to the scattering amplitude approach that we make in \cref{sec:AmplitudesConnection}.
Finally, the technology developed in this paper is applied in \cref{sec:Observables} to produce novel corrections to observables for two spinning black holes at $\cO(G^{3})$ and spin orders $S_{1}^{n_{1}}S_{2}^{n_{2}}$ for $n_{1}+n_{2}\leq4$.

\section{On-shell action in WQFT}\label{sec:onShellAction}

We begin by introducing and motivating the on-shell action in WQFT,
as it applies to spinning particles interacting via a gravitational field
$g_{\mu\nu}=\eta_{\mu\nu}+\kappa h_{\mu\nu}$.
For this purpose we adopt an operator-based quantum formalism,
rather than a path-integral description as we have used in the past.
The fields obey equal-time commutation relations:
\begin{subequations}\label{eq:commutators}
\begin{align}
    [\hat{x}_i^\mu(\tau_i),\hat{p}_{j,\nu}(\tau_i)]&=-\i\hbar\delta_{ij}\delta^\mu_\nu\,,\\
    [\hat\alpha_i^a(\tau_i),\hat{\bar{\alpha}}_j^b(\tau_i)]&=
    \s{1}
     \hbar\delta_{ij}\eta^{ab}/m_i\,,\label{oscillatorBracket}\\
    [\hat{h}_{\mu\nu}(t,\mathbf{x}),\hat\pi^{\rho\sigma}(t,\mathbf{y})]&=
    -\i\hbar\delta_{(\mu}^\rho\delta_{\nu)}^\sigma\delta(\mathbf{x}-\mathbf{y})\,.
\end{align}
\end{subequations}
We have introduced the position $\hat{x}_i^\mu$ and momentum $\hat{p}_{i,\mu}$
operators, parametrised by the time coordinates of each particle $\tau_i$ and
we (anti)--symmetrise with unit weight.
The complex bosonic oscillators $\hat{\alpha}_i^a(\tau)$,
introduced in ref.~\cite{Haddad:2024ebn},
are defined in the locally flat frame, whose indices are represented with Latin letters, and are translated to the full spacetime by means of vierbeins, $\hat\alpha_i^{\mu}(\tau)=e^\mu_a\hat\alpha_i^a(\tau)$.
These oscillators encode spin degrees of freedom through the spin tensor
\begin{align}
    \hat{S}_{i}^{ab}(\tau)= \s{2}
     2\i m_{i}\hat{\bar\alpha}^{[a}_{i}(\tau)\hat\alpha_{i}^{b]}(\tau)\,.
\end{align}
The commutator between bosonic oscillators~\eqref{oscillatorBracket}
then naturally implies the $\mathfrak{so}(1,3)$ algebra between spin tensors:
\begin{align}
    [\hat{S}_i^{ab},\hat{S}_i^{cd}]=
    \s{1} \s{2}\i\hbar  (
       \eta^{bc}\hat{S}_i^{ad}+\eta^{ad}\hat{S}_i^{bc}
        -\eta^{bd}\hat{S}_i^{ac}-\eta^{ac}\hat{S}_i^{bd})\,.
\end{align}
The WQFT's central feature, illustrated here,
is to combine first-quantised operators (particles) with second-quantised operators (fields).

In this operator-based language, scattering observables may be computed as
differences between observables $\hat{\cal O}$ at future and past infinity:
\begin{align}\label{eq:deltaO}
    \Delta\cO=\langle0|\hat{S}^\dagger\hat\cO\hat{S}|0\rangle-\langle0|\hat\cO|0\rangle\,,
\end{align}
where $|0\rangle$ is the vacuum at past infinity and $\hat S$ denotes the $S$-matrix 
performing the time-evolution from past to future infinity.
As has been noted by several authors~\cite{Damgaard:2021ipf,Damgaard:2023ttc,Caron-Huot:2023vxl,Kim:2024svw,Gonzo:2024xjk},
it is convenient to introduce an exponential representation of the $S$-matrix:
\begin{align}\label{eq:exponential}
    \hat{S}=e^{\ihbar \hat N}\,.
\end{align}
The unitarity constraint $\hat{S}^\dagger\hat{S}=1$ implies that $\hat{N}$ is Hermitian, $\hat{N}=\hat{N}^\dagger$.
Inserting this into Eq.~\eqref{eq:deltaO} yields
\begin{align}\label{eq:observables}
\begin{aligned}
    \Delta\cO&=\langle0|e^{-\ihbar\hat N}\hat\cO e^{\ihbar\hat N}|0\rangle-\langle0|\hat\cO|0\rangle\,,\\
    &=\langle0|e^{-\ihbar[\hat{N},\bullet]}\hat\cO|0\rangle-\langle0|\hat\cO|0\rangle\,,\\
    &=-\ihbar\langle0|[\hat{N},\hat\cO]|0\rangle-\sfrac1{2! \hbar^{2}}\langle0|[\hat{N},[\hat{N},\hat\cO]]|0\rangle+\cdots\,.
\end{aligned}
\end{align}
Thus, in order to compute the shift of $\mathcal{O}$ we apply commutation relations on $\hat{N}$
and subsequently take the matrix element on $|0\rangle$.

The on-shell action $N$ is defined by the incoming vacuum expectation value of the
$\hat N$-operator 
\begin{align}\label{eq:radialAction}
    N(b_{i}^\mu,v_i^\mu,\alpha_{i}^a,\bar{\alpha}_{i}^a):=\langle0|\hat{N}|0\rangle\, ,
\end{align}
where the arguments $b_{i}^\mu,v_i^\mu,\alpha_{i}^a,\bar{\alpha}_{i}^a$ are the background field values
of the worldline fields.
Using Eq.~\eqref{eq:observables},
this may be used to derive scattering observables.
However, as we have already taken the matrix element on $|0\rangle$ it is \emph{a priori}
unclear how one may still act with the commutation relations~\eqref{eq:commutators}.
Our solution relies on the fact that each worldline field is
expanded around its value at past infinity:
\begin{subequations}\label{backgroundExpansions}
\begin{align}
    \hat{x}_i^\mu(\tau_i)&=b_i^\mu+\tau_i v_i^\mu+\hat{z}_i^\mu(\tau_i)\,,\\
    \hat{p}_i^\mu(\tau_i)&=m_iv_i^\mu+\hat{p}_i^{\prime\mu}(\tau_i)\,,\\
    \hat{\alpha}_i^a(\tau_i)&=\alpha_{i}^a+\hat\alpha_i^{\prime a}(\tau_i)\,,\\
    \hat{\bar\alpha}_i^a(\tau_i)&=\bar\alpha_{i}^a+\hat{\bar\alpha}_i^{\prime a}(\tau_i)\,.
\end{align}
\end{subequations}
Note that we distinguish the constant background $\alpha_{i}$ from the worldline
field $\hat\alpha_{i}(\tau)$ by the explicit $\tau$ dependence. 
Thus, ignoring for now the crucial issue of constraints on the background parameters ---
which we will discuss in \cref{sec:symmetries} ---
the dependence within $\hat{N}$ on the full worldline operators
$\hat{x}_i^\mu(\tau)$, $\hat{p}_i^\mu(\tau)$, $\hat{\alpha}_i^a(\tau)$ and $\hat{\bar\alpha}_i^a(\tau)$
is captured by its dependence on the background variables
$b_i^\mu$, $v_i^\mu$, $\alpha_{i}^a$ and $\bar\alpha_{i}^a$.
To exploit $N$'s dependence on the background parameters,
we introduce a new set of background Poisson brackets:
\begin{subequations}\label{PBbv}
\begin{align}
    \{b_i^\mu,v_{j}^{\nu}\}&=
    -
    \frac{\delta_{ij}}{m_{i}}\eta^\mn \,,\\
    \{ \alpha_{i}^{a}, \bar\alpha_{j}^{b} \} &= - \s{1} \i \frac{ \,\delta_{ij}}{m_{i}}\, \eta^{ab}\, .
\end{align}
\end{subequations}
These brackets replicate the full quantum brackets~\eqref{eq:commutators}
at the level of the background parameters.
As these are also classical objects, we have replaced
$[\hat A,\hat B]\to i\hbar\{A,B\}$.
We may act with these classical brackets on $N$,\footnote{An alternative approach where a Routhian, rather than the on-shell action, generates worldline scattering dynamics has been employed in ref.~\cite{Liu:2021zxr}.}
\begin{align}\label{expObservables}
\begin{aligned}
    \Delta {\cal O}_{N}
    &= e^{ \{ N, \bullet \} } {\cal O} - {\cal O}\\
    &=
    \{
        N,
        {\cal O}
    \}
    \!+\!
    \sfrac1{2!} 
    \big\{N,
        \{N,{\cal O}
    \}
    \big\}
    \!+\!
    \sfrac1{3!} 
    \big\{N,\big\{N,
        \{N,{\cal O}
    \}
    \big\}\big\}
    +
    \cdots\,,
\end{aligned}
\end{align}
and thus derive observable quantities. 

An important caveat in this logic is that we have
ignored brackets involving the gravitational field operator $\hat{h}_{\mu\nu}$.
The absence of a corresponding background field
(other than the Minkowski metric $\eta_{\mu\nu}$)
means that corresponding background brackets cannot be introduced.
Technically, in our subsequent analysis, we will therefore be unable to distinguish
between different causality prescriptions of graviton propagators.
More generally, omitting effects due to brackets of the gravitational field background implies that observables generated from $N$ (labelled with an $N$ subscript) do not include flux effects which give rise to loss of total linear or angular momentum, starting at 3PM and 2PM respectively.
However, they will capture all conservative effects and ``conservative-like'' dissipative effects ---
to be discussed in \cref{obsFromAction}.
In order to bypass this restriction, a corresponding gravitational background variable was
recently introduced by Kim~\cite{Kim:2025hpn} ---
see also \rcite{Alessio:2025flu} for recent work on deriving observables from
the $\hat{N}$ operator in the presence of radiation.

\section{Spinning worldline action and its symmetries}\label{sec:symmetries}

Let us now review the worldline action of a spinning particle
and its underlying symmetries.
The action involving non-minimal couplings
of the bosonic oscillators $\alpha_i^\mu(\tau)$ reads~\cite{Haddad:2024ebn} 
\begin{subequations}\label{eq:sWQFT2}
\begin{align}
     S&=-m\int\d\tau\,\left(\sfrac{1}{2}g_{\mu\nu}\dot{x}^{\mu}(\tau)\dot{x}^{\nu}(\tau)-
    \i\bar{\alpha}_{\mu}(\tau)\frac{{\rm D}\alpha^{\mu}(\tau)}{\d\tau}-\cL_{\rm nm}\right)\,,
\end{align}
for brevity omitting the particle label $i$. 
The non-minimal spin-terms terms for Kerr black holes at linear order in curvature
 couple to the electric and magnetic Weyl tensors
	$E_{\mu\nu}$ and $B_{\mu\nu}$ via
\begin{align}\label{eq:Lnm}
    &\cL_{\rm nm}=-\frac{1}{m}B_{a(\tau)Z(\tau)}+\left(1+\frac{\nabla_{Z(\tau)}}{m|\dot{x}(\tau)|}\right)\sum_{n=1}^{\infty}\left(\i\nabla_{a(\tau)}\right)^{2n-2}\left[-\frac{E_{a(\tau)a(\tau)}}{(2n)!}+\frac{\nabla_{a(\tau)}B_{a(\tau)a(\tau)}}{(2n+1)!}\right].
\end{align}
\end{subequations}
Quadratic-in-curvature operators which are relevant to Kerr only play a role starting from fifth order in the spin tensor, which is beyond the scope of our present work \cite{Haddad:2024ebn}.
Here we have adopted the Schoonschip notation,
where e.g.~$B_{a(\tau)Z(\tau)}:=B_{\mu\nu} a^{\mu}(\tau)Z^{\nu}(\tau)$, and
\begin{align}\label{aZdefswithosc}
    &a^{\mu}(\tau):=\s{2}\i\frac{\varepsilon^{\mu\dot{x}(\tau)\bar{\alpha}(\tau)\alpha(\tau)}}{|\dot x(\tau)|}\, ,&
    &Z^{\mu}(\tau):=m\s{2}\i\frac{2\bar{\alpha}^{[\mu}(\tau)\alpha^{\nu]}(\tau)\dot{x}_{\nu}(\tau)}{|\dot x(\tau)|} \, . %new sign on spin tensor
\end{align}
Here $|\dot{x}(\tau)|=\sqrt{g_{\mu\nu}\dot{x}^{\mu}(\tau)\dot{x}^{\nu}(\tau)}$ and $\varepsilon^{\mu\nu\rho\tau}=\epsilon^{\mu\nu\rho\tau}/\sqrt{-g}$ defines the totally antisymmetric Levi-Civita tensor in terms of the Levi-Civita symbol and the metric determinant.

The symmetries of the action were discussed at great length in ref.~\cite{Haddad:2024ebn}.
Motivated by our introduction of brackets on the background parameters~\eqref{PBbv},
we focus here on the asymptotic symmetries of the worldline action~\eqref{eq:sWQFT2}
when the curvature of spacetime vanishes.
In flat space the worldline action reduces to
\be
S_{0}=-m\int\!\d\tau
 \Bigl (\sfrac{1}{2} \dot x^{2}(\tau) - \i \bar\alpha_{\mu}(\tau) \dot \alpha^{\mu}(\tau)
 \Bigr ) \, .
 \ee
This action is invariant under the rigid symmetries~\cite{Jakobsen:2023oow}
\begin{align}\label{busy}
\begin{aligned}
\delta x^{\mu}(\tau)&= \beta \dot{x}^\mu(\tau) + \i\bar\epsilon \alpha^{\mu}(\tau) -\i \epsilon \bar\alpha^{\mu}(\tau)\,, \\
\delta\alpha^{\mu}(\tau)&=\epsilon \dot x^{\mu}(\tau)\, , \qquad \delta\bar\alpha^{\mu}(\tau)=\bar\epsilon \dot x^{\mu}(\tau)\, ,
\end{aligned}
\end{align}
of a translation, parametrised by $\beta$,
and a ``bosonic supersymmetry’’ (``BUSY’’) parametrised by the complex constant
parameter $\epsilon$. 
The translation is generated by the proper-time Hamiltonian $H=p^2/2m$, where $p^\mu(\tau)=m\dot{x}^\mu(\tau)$,
and the BUSY is generated by the charges $Q=\alpha\cdot p$ and  $\bar Q=\bar\alpha\cdot p$.
These symmetries 
constitute the asymptotic symmetries of the  background variables
$b_i^\mu$, $v_i^\mu$, $\alpha_{i}^a$ and $\bar\alpha_{i}^a$
in the context of the full problem in a curved spacetime background. 

At the level of the Poisson brackets for the background variables
of eqs.~\eqref{PBbv} for the two-body system, with particle labels $i=1,2$,
we have
\begin{align}
Q_{i}&=m_{i}\,\alpha_{i}\cdot v_{i}\,,&
H_i&=\frac{m_i}{2}v_i^2\,.
\end{align}
These charges enjoy an algebra:
\begin{align}
    \{Q_i,\bar{Q}_i\}&=-2\i H_i\,, &
    \{H_i,Q_i\}&=0\,, &
    \{H_i,\bar{Q}_i\}&=0\,.
\end{align}
They may also be used to generate the shifts in background parameters,
with
\be
\mathcal{Q}:=\sum_{i=1}^{2} \left (\i \bar\epsilon_{i}  Q_{i} - \i\epsilon_{i}\bar{Q}_{i}
+\beta_i H_i\right )\,.
\ee 
One then easily finds that
\begin{subequations} \label{busytrans}
\begin{align}
\delta b_{i}^{\mu} &= \{ \mathcal{Q}, b_{i}^{\mu}\} = \i \bar \epsilon_{i} \alpha_{i}^{\mu} - \i \epsilon_{i}\bar\alpha_{i }^{\mu}+\beta_iv_i^\mu\,, \\
\delta v_{i}^{\mu} &=\{ \mathcal{Q}, v_{i}^{\mu}\} = 0 \, ,\\
\delta \alpha_{i }^{\mu} &= \{ \mathcal{Q}, \alpha_{i }^{\mu}\} =  \s{1}\epsilon_{i} v_{i}^{\mu}\,,\\
\delta \bar\alpha_{i }^{\mu} &= \{  \mathcal{Q}, \bar\alpha_{i }^{\mu}\} =  \s{1}\bar\epsilon_{i} v_{i}^{\mu}\,.
\end{align}
\end{subequations}
Gauge invariant vectors are then the velocities $v_{i}^{\mu}$, that are constrained to
$v_{i}^{2}=1$ as a consequence of gauge fixing the worldline einbein, as well as the
 Pauli-Lubanski vector $S_{i}^{\mu}$ 
 \be
S_{i}^{\mu}
=
m_i a_i^\mu
=
\sfrac{1}{2}\epsilon^{\mu}{}_{\nu\rho\sigma}v_{i}^{\nu}S_{i}^{\rho\sigma} 
=
\s{2} \i m_{i} \epsilon^{\mu v_{i}\bar\alpha_{i } \alpha_{i }}
\, .
\ee
and the invariant impact parameter\footnote{As our action \cref{eq:sWQFT2} truncates at linear order in the SSC vector, it is not obvious that SSC vectors will only shift the impact parameters in perturbative calculations. This is proven in appendix~\ref{app:SSC}.} 
\be
b^{\mu}=P^{\mu}{}_{\nu}(b_{2}^{\nu}-b_{1}^{\nu}+Z_{2}^{\nu}/m_{2}-Z_{1}^{\nu}/m_{1}),\label{eq:GeneralisedIP}
\ee
involving the SSC vectors $Z_{i}^{\mu}:= S_{i}^{\mu\nu}v_{i,\nu}$.  
Here we have introduced the projector off the $v_{1}-v_{2}$ plane
\begin{align}
P^{\mu\nu}=\eta^{\mu\nu}-v_{1}^{\mu}\bar v_{1}^{\nu}-v_{2}^{\mu}\bar v_{2}^{\nu}
= \eta^{\mu\nu}+ \frac{v_{1}^{\mu}v_{1}^{\nu}  + v_{2}^{\mu}v_{2}^{\nu}
-2\gamma v_{1}^{(\mu}v_{2}^{\nu)} }{\gamma^{2}-1}\, ,
\end{align}
with $\gamma=v_{1}\cdot v_{2}$, and introduced the dual velocities
\be
\bar v_{1}^\mu
=
-\frac{v_{1}^\mu-\gamma v_{2}^\mu}{\gamma^{2}-1}\, , \quad
\bar v_{2}^\mu
=
-\frac{v_{2}^\mu-\gamma v_{1}^\mu}{\gamma^{2}-1}\, , \quad v_{i}\cdot \bar v_{j}=\delta_{ij}\, .
\ee
While the invariances $\{ \mathcal{Q}, v_{i}^{\mu} \} =\{ \mathcal{Q}, S_{i}^{\mu} \}=0$ are already manifest,
the invariance $\{ \mathcal{Q}, b^{\mu} \} =0$ must be confirmed explicitly.
It follows from the transformation of the SSC vector:
\begin{align}
 \frac{\delta Z_{i}^{\mu}}{m_{i}}&= -\delta b_{i}^{\mu}+\beta_{i} v_{i}^{\mu}- \i v_{i}^{\mu} (\epsilon_{i} \bar\alpha_{i}\cdot v_{i} - \bar\epsilon_{i} \alpha_{i}\cdot v_{i} )\,. 
\end{align}
In the variation of $b^{\mu}$
from \eqn{eq:GeneralisedIP}, the variations $\delta b_{2}^{\mu}
-\delta b_{1}^{\mu}$ are canceled by $\delta Z_{2}^{\mu}-\delta Z_{1}^{\mu}$,
while the terms proportional to $v_{i}^{\mu}$ in $\delta Z_{i}^{\mu}$ are projected out by $P_{\mu\nu}$.
In fact, the covariant BUSY gauge choice $\alpha_{i}\cdot v_{i}=0=
\bar\alpha_{i}\cdot v_{i}$ sets the SSC vectors $Z_{i}^\mu=S_{i}^{\mu\nu}v_{i,\nu}$ to zero.
Notice that upon imposing the usual constraints, $(b_{2}-b_{1})\cdot v_{i}=0=Z^{\mu}_{i}$ the invariant impact parameter $b^{\mu}$ of \eqn{eq:GeneralisedIP} reduces to
the familiar $b^{\mu}_{2}-b^{\mu}_{1}$.

Hence, all observables will depend only on the BUSY-invariant vectors $\{v_{i}^{\mu}, b^{\mu}, S_{i}^{\mu} \}$
and the invariant scalars 
\begin{align}
\{m_{i}, \gamma, b^{2}, b\cdot S_{i}, S_{i}\cdot S_{j}, v_{1}\cdot S_{2},
v_{2}\cdot S_{1} \}\, .
\end{align}
Using eqs.~\eqref{PBbv},
we may straightforwardly determine the Poisson brackets of these vectors as ($i,j=1,2$)
\begin{subequations} \label{PBgi}
\begin{flalign}
\{ b^{\mu}, v_{i}^{\nu}\} 
&=
-
\frac{(-1)^{i}}{m_{i}} P^{\mu\nu}\,  , 
\\
\{ S_{i}^{\mu}, S_{j}^{\nu}\} 
&= 
\delta_{ij} \epsilon^{\mu\nu v_{i} S_{i}} \, ,
\\
\{ b^{\mu}, S_{i}^{\nu}\} 
&=
\frac{(-1)^{i}}{m_{i}}P^{\mu S_{i}} v_{i}^{\nu}\, , 
\\
\label{25c}
\{ b^{\mu}, b^{\nu}\} 
&=
2b^{[\mu}\left ( \frac{\bar v_{2}^{\nu]}}{m_{2}} -  \frac{\bar v_{1}^{\nu]}}{m_{1}}\right ) 
+ 
P^\mu{}_\rho P^\nu{}_\sigma
\Biggl ( 
\frac{\epsilon^{\rho\sigma v_{1} S_{1}}}{m_{1}^{2}} + \frac{\epsilon^{\rho\sigma  v_{2} S_{2}}}{m_{2}^{2}} \, \Biggr )\, .
\end{flalign}
\end{subequations}
We have neither used the BUSY constraints $\alpha_{i}\cdot v_{i}=\bar\alpha_{i}\cdot v_{i}=0$
nor imposed an orthogonality  $v_{i}\cdot (b_{2}-b_{1})=0$ on the right-hand-side ---
only the mass-shell condition $v_{i}^{2}=1$ has been used,
which corresponds with our choice of proper time gauge.
Yet our result is fully equivalent to the Dirac bracket analysis of refs.~\cite{Gonzo:2024zxo,Akpinar:2025bkt} that
studied the constrained system with $Z_{i}^{\mu}=0$, $v_{i}^{2}=1$ and $v_{i}\cdot (b_{2}-b_{1})=0$ imposed\footnote{ For comparisons with~\cite{Gonzo:2024zxo,Akpinar:2025bkt,Alessio:2025flu}, note that
$[b^{\mu}b^{\nu}+ l^{\mu}l^{\nu}/(\gamma^{2}-1)]/b^{2} = - P^{\mu\nu}$, where $l^{\mu}:=\epsilon^{\mu b v_{1} v_{2}}$ is proportional to the orbital angular momentum vector.}.
Indeed, one can check that using the above brackets one has
\be
\{Z_{i}^{\mu}, b^{\nu}\} =  \{Z_{i}^{\mu}, v_{j}^{\nu}\} = \{Z_{i}^{\mu}, S_{j}^{\nu}\} = 0
\,  ,
\ee
which is consistent with the invariance of $b^{\mu},v^{\mu}_{i}, S^{\mu}_{i}$.

The fact that we have not applied any constraints implies that our brackets are entirely
equivalent to those in \eqn{PBbv}, and we may continue to use them freely.
We may also now perform perturbative calculation of all observables with the \emph{constrained} variables,
i.e.~imposing $\alpha_{i}\cdot v_{i}=\bar\alpha_{i}\cdot v_{i}=v_{i}\cdot (b_{2}-b_{1})=0$ throughout.
Variables may then be lifted in the final result to their appropriate gauge-invariant partners:
\begin{subequations} \label{Free2Gauge}
\begin{flalign}
b_{2}^{\mu}-b_{1}^\mu &\to b^{\mu}=P^{\mu}{}_{{}\nu}(b_{2}^{\nu}-b_{1}^{\nu}+Z_{2}^{\nu}-Z_{1}^{\nu})\, ,
\\
\s{2}2m_{i}\i\, \bar\alpha^{[\mu}_{i}\alpha^{\nu]}_{i} &\to
 S_{i}^{\mu\nu} = \epsilon^{\mu\nu \rho\sigma}v_{i,\rho}S_{i,\sigma} + m_{i}Z_{i}^{\mu}{v_{i}}^{\nu} -  m_{i}Z_{i}^{\nu}{v_{i}}^{\mu}\, ,
\end{flalign}
\end{subequations}
and thereafter manipulated using the unconstrained brackets of \eqn{PBbv}.
This gives considerable flexibility for our subsequent analysis.

\section{Observables from WQFT and the on-shell action}\label{sec:diagObservables}

The spinning WQFT action~\eqref{eq:sWQFT2} gives rise to propagators for the deflection
modes $z^{\mu}_{i}(\omega)$, $\alpha_{i}^{\prime\mu}(\omega)$ and $\bar\alpha_{i}^{\prime\mu}(\omega)$ introduced via the background field expansion~\eqref{backgroundExpansions}.
These we conveniently combine into a composite vector \cite{Haddad:2024ebn}:\footnote{We do not distinguish between locally flat and curved spacetime indices in this section, as we are perturbing around Minkowski space. All indices maybe therefore be considered as flat.}
\begin{align}\label{Zcaldef}
 \Z^{\mu}_{I}(\omega)=\{z^{\mu}(\omega), \alpha^{\prime \mu}(\omega),
\bar\alpha^{\prime \mu}(\omega)\}
\, ,
\end{align}
with flavour index $I=1,2,3$.
The retarded propagator of this flavoured worldline field is 
\begin{align}\label{WLprops}
 \begin{tikzpicture}[baseline={(current bounding box.center)}]
    \coordinate (in) at (-0.6,0);
    \coordinate (out) at (1.4,0);
    \coordinate (x) at (-.2,0);
    \coordinate (y) at (1.0,0);
    \draw [zParticle] (x) -- (y) node [midway, below] {$\omega$};
    \draw [worldlineStatic] (in) -- (x);
    \draw [worldlineStatic] (y) -- (out);
    %\draw [dotted] (in) -- (x);
    %\draw [dotted] (y) -- (out);
    \draw [fill] (x) circle (.06) node [above] {$\mu$} node [below] {$I$};
    \draw [fill] (y) circle (.06) node [above] {$\nu$} node [below] {$J$};
  \end{tikzpicture}=%\langle \Z_{I}^{\mu}(-\omega) \Z^{\nu}_{J}(\omega) \rangle_{0}=
-\i\frac{\eta^{\mu\nu}}{m}
 \left (\begin{matrix}
  \frac1{(\omega+\i0^{+})^{2}} & 0 & 0 \\
  0 & 0 &   -\frac1{\omega+\i 0^{+}} \\
0 & \frac1{\omega+\i0^{+}} & 0
 \end{matrix}\right )_{IJ}.
\end{align}
As dictated by causality and the in-in formalism \cite{Jakobsen:2022psy,Kalin:2022hph},
we exclusively use retarded worldline propagators.
The arrow indicates causality flow and
the energy $\omega$ travels in the same direction.
The Feynman vertices emerging from \eqn{eq:sWQFT2} have been
collected in refs.~\cite{Mogull:2020sak,Driesse:2024feo} for the non-spinning case and in ref.~\cite{Haddad:2024ebn} for the spinning case.
On the worldline they take the generic form 
\begin{align}\label{eq:WLvertexrule}
%\begin{aligned}
\raisebox{-0pt}{$V_{m|n} =$\!\!\!
\resizebox{0.2\textwidth}{!}{
\begin{tikzpicture}[baseline={(current bounding box.center)},scale=1.4]
  \coordinate (in) at (-1,0);
  \coordinate (out1) at (1,0);
  \coordinate (out2) at (1,0.5);
  \coordinate (out3) at (1,0.9);
  \coordinate (out4) at (1,1.4);
  \coordinate (x) at (0,0);
   \node (k1) at (-.8,-1.3) [below] {$k_{1}$};
    \node (k2) at (0,-1.2) {};
    \node (k3) at (.8,-1.3) [below] {$k_{l}$};
  \draw (out2) node [right] {$\!\!\!\vdots$};
  \draw (out4) node [right] {$\omega_{n}$};
   \draw (out3) node [right] {$\omega_{n-1}$};
   \draw (out1) node [right] {$\omega_{1}$};
  \draw [worldlineStatic] (in) -- (x);
  \draw [zUndirected] (x) -- (out1);
  \draw [zUndirected] (x) to[out=30,in=180] (out3);
  \draw [zUndirected] (x) to[out=60,in=180] (out4);
   \draw [graviton] (x) -- (k1);
    \draw (k2) node [above] {$\dots$};
    \draw [graviton] (x) -- (k3);
  \draw [fill] (x) circle (.06);
  \end{tikzpicture}}$\!\!\!\sim$ }
  &
   m {\sqrt{G}}^{\,l}\, e^{\i k\cdot b} \delta\bigg(k\cdot 
    v+\sum_{i=1}^n\omega_i\bigg) 
%    \\[-35pt] 
%    & \times 
   ( \text{polynomial in $\omega_{i},k_{j}$}
   )\,,
%\end{aligned}
\end{align}
coupling $m$ gravitons to $n$ worldline deflections,
where $k^{\mu}=\sum_{j=1}^{l}k_{j}^{\mu}$ is the total outflowing four-momentum.
The dotted line symbolises the black hole background parameters.

These Feynman rules involving the worldlines are taken in combination with bulk gravitons,
whose dynamics arise from the usual Einstein-Hilbert action.
The graviton propagator is denoted by a wavy line; in de Donder gauge,
\begin{align}\label{eq:gravProp}
	\begin{tikzpicture}[baseline={(current bounding box.center)}]
	\coordinate (x) at (-.7,0);
	\coordinate (y) at (0.5,0);
	\draw [graviton] (x) -- (y) node [midway, below] {$k$} ;
	\draw [fill] (x) circle (.06) node [above] {$\mu\nu$};
	\draw [fill] (y) circle (.06) node [above] {$\alpha\beta$};
	\end{tikzpicture}
        &=\frac{\i  P_{\mu\nu;\alpha\beta}
          }{k^{2}+ \text{sgn}{(k^{0})}\i 0^{+}}\, , \qquad
          P_{\mu\nu;\alpha\beta}:=\eta_{\mu(\alpha}\eta_{\beta)\nu}-\sfrac12\eta_{\mu\nu}\eta_{\alpha\beta}\, ,
          \end{align}
We justify restricting the propagator to four dimensions in the discussion under \cref{eq:LoopMeasure}. 
Again, due to the in-in formalism this has a retarded $\iO$ prescription
\cite{Jakobsen:2022psy,Kalin:2022hph}.
Self-interaction vertices of arbitrary graviton multiplicities are also present,
as determined by the Einstein-Hilbert action ---
see ref.~\cite{Driesse:2024feo} for explicit expressions up to six gravitons.

\subsection{Observables from WQFT}

Observables emerge from tree-level one-point functions of the dynamical perturbations in the WQFT action \cite{Mogull:2020sak,Jakobsen:2021smu,Jakobsen:2021lvp,Jakobsen:2021zvh}.
Specifically, the linear impulse and spin kick depend on the worldline and oscillator perturbations respectively through 
\begin{subequations}\label{eq:ImpulseSpinKick}
\begin{align}
    \Delta p^{\mu}_{i}&=-m_{i}\omega^{2}\left.\langle z^{\mu}_{i}(\omega)\rangle\right|_{\omega\rightarrow0}, \\
    \Delta S^{\mu\nu}_{i}&=-\s{2}2m_i\omega\left[\langle\bar\alpha^{\prime\,[\mu}_{i}(\omega)\rangle\alpha^{\nu]}_{i }+\bar{\alpha}^{[\mu}_{i }\langle\alpha^{\prime\,\nu]}_{i}(\omega)
    \rangle
    +\i\omega\langle\bar\alpha^{\prime\,[\mu}_{i}(\omega)\rangle\langle\alpha^{\prime\,\nu]}_{i}(\omega)\rangle\right]_{\omega\rightarrow0}.
\end{align}
\end{subequations}
In a two-body context, the one-point function of the composite worldline deflection
vector $\langle \Z_{I,i}^{\mu}(\omega)\rangle$ of \eqn{Zcaldef} is computed perturbatively via the diagram \cite{Mogull:2020sak}
\begin{align}\label{eq:1PtFns}
    \langle \Z_{I,i}^\mu(\omega)\rangle=\vcenter{\hbox{\begin{tikzpicture}[baseline={(current bounding box.center)}]
        \node (A1) at (-1.33,.0) {$\Z_{I,i}^\mu(\omega)$};
        %\node (A2) at (-2.5,-.3) {$\omega\rightarrow$};
        \path [draw=black, worldlineStatic] (-4.7,0) -- (-3.83,0);
        \path [draw=black, zParticle] (-3.17,0) -- (-2.2,0);
        \path [draw=black, worldlineStatic] (-4.7,-1.5) -- (-3.83,-1.5);
        \path [draw=black, worldlineStatic] (-3.17,-1.5) -- (-2.2,-1.5);
        %\path [draw=black, zUndirected] (-4.6,0) -- (-3.4,0);
        \draw[pattern=north east lines] (-3.5,-0.75) ellipse (.55cm and .9cm);
    \end{tikzpicture}}},
\end{align}
which represents a sum over all possible connected tree-level diagrams.
The factors of $\omega$ in \cref{eq:ImpulseSpinKick} cancel the propagators of the external perturbations, and thus amount to amputations of the outgoing leg.
Within the ``blob’’, all propagators are directed from the (non-fluctuating) worldlines to the outgoing $\mathcal{Z}^{\mu}_{I,i}(\omega)$ line, encoding the causality flow within each diagram.

\Cref{eq:1PtFns} provides a direct means for accessing scattering observables.
However, a unified object exists from which we can extract quantities of interest: the on-shell action.
Working backwards from \cref{eq:1PtFns} -- by adapting \cref{PBbv} to collections of diagrams -- will enable us to work out a diagrammatic representation of the on-shell action for WQFT up to the formal 3PM order.
The former's unambiguous causality flow will be inherited by the latter.

\subsection{Brackets on diagrams and causality cuts}

Let us now explain the interplay between Poisson brackets and WQFT Feynman diagrams.
We begin with the realisation, previously made in refs.~\cite{Mogull:2020sak,Jakobsen:2021zvh},
that the WQFT vertices satisfy iterative relationships:
\begin{subequations}
\begin{align}
\allowdisplaybreaks
  \left[\quad\begin{tikzpicture}[baseline={(current bounding box.center)},scale=.9]
  \coordinate (in) at (-1,0);
  \coordinate (out1) at (1,0);
  \coordinate (out2) at (1,0.5);
  \coordinate (out3) at (1,0.9);
  \coordinate (out4) at (1,1.4);
  \coordinate (x) at (0,0);
   \node (k1) at (-.8,-1.3) {};
    \node (k2) at (0,-1.2) {};
    \node (k3) at (.8,-1.3) {};
  \draw (out2) node [right] {$\!\!\!\vdots$};
  \draw (out4) node [right] {$z_i^\mu(\omega)$};
  \draw [worldlineStatic] (in) -- (x);
  \draw [zUndirected] (x) -- (out1);
  \draw [zUndirected] (x) to[out=30,in=180] (out3);
  \draw [zUndirected] (x) to[out=60,in=180] (out4);
   \draw [graviton] (x) -- (k1);
    \draw (k2) node [above] {$\dots$};
    \draw [graviton] (x) -- (k3);
  \draw [fill] (x) circle (.06);
  \end{tikzpicture}\right]_{\omega=0}&=
  \frac{\partial}{\partial b_i^{\mu}}\quad\,
  \begin{tikzpicture}[baseline={(current bounding box.center)},scale=.9]
    \coordinate (in) at (-1,0);
    \coordinate (out1) at (1,0);
    \coordinate (out2) at (1,0.5);
    \coordinate (out3) at (1,0.9);
    \coordinate (x) at (0,0);
     \node (k1) at (-.8,-1.3) {};
    \node (k2) at (0,-1.2) {};
    \node (k3) at (.8,-1.3) {};
    \draw (out1) node [right] {};
    \draw (out2) node [right] {$\!\!\!\vdots$};
    \draw (out3) node [right] {};
    \draw [worldlineStatic] (in) -- (x);
    \draw [zUndirected] (x) -- (out1);
    \draw [zUndirected] (x) to[out=30,in=180] (out3);
     \draw [graviton] (x) -- (k1);
    \draw (k2) node [above] {$\dots$};
    \draw [graviton] (x) -- (k3);
    \draw [fill] (x) circle (.06);
    \end{tikzpicture}\,,  \\
    \left[\quad\begin{tikzpicture}[baseline={(current bounding box.center)},scale=.9]
  \coordinate (in) at (-1,0);
  \coordinate (out1) at (1,0);
  \coordinate (out2) at (1,0.5);
  \coordinate (out3) at (1,0.9);
  \coordinate (out4) at (1,1.4);
  \coordinate (x) at (0,0);
   \node (k1) at (-.8,-1.3) {};
    \node (k2) at (0,-1.2) {};
    \node (k3) at (.8,-1.3) {};
  \draw (out2) node [right] {$\!\!\!\vdots$};
  \draw (out4) node [right] {$\alpha_i^{\prime\mu}(\omega)$};
  \draw [worldlineStatic] (in) -- (x);
  \draw [zUndirected] (x) -- (out1);
  \draw [zUndirected] (x) to[out=30,in=180] (out3);
  \draw [zUndirected] (x) to[out=60,in=180] (out4);
   \draw [graviton] (x) -- (k1);
    \draw (k2) node [above] {$\dots$};
    \draw [graviton] (x) -- (k3);
  \draw [fill] (x) circle (.06);
  \end{tikzpicture}\right]_{\omega=0}&=
  \frac{\partial}{\partial \alpha_i^{\mu}}\quad\,
  \begin{tikzpicture}[baseline={(current bounding box.center)},scale=.9]
    \coordinate (in) at (-1,0);
    \coordinate (out1) at (1,0);
    \coordinate (out2) at (1,0.5);
    \coordinate (out3) at (1,0.9);
    \coordinate (x) at (0,0);
     \node (k1) at (-.8,-1.3) {};
    \node (k2) at (0,-1.2) {};
    \node (k3) at (.8,-1.3) {};
    \draw (out1) node [right] {};
    \draw (out2) node [right] {$\!\!\!\vdots$};
    \draw (out3) node [right] {};
    \draw [worldlineStatic] (in) -- (x);
    \draw [zUndirected] (x) -- (out1);
    \draw [zUndirected] (x) to[out=30,in=180] (out3);
    \draw [graviton] (x) -- (k1);
  \draw (k2) node [above] {$\dots$};
  \draw [graviton] (x) -- (k3);
    \draw [fill] (x) circle (.06);
    \end{tikzpicture}\,,\\
  \left[-\i\frac{\partial}{\partial\omega}\quad\begin{tikzpicture}[baseline={(current bounding box.center)},scale=.9]
  \coordinate (in) at (-1,0);
  \coordinate (out1) at (1,0);
  \coordinate (out2) at (1,0.5);
  \coordinate (out3) at (1,0.9);
  \coordinate (out4) at (1,1.4);
  \coordinate (x) at (0,0);
   \node (k1) at (-.8,-1.3) {};
    \node (k2) at (0,-1.2) {};
    \node (k3) at (.8,-1.3) {};
  \draw (out2) node [right] {$\!\!\!\vdots$};
  \draw (out4) node [right] {$z_i^{\mu}(\omega)$};
  \draw [worldlineStatic] (in) -- (x);
  \draw [zUndirected] (x) -- (out1);
  \draw [zUndirected] (x) to[out=30,in=180] (out3);
  \draw [zUndirected] (x) to[out=60,in=180] (out4);
  \draw [graviton] (x) -- (k1);
  \draw (k2) node [above] {$\dots$};
  \draw [graviton] (x) -- (k3);
  \draw [fill] (x) circle (.06);
  \end{tikzpicture}\right]_{\omega=0}&=
  \frac{\partial}{\partial v_i^{\mu}}\quad
  \begin{tikzpicture}[baseline={(current bounding box.center)},scale=.9]
    \coordinate (in) at (-1,0);
    \coordinate (out1) at (1,0);
    \coordinate (out2) at (1,0.5);
    \coordinate (out3) at (1,0.9);
    \coordinate (x) at (0,0);
    \node (k1) at (-.8,-1.3) {};
    \node (k2) at (0,-1.2) {};
    \node (k3) at (.8,-1.3) {};
    \draw (out2) node [right] {$\!\!\!\vdots$};
    \draw [worldlineStatic] (in) -- (x);
    \draw [zUndirected] (x) -- (out1);
    \draw [zUndirected] (x) to[out=30,in=180] (out3);
    \draw [graviton] (x) -- (k1);
    \draw (k2) node [above] {$\dots$};
    \draw [graviton] (x) -- (k3);
    \draw [fill] (x) circle (.06);
    \end{tikzpicture}\,.
\end{align}
\end{subequations}
The third relationship involving a derivative with respect to $v_i^\rho$, while not previously reported,
is easily checked for the WQFT Feynman rules in any particular theory (see, however, refs.~\cite{Jakobsen:2023oow,Haddad:2024ebn,Kim:2024svw} for similar relations). All of these properties follow directly from the dependence of the worldline action in the frequency domain on
$x_i^\mu(\omega)=\dd(\omega)b_i^\mu-\i\dd^{\,\prime}(\omega)v_i^\mu+z_i^\mu(\omega)$
and $\alpha_i^\mu(\omega)=\dd(\omega)\alpha_{i }^\mu+\alpha_i^{\prime\mu}(\omega)$.
As the action depends on the propagating fields $z_i^\mu(\omega)$ and $\alpha_i^{\prime\mu}(\omega)$
via their true dependence on $x_i^\mu(\omega)$ and $\alpha_i(\omega)$,
functional derivatives are equivalent to partial derivatives with respect to the corresponding 
background fields.
This logic is entirely equivalent to that used in \cref{sec:onShellAction}
to argue that Dirac brackets on quantum operators may be replaced by corresponding brackets on the backgrounds fields.

Using these vertex rules, we can understand how the brackets act on diagrams.
For this purpose, it is helpful to work with the unconstrained brackets~\eqref{PBbv}:
\begin{subequations}
\begin{align}
    \left\{\,\,\vcenter{\hbox{\begin{tikzpicture}[baseline={([yshift=-.5ex](.5,-.5))},scale=.6]
        \path [draw=black, worldlineStatic] (-4.7,0) -- (-3.83,0);
        \path [draw=black, worldlineStatic] (-3.17,0) -- (-2.2,0);
        \path [draw=black, worldlineStatic] (-4.7,-1.5) -- (-3.83,-1.5);
        \path [draw=black, worldlineStatic] (-3.17,-1.5) -- (-2.2,-1.5);
        %\path [draw=black, zUndirected] (-4.6,0) -- (-3.4,0);
        \draw[pattern=north east lines] (-3.5,-0.75) ellipse (.55cm and .9cm);
    \end{tikzpicture}}}\,\,,\,p_1^\mu\right\}&=
    \{b_1^\nu,p_1^\mu\}\,\,\,\,\frac{\partial}{\partial b_1^\nu}\,\vcenter{\hbox{\begin{tikzpicture}[baseline={([yshift=-.5ex](.5,-.5))},scale=.6]
        \path [draw=black, worldlineStatic] (-4.7,0) -- (-3.83,0);
        \path [draw=black, worldlineStatic] (-3.17,0) -- (-2.2,0);
        \path [draw=black, worldlineStatic] (-4.7,-1.5) -- (-3.83,-1.5);
        \path [draw=black, worldlineStatic] (-3.17,-1.5) -- (-2.2,-1.5);
        %\path [draw=black, zUndirected] (-4.6,0) -- (-3.4,0);
        \draw[pattern=north east lines] (-3.5,-0.75) ellipse (.55cm and .9cm);
    \end{tikzpicture}}}=\,\, -
    \left(\vcenter{\hbox{\begin{tikzpicture}[baseline={([yshift=-.5ex](.5,-.5))},scale=.6]
        \path [draw=black, worldlineStatic] (-4.7,0) -- (-3.83,0);
        \path [draw=black, zUndirected] (-3.17,0) -- (-2.2,0);
        \path [draw=black, worldlineStatic] (-4.7,-1.5) -- (-3.83,-1.5);
        \path [draw=black, worldlineStatic] (-3.17,-1.5) -- (-2.2,-1.5);
        %\path [draw=black, zUndirected] (-4.6,0) -- (-3.4,0);
        \draw[pattern=north east lines] (-3.5,-0.75) ellipse (.55cm and .9cm);
        \draw (-2.2,-.3) node [right] {$z_1^\mu(\omega)$};
    \end{tikzpicture}}}\right)_{\omega=0}\,,\\
    \left\{\,\,\vcenter{\hbox{\begin{tikzpicture}[baseline={([yshift=-.5ex](.5,-.5))},scale=.6]
        \path [draw=black, worldlineStatic] (-4.7,0) -- (-3.83,0);
        \path [draw=black, worldlineStatic] (-3.17,0) -- (-2.2,0);
        \path [draw=black, worldlineStatic] (-4.7,-1.5) -- (-3.83,-1.5);
        \path [draw=black, worldlineStatic] (-3.17,-1.5) -- (-2.2,-1.5);
        %\path [draw=black, zUndirected] (-4.6,0) -- (-3.4,0);
        \draw[pattern=north east lines] (-3.5,-0.75) ellipse (.55cm and .9cm);
    \end{tikzpicture}}}\,\,,\,\bar\alpha_1^\mu\right\}&=
    \{{\alpha}_1^\nu,\bar{\alpha}_1^\mu\}\,\,\frac{\partial}{\partial\alpha_1^\nu}\,\vcenter{\hbox{\begin{tikzpicture}[baseline={([yshift=-.5ex](.5,-.5))},scale=.6]
        \path [draw=black, worldlineStatic] (-4.7,0) -- (-3.83,0);
        \path [draw=black, worldlineStatic] (-3.17,0) -- (-2.2,0);
        \path [draw=black, worldlineStatic] (-4.7,-1.5) -- (-3.83,-1.5);
        \path [draw=black, worldlineStatic] (-3.17,-1.5) -- (-2.2,-1.5);
        %\path [draw=black, zUndirected] (-4.6,0) -- (-3.4,0);
        \draw[pattern=north east lines] (-3.5,-0.75) ellipse (.55cm and .9cm);
    \end{tikzpicture}}}=-\i
    \left(\vcenter{\hbox{\begin{tikzpicture}[baseline={([yshift=-.5ex](.5,-.5))},scale=.6]
        \path [draw=black, worldlineStatic] (-4.7,0) -- (-3.83,0);
        \path [draw=black, zUndirected] (-3.17,0) -- (-2.2,0);
        \path [draw=black, worldlineStatic] (-4.7,-1.5) -- (-3.83,-1.5);
        \path [draw=black, worldlineStatic] (-3.17,-1.5) -- (-2.2,-1.5);
        %\path [draw=black, zUndirected] (-4.6,0) -- (-3.4,0);
        \draw[pattern=north east lines] (-3.5,-0.75) ellipse (.55cm and .9cm);
        \draw (-2.2,-.3) node [right] {$\alpha_1^{\prime \mu}(\omega)$};
    \end{tikzpicture}}}\right)_{\omega=0}\,\, .
\end{align}
\end{subequations}
We can also determine the bracket between collections of diagrams.
For simplicity of the argument, let us focus only on the first worldline (WL1),
and restrict our attention to brackets involving $b_1^\mu$ and $v_1^\mu$ only ---
generalisations to include spin and the second worldline are straightforward and follow the same pattern.
The bracket between two collections of diagrams (A/B) is
\begin{align}
    &\left\{\,\,\vcenter{\hbox{\begin{tikzpicture}[baseline={([yshift=-.5ex](.5,-.5))},scale=.6]
        \path [draw=black, worldlineStatic] (-4.7,0) -- (-3.83,0);
        \path [draw=black, worldlineStatic] (-3.17,0) -- (-2.2,0);
        \path [draw=black, worldlineStatic] (-4.7,-1.5) -- (-3.83,-1.5);
        \path [draw=black, worldlineStatic] (-3.17,-1.5) -- (-2.2,-1.5);
        %\path [draw=black, zUndirected] (-4.6,0) -- (-3.4,0);
        \draw[pattern=north east lines] (-3.5,-0.75) ellipse (.55cm and .9cm);
        \filldraw[fill=white,color=white] (-3.5,-0.75) circle (9pt);
        \node (middle) at (-3.5,-0.75) {A}; 
    \end{tikzpicture}}}\,\,,\,\,
    \vcenter{\hbox{\begin{tikzpicture}[baseline={([yshift=-.5ex](.5,-.5))},scale=.6]
        \path [draw=black, worldlineStatic] (-4.7,0) -- (-3.83,0);
        \path [draw=black, worldlineStatic] (-3.17,0) -- (-2.2,0);
        \path [draw=black, worldlineStatic] (-4.7,-1.5) -- (-3.83,-1.5);
        \path [draw=black, worldlineStatic] (-3.17,-1.5) -- (-2.2,-1.5);
        %\path [draw=black, zUndirected] (-4.6,0) -- (-3.4,0);
        \draw[pattern=north east lines] (-3.5,-0.75) ellipse (.55cm and .9cm);
        \filldraw[fill=white,color=white] (-3.5,-0.75) circle (9pt);
        \node (middle) at (-3.5,-0.75) {B}; 
    \end{tikzpicture}}}\right\}_{\rm{WL1}}\\
    &=\{b_1^\mu,v_1^\nu\}\left[
    \left(\frac{\partial}{\partial b_1^\mu}\,\vcenter{\hbox{\begin{tikzpicture}[baseline={([yshift=-.5ex](.5,-.5))},scale=.6]
        \path [draw=black, worldlineStatic] (-4.7,0) -- (-3.83,0);
        \path [draw=black, worldlineStatic] (-3.17,0) -- (-2.2,0);
        \path [draw=black, worldlineStatic] (-4.7,-1.5) -- (-3.83,-1.5);
        \path [draw=black, worldlineStatic] (-3.17,-1.5) -- (-2.2,-1.5);
        %\path [draw=black, zUndirected] (-4.6,0) -- (-3.4,0);
        \draw[pattern=north east lines] (-3.5,-0.75) ellipse (.55cm and .9cm);
        \filldraw[fill=white,color=white] (-3.5,-0.75) circle (9pt);
        \node (middle) at (-3.5,-0.75) {A}; 
    \end{tikzpicture}}}\right)
    \left(\frac{\partial}{\partial v_1^\nu}\,\vcenter{\hbox{\begin{tikzpicture}[baseline={([yshift=-.5ex](.5,-.5))},scale=.6]
        \path [draw=black, worldlineStatic] (-4.7,0) -- (-3.83,0);
        \path [draw=black, worldlineStatic] (-3.17,0) -- (-2.2,0);
        \path [draw=black, worldlineStatic] (-4.7,-1.5) -- (-3.83,-1.5);
        \path [draw=black, worldlineStatic] (-3.17,-1.5) -- (-2.2,-1.5);
        %\path [draw=black, zUndirected] (-4.6,0) -- (-3.4,0);
        \draw[pattern=north east lines] (-3.5,-0.75) ellipse (.55cm and .9cm);
        \filldraw[fill=white,color=white] (-3.5,-0.75) circle (9pt);
        \node (middle) at (-3.5,-0.75) {B}; 
    \end{tikzpicture}}}\right)-
    \left(\frac{\partial}{\partial v_1^\mu}\,\vcenter{\hbox{\begin{tikzpicture}[baseline={([yshift=-.5ex](.5,-.5))},scale=.6]
        \path [draw=black, worldlineStatic] (-4.7,0) -- (-3.83,0);
        \path [draw=black, worldlineStatic] (-3.17,0) -- (-2.2,0);
        \path [draw=black, worldlineStatic] (-4.7,-1.5) -- (-3.83,-1.5);
        \path [draw=black, worldlineStatic] (-3.17,-1.5) -- (-2.2,-1.5);
        %\path [draw=black, zUndirected] (-4.6,0) -- (-3.4,0);
        \draw[pattern=north east lines] (-3.5,-0.75) ellipse (.55cm and .9cm);
        \filldraw[fill=white,color=white] (-3.5,-0.75) circle (9pt);
        \node (middle) at (-3.5,-0.75) {A}; 
    \end{tikzpicture}}}\right)
    \left(\frac{\partial}{\partial b_1^\nu}\,\vcenter{\hbox{\begin{tikzpicture}[baseline={([yshift=-.5ex](.5,-.5))},scale=.6]
        \path [draw=black, worldlineStatic] (-4.7,0) -- (-3.83,0);
        \path [draw=black, worldlineStatic] (-3.17,0) -- (-2.2,0);
        \path [draw=black, worldlineStatic] (-4.7,-1.5) -- (-3.83,-1.5);
        \path [draw=black, worldlineStatic] (-3.17,-1.5) -- (-2.2,-1.5);
        %\path [draw=black, zUndirected] (-4.6,0) -- (-3.4,0);
        \draw[pattern=north east lines] (-3.5,-0.75) ellipse (.55cm and .9cm);
        \filldraw[fill=white,color=white] (-3.5,-0.75) circle (9pt);
        \node (middle) at (-3.5,-0.75) {B}; 
    \end{tikzpicture}}}\right)\right]\nn\\
    &=-\frac{\eta_{\mu\nu}}{m_1}\left[
    \left(\,\vcenter{\hbox{\begin{tikzpicture}[baseline={([yshift=-.5ex](.5,-.5))},scale=.6]
        \path [draw=black, worldlineStatic] (-4.7,0) -- (-3.83,0);
        \path [draw=black, zUndirected] (-3.17,0) -- (-2.2,0);
        \path [draw=black, worldlineStatic] (-4.7,-1.5) -- (-3.83,-1.5);
        \path [draw=black, worldlineStatic] (-3.17,-1.5) -- (-2.2,-1.5);
        %\path [draw=black, zUndirected] (-4.6,0) -- (-3.4,0);
        \draw[pattern=north east lines] (-3.5,-0.75) ellipse (.55cm and .9cm);
        \draw (-2.2,-0.2) node [right] {$z_1^\mu(\omega)$};
        \filldraw[fill=white,color=white] (-3.5,-0.75) circle (9pt);
        \node (middle) at (-3.5,-0.75) {A}; 
    \end{tikzpicture}}}\right)
    \left(-\i\frac{\partial}{\partial(-\omega)}
    \vcenter{\hbox{\begin{tikzpicture}[baseline={([yshift=-.5ex](.5,-.5))},scale=.6]
        \path [draw=black, zUndirected] (-4.7,0) -- (-3.83,0);
        \path [draw=black, worldlineStatic] (-3.17,0) -- (-2.2,0);
        \path [draw=black, worldlineStatic] (-4.7,-1.5) -- (-3.83,-1.5);
        \path [draw=black, worldlineStatic] (-3.17,-1.5) -- (-2.2,-1.5);
        %\path [draw=black, zUndirected] (-4.6,0) -- (-3.4,0);
        \draw[pattern=north east lines] (-3.5,-0.75) ellipse (.55cm and .9cm);
        \draw (-4.7,-0.2) node [left] {$z_1^\nu(-\omega)$};
        \filldraw[fill=white,color=white] (-3.5,-0.75) circle (9pt);
        \node (middle) at (-3.5,-0.75) {B}; 
    \end{tikzpicture}}}\right)\right.\nn\\
    &\qquad\qquad-\left.
     \left(-\i\frac{\partial}{\partial\omega}\,\vcenter{\hbox{\begin{tikzpicture}[baseline={([yshift=-.5ex](.5,-.5))},scale=.6]
        \path [draw=black, worldlineStatic] (-4.7,0) -- (-3.83,0);
        \path [draw=black, zUndirected] (-3.17,0) -- (-2.2,0);
        \path [draw=black, worldlineStatic] (-4.7,-1.5) -- (-3.83,-1.5);
        \path [draw=black, worldlineStatic] (-3.17,-1.5) -- (-2.2,-1.5);
        %\path [draw=black, zUndirected] (-4.6,0) -- (-3.4,0);
        \draw[pattern=north east lines] (-3.5,-0.75) ellipse (.55cm and .9cm);
        \draw (-2.2,-0.2) node [right] {$z_1^\mu(\omega)$};
        \filldraw[fill=white,color=white] (-3.5,-0.75) circle (9pt);
        \node (middle) at (-3.5,-0.75) {A}; 
    \end{tikzpicture}}}\right)
    \left(
    \vcenter{\hbox{\begin{tikzpicture}[baseline={([yshift=-.5ex](.5,-.5))},scale=.6]
        \path [draw=black, zUndirected] (-4.7,0) -- (-3.83,0);
        \path [draw=black, worldlineStatic] (-3.17,0) -- (-2.2,0);
        \path [draw=black, worldlineStatic] (-4.7,-1.5) -- (-3.83,-1.5);
        \path [draw=black, worldlineStatic] (-3.17,-1.5) -- (-2.2,-1.5);
        %\path [draw=black, zUndirected] (-4.6,0) -- (-3.4,0);
        \draw[pattern=north east lines] (-3.5,-0.75) ellipse (.55cm and .9cm);
        \draw (-4.7,-0.2) node [left] {$z_1^\nu(-\omega)$};
        \filldraw[fill=white,color=white] (-3.5,-0.75) circle (9pt);
        \node (middle) at (-3.5,-0.75) {B}; 
    \end{tikzpicture}}}\right)\right]_{\omega=0}\nn\\
    &=\frac{\i}{m_1}\eta_{\mu\nu}\int_\omega\dd(\omega)\left[
    -\left(\,\vcenter{\hbox{\begin{tikzpicture}[baseline={([yshift=-.5ex](.5,-.5))},scale=.6]
        \path [draw=black, worldlineStatic] (-4.7,0) -- (-3.83,0);
        \path [draw=black, zUndirected] (-3.17,0) -- (-2.2,0);
        \path [draw=black, worldlineStatic] (-4.7,-1.5) -- (-3.83,-1.5);
        \path [draw=black, worldlineStatic] (-3.17,-1.5) -- (-2.2,-1.5);
        %\path [draw=black, zUndirected] (-4.6,0) -- (-3.4,0);
        \draw[pattern=north east lines] (-3.5,-0.75) ellipse (.55cm and .9cm);
        \draw (-2.2,-0.2) node [right] {$z_1^\mu(\omega)$};
        \filldraw[fill=white,color=white] (-3.5,-0.75) circle (9pt);
        \node (middle) at (-3.5,-0.75) {A}; 
    \end{tikzpicture}}}\right)
    \left(\frac{\partial}{\partial\omega}
    \vcenter{\hbox{\begin{tikzpicture}[baseline={([yshift=-.5ex](.5,-.5))},scale=.6]
        \path [draw=black, zUndirected] (-4.7,0) -- (-3.83,0);
        \path [draw=black, worldlineStatic] (-3.17,0) -- (-2.2,0);
        \path [draw=black, worldlineStatic] (-4.7,-1.5) -- (-3.83,-1.5);
        \path [draw=black, worldlineStatic] (-3.17,-1.5) -- (-2.2,-1.5);
        %\path [draw=black, zUndirected] (-4.6,0) -- (-3.4,0);
        \draw[pattern=north east lines] (-3.5,-0.75) ellipse (.55cm and .9cm);
        \draw (-4.7,-0.2) node [left] {$z_1^\nu(-\omega)$};
        \filldraw[fill=white,color=white] (-3.5,-0.75) circle (9pt);
        \node (middle) at (-3.5,-0.75) {B}; 
    \end{tikzpicture}}}\right)\right.\nn\\
    &\qquad\qquad\qquad\qquad\quad-\left.
     \left(\frac{\partial}{\partial\omega}\,\vcenter{\hbox{\begin{tikzpicture}[baseline={([yshift=-.5ex](.5,-.5))},scale=.6]
        \path [draw=black, worldlineStatic] (-4.7,0) -- (-3.83,0);
        \path [draw=black, zUndirected] (-3.17,0) -- (-2.2,0);
        \path [draw=black, worldlineStatic] (-4.7,-1.5) -- (-3.83,-1.5);
        \path [draw=black, worldlineStatic] (-3.17,-1.5) -- (-2.2,-1.5);
        %\path [draw=black, zUndirected] (-4.6,0) -- (-3.4,0);
        \draw[pattern=north east lines] (-3.5,-0.75) ellipse (.55cm and .9cm);
        \draw (-2.2,-0.2) node [right] {$z_1^\mu(\omega)$};
        \filldraw[fill=white,color=white] (-3.5,-0.75) circle (9pt);
        \node (middle) at (-3.5,-0.75) {A}; 
    \end{tikzpicture}}}\right)
    \left(
    \vcenter{\hbox{\begin{tikzpicture}[baseline={([yshift=-.5ex](.5,-.5))},scale=.6]
        \path [draw=black, zUndirected] (-4.7,0) -- (-3.83,0);
        \path [draw=black, worldlineStatic] (-3.17,0) -- (-2.2,0);
        \path [draw=black, worldlineStatic] (-4.7,-1.5) -- (-3.83,-1.5);
        \path [draw=black, worldlineStatic] (-3.17,-1.5) -- (-2.2,-1.5);
        %\path [draw=black, zUndirected] (-4.6,0) -- (-3.4,0);
        \draw[pattern=north east lines] (-3.5,-0.75) ellipse (.55cm and .9cm);
        \draw (-4.7,-0.2) node [left] {$z_1^\nu(-\omega)$};
        \filldraw[fill=white,color=white] (-3.5,-0.75) circle (9pt);
        \node (middle) at (-3.5,-0.75) {B}; 
    \end{tikzpicture}}}\right)\right]\nn\\
    &=\frac{\i}{m_1}\eta_{\mu\nu}\int_\omega\dd^{\,\prime}(\omega)
    \left(\,\vcenter{\hbox{\begin{tikzpicture}[baseline={([yshift=-.5ex](.5,-.5))},scale=.6]
        \path [draw=black, worldlineStatic] (-4.7,0) -- (-3.83,0);
        \path [draw=black, zUndirected] (-3.17,0) -- (-2.2,0);
        \path [draw=black, worldlineStatic] (-4.7,-1.5) -- (-3.83,-1.5);
        \path [draw=black, worldlineStatic] (-3.17,-1.5) -- (-2.2,-1.5);
        %\path [draw=black, zUndirected] (-4.6,0) -- (-3.4,0);
        \draw[pattern=north east lines] (-3.5,-0.75) ellipse (.55cm and .9cm);
        \draw (-2.2,-0.2) node [right] {$z_1^\mu(\omega)$};
        \filldraw[fill=white,color=white] (-3.5,-0.75) circle (9pt);
        \node (middle) at (-3.5,-0.75) {A}; 
    \end{tikzpicture}}}\right)
    \left(
    \vcenter{\hbox{\begin{tikzpicture}[baseline={([yshift=-.5ex](.5,-.5))},scale=.6]
        \path [draw=black, zUndirected] (-4.7,0) -- (-3.83,0);
        \path [draw=black, worldlineStatic] (-3.17,0) -- (-2.2,0);
        \path [draw=black, worldlineStatic] (-4.7,-1.5) -- (-3.83,-1.5);
        \path [draw=black, worldlineStatic] (-3.17,-1.5) -- (-2.2,-1.5);
        %\path [draw=black, zUndirected] (-4.6,0) -- (-3.4,0);
        \draw[pattern=north east lines] (-3.5,-0.75) ellipse (.55cm and .9cm);
        \draw (-4.7,0) node [left] {$z_1^\nu(-\omega)$};
        \filldraw[fill=white,color=white] (-3.5,-0.75) circle (9pt);
        \node (middle) at (-3.5,-0.75) {B}; 
    \end{tikzpicture}}}\right)\,.\nn
\end{align}
In the second to last step we have inserted an integral over the energy $\omega$ with $\delta(\omega)$, using
\be
\int_{\omega}\ldots := \int_{-\infty}^{\infty} \frac{d\omega}{2\pi}\ldots\, , 
\qquad \dd(\omega):=2\pi\, \delta(\omega)\, ,
\ee
which ensures the condition $\omega=0$, and then integrated by parts.
The result is a connection between the two diagrams, with a cut worldline propagator
integrated over its energy.
This can also be interpreted as a difference of propagators with opposite $\iO$ prescriptions, as
\begin{align}
    \dd^{\,\prime}(\omega)=-\i\left(\frac1{(\omega+\iO)^2}-\frac1{(\omega-\iO)^2}\right)\,.
\end{align}
This cut propagator generalises naturally to include the spin lines with $\alpha_i^\mu$:
\begin{align}
 \begin{tikzpicture}[baseline={(0,-.1)}]
    \coordinate (in) at (-0.6,0);
    \coordinate (out) at (1.4,0);
    \coordinate (x) at (-.2,0);
    \coordinate (y) at (1.0,0);
    \draw [worldlineCut2] (x) -- (y) node [midway, below] {$\omega$};
    \draw [worldlineStatic] (in) -- (x);
    \draw [worldlineStatic] (y) -- (out);
    \draw [fill] (x) circle (.06) node [above] {$\mu$} node [below] {$I$};
    \draw [fill] (y) circle (.06) node [above] {$\nu$} node [below] {$J$};
  \end{tikzpicture}=
   \begin{tikzpicture}[baseline={(0,-.1)}]
    \coordinate (in) at (-0.6,0);
    \coordinate (out) at (1.4,0);
    \coordinate (x) at (-.2,0);
    \coordinate (y) at (1.0,0);
    \draw [zParticle] (x) -- (y) node [midway, below] {$\omega$};
    \draw [worldlineStatic] (in) -- (x);
    \draw [worldlineStatic] (y) -- (out);
    \draw [fill] (x) circle (.06) node [above] {$\mu$} node [below] {$I$};
    \draw [fill] (y) circle (.06) node [above] {$\nu$} node [below] {$J$};
  \end{tikzpicture}
  -
   \begin{tikzpicture}[baseline={(0,-.1)}]
    \coordinate (in) at (-0.6,0);
    \coordinate (out) at (1.4,0);
    \coordinate (x) at (-.2,0);
    \coordinate (y) at (1.0,0);
    \draw [zParticle] (y) -- (x) node [midway, below] {$-\omega$};
    \draw [worldlineStatic] (in) -- (x);
    \draw [worldlineStatic] (y) -- (out);
    \draw [fill] (x) circle (.06) node [above] {$\mu$} node [below] {$I$};
    \draw [fill] (y) circle (.06) node [above] {$\nu$} node [below] {$J$};
  \end{tikzpicture}=
  %\langle \Z_{I}^{\mu}(-\omega) \Z^{\nu}_{J}(\omega) \rangle_{0}=
\frac{\eta^{\mu\nu}}{m}
 \left (\begin{matrix}
  \dd^{\,\prime}(\omega) & 0 & 0 \\
  0 & 0 &   -\dd(\omega) \\
  0 & \dd(\omega) & 0
 \end{matrix}\right )_{IJ} \, ,
\end{align}
Ultimately, we learn that the bracket between two arbitrary collections of diagrams is given by
\begin{align}\label{diagBracket}
\left\{\,\,\vcenter{\hbox{\begin{tikzpicture}[baseline={([yshift=-.5ex](.5,-.5))},scale=.6]
        \path [draw=black, worldlineStatic] (-4.7,0) -- (-3.83,0);
        \path [draw=black, worldlineStatic] (-3.17,0) -- (-2.2,0);
        \path [draw=black, worldlineStatic] (-4.7,-1.5) -- (-3.83,-1.5);
        \path [draw=black, worldlineStatic] (-3.17,-1.5) -- (-2.2,-1.5);
        %\path [draw=black, zUndirected] (-4.6,0) -- (-3.4,0);
        \draw[pattern=north east lines] (-3.5,-0.75) ellipse (.55cm and .9cm);
        % \coordinate (label) at (0,0);
        \filldraw[fill=white,color=white] (-3.5,-0.75) circle (9pt);
        \node (middle) at (-3.5,-0.75) {A}; 
    \end{tikzpicture}}}\,\,,\,\,
    \vcenter{\hbox{\begin{tikzpicture}[baseline={([yshift=-.5ex](.5,-.5))},scale=.6]
        \path [draw=black, worldlineStatic] (-4.7,0) -- (-3.83,0);
        \path [draw=black, worldlineStatic] (-3.17,0) -- (-2.2,0);
        \path [draw=black, worldlineStatic] (-4.7,-1.5) -- (-3.83,-1.5);
        \path [draw=black, worldlineStatic] (-3.17,-1.5) -- (-2.2,-1.5);
        %\path [draw=black, zUndirected] (-4.6,0) -- (-3.4,0);
        \draw[pattern=north east lines] (-3.5,-0.75) ellipse (.55cm and .9cm);
        \filldraw[fill=white,color=white] (-3.5,-0.75) circle (9pt);
        \node (middle) at (-3.5,-0.75) {B}; 
\end{tikzpicture}}}\right\}=\i\left(\,\vcenter{\hbox{\begin{tikzpicture}[baseline={([yshift=-.5ex](.5,-.5))},scale=.6]
        \path [draw=black, worldlineStatic] (-4.7,0) -- (-3.83,0);
        \path [draw=black, worldlineCut2] (-3.17,0) -- (-2.2,0);
        \path [draw=black, worldlineStatic] (-1.54,0) -- (-0.67,0);
        \path [draw=black, worldlineStatic] (-4.7,-1.5) -- (-3.83,-1.5);
        \path [draw=black, worldlineStatic] (-3.17,-1.5) -- (-2.2,-1.5);
        \path [draw=black, worldlineStatic] (-1.54,-1.5) -- (-0.67,-1.5);
        %\path [draw=black, zUndirected] (-4.6,0) -- (-3.4,0);
        \draw[pattern=north east lines] (-3.5,-0.75) ellipse (.55cm and .9cm);
        \draw[pattern=north east lines] (-1.87,-0.75) ellipse (.55cm and .9cm);
        \filldraw[fill=white,color=white] (-3.5,-0.75) circle (9pt);
        \node (A) at (-3.5,-0.75) {A};
        \filldraw[fill=white,color=white] (-1.87,-0.75) circle (9pt);
        \node (B) at (-1.87,-0.75) {B}; 
    \end{tikzpicture}}}\,+\,\vcenter{\hbox{\begin{tikzpicture}[baseline={([yshift=-.5ex](.5,-.5))},scale=.6]
        \path [draw=black, worldlineStatic] (-4.7,0) -- (-3.83,0);
        \path [draw=black, worldlineStatic] (-3.17,0) -- (-2.2,0);
        \path [draw=black, worldlineStatic] (-1.54,0) -- (-0.67,0);
        \path [draw=black, worldlineStatic] (-4.7,-1.5) -- (-3.83,-1.5);
        \path [draw=black, worldlineCut2] (-3.17,-1.5) -- (-2.2,-1.5);
        \path [draw=black, worldlineStatic] (-1.54,-1.5) -- (-0.67,-1.5);
        %\path [draw=black, zUndirected] (-4.6,0) -- (-3.4,0);
        \draw[pattern=north east lines] (-3.5,-0.75) ellipse (.55cm and .9cm);
        \draw[pattern=north east lines] (-1.87,-0.75) ellipse (.55cm and .9cm);
        \filldraw[fill=white,color=white] (-3.5,-0.75) circle (9pt);
        \node (A) at (-3.5,-0.75) {A};
        \filldraw[fill=white,color=white] (-1.87,-0.75) circle (9pt);
        \node (B) at (-1.87,-0.75) {B}; 
    \end{tikzpicture}}}\,\right)\,.
\end{align}
This involves summing over all possible ways the directed cut can be attached to both collections of graphs.
The relationship, known in ref.~\cite{Kim:2024svw} as a causality cut,
plays a crucial role in understanding how scattering observables
are derived from the on-shell action $N$.

\subsection{Observables from the on-shell action }\label{obsFromAction}

Now that we understand how the brackets act on collections of diagrams,
we can determine expressions for the conservative on-shell action $N$.
In general, these are given by sums of graphs without any outgoing lines ---
however, the relevant $\i 0^{+}$ prescriptions will be a crucial issue to resolve.
Up to 2PM order we have
\begin{align}
    \i N^{(1)}&=\,\,\begin{tikzpicture}[baseline={([yshift=-.5ex](.5,-.5))}]
        %%%%%%%%%%%%%%%%%%%%%%
        \path [draw=black, worldlineStatic] (-0.6,0) -- (0,0);
        \path [draw=black, worldlineStatic] (0,0) -- (0.6,0);
        \path [draw=black, worldlineStatic] (-0.6,-1.) -- (0.6,-1.);
        \path [draw=black, photon] (0,-1.) -- (0,0);
        \filldraw[fill=black] (0,0) circle (.06);
        \filldraw[fill=black] (0,-1.) circle (.06);
    \end{tikzpicture}\,,\label{eq:N1}\\
    \i N^{(2)}&=\,\,
        \frac12\,
        \begin{tikzpicture}[baseline={([yshift=-.5ex](.5,-3.5))}]
            %%%%%%%%%%%%%%%%%%%%%%
            \path [draw=black, worldlineStatic] (-3.,-3) -- (-1.,-3);
            \path [draw=black, worldlineStatic] (-3.,-4.) -- (-1.,-4.);
            \path [draw=black, photon] (-2.4,-4.) -- (-2,-3);
            \path [draw=black, photon] (-1.6,-4.) -- (-2,-3);
            \filldraw[fill=white, draw=black,thick] (-2.25,-3.2) rectangle (-1.75,-2.8);
            \node (A) at (-2,-3.02) {$\rightarrow$};
            \filldraw[fill=black] (-2.4,-4.) circle (.06);
            \filldraw[fill=black] (-1.6,-4.) circle (.06);
        \end{tikzpicture}
        +\frac12\,
        \begin{tikzpicture}[baseline={([yshift=-.5ex](.5,-3.5))}]
            %%%%%%%%%%%%%%%%%%%%%%
            \path [draw=black, worldlineStatic] (-3.,-3) -- (-1.,-3);
            \path [draw=black, worldlineStatic] (-3.,-4.) -- (-1.,-4.);
            \path [draw=black, photon] (-2.4,-3) -- (-2,-4);
            \path [draw=black, photon] (-1.6,-3) -- (-2,-4);
            \filldraw[fill=white, draw=black,thick] (-2.25,-4.2) rectangle (-1.75,-3.8);
            \node (A) at (-2,-4.02) {$\rightarrow$};
            \filldraw[fill=black] (-2.4,-3) circle (.06);
            \filldraw[fill=black] (-1.6,-3) circle (.06);
        \end{tikzpicture}\,.\label{eq:N2}
\end{align}
Here we have introduced collections of diagrams,
with the direction of contained worldline propagators denoted by an arrow:
\begin{subequations}\label{eq:ComptonDiagrams}
\begin{align}
        \begin{tikzpicture}[baseline={([yshift=-.5ex](.5,-3.5))}]
        %%%%%%%%%%%%%%%%%%%%%%
        \path [draw=black, worldlineStatic] (-2.8,-3) -- (-1.2,-3);
        \path [draw=black, photon] (-2.4,-4.) -- (-2,-3);
        \path [draw=black, photon] (-1.6,-4.) -- (-2,-3);
        \filldraw[fill=white, draw=black,thick] (-2.25,-3.2) rectangle (-1.75,-2.8);
        \node (A) at (-2,-3.02) {$\rightarrow$};
    \end{tikzpicture}&:=
        \begin{tikzpicture}[baseline={([yshift=-.5ex](.5,-3.5))}]
            %%%%%%%%%%%%%%%%%%%%%%
            \path [draw=black, worldlineStatic] (-2.8,-3) -- (-1.2,-3);
            \path [draw=black, photon] (-2.4,-4.) -- (-2,-3);
            \path [draw=black, photon] (-1.6,-4.) -- (-2,-3);
            \filldraw[fill=black] (-2,-3) circle (.06);
        \end{tikzpicture}
        +
        \begin{tikzpicture}[baseline={([yshift=-.5ex](.5,-3.5))}]
            %%%%%%%%%%%%%%%%%%%%%%
            \path [draw=black, worldlineStatic] (1.2,-3) -- (2.8,-3);
            \path [draw=black, photon] (2,-3.5) -- (2,-3);
            \path [draw=black, photon] (1.6,-4.) -- (2,-3.5);
            \path [draw=black, photon] (2.4,-4.) -- (2,-3.5);
            \filldraw[fill=black] (2,-3) circle (.06);
            \filldraw[fill=black] (2,-3.5) circle (.06);
        \end{tikzpicture}
        +
    \begin{tikzpicture}[baseline={([yshift=-.5ex](.5,-.5))}]
            %%%%%%%%%%%%%%%%%%%%%%
            \path [draw=black, worldlineStatic] (-0.8,0) -- (-0.2,0);
            \path [draw=black, worldlineStatic] (0.2,0) -- (0.8,0);
            \path [draw=black, zParticle2] (-.3,0) -- (.3,0);
            \path [draw=black, photon] (-0.3,-1.) -- (-0.3,0);
            \path [draw=black, photon] (0.3,-1.) -- (0.3,0);
            \filldraw[fill=black] (-0.3,0) circle (.06);
            \filldraw[fill=black] (0.3,0) circle (.06);
        \end{tikzpicture}\,, 
        \\
        \begin{tikzpicture}[baseline={([yshift=-.5ex](.5,-3.5))}]
        %%%%%%%%%%%%%%%%%%%%%%
        \path [draw=black, worldlineStatic] (-2.8,-3) -- (-1.2,-3);
        \path [draw=black, photon] (-2.4,-4.) -- (-2,-3);
        \path [draw=black, photon] (-1.6,-4.) -- (-2,-3);
        \filldraw[fill=white, draw=black,thick] (-2.25,-3.2) rectangle (-1.75,-2.8);
        \node (A) at (-2,-3.02) {$\leftarrow$};
    \end{tikzpicture}&:=
        \begin{tikzpicture}[baseline={([yshift=-.5ex](.5,-3.5))}]
            %%%%%%%%%%%%%%%%%%%%%%
            \path [draw=black, worldlineStatic] (-2.8,-3) -- (-1.2,-3);
            \path [draw=black, photon] (-2.4,-4.) -- (-2,-3);
            \path [draw=black, photon] (-1.6,-4.) -- (-2,-3);
            \filldraw[fill=black] (-2,-3) circle (.06);
        \end{tikzpicture}
        +
        \begin{tikzpicture}[baseline={([yshift=-.5ex](.5,-3.5))}]
            %%%%%%%%%%%%%%%%%%%%%%
            \path [draw=black, worldlineStatic] (1.2,-3) -- (2.8,-3);
            \path [draw=black, photon] (2,-3.5) -- (2,-3);
            \path [draw=black, photon] (1.6,-4.) -- (2,-3.5);
            \path [draw=black, photon] (2.4,-4.) -- (2,-3.5);
            \filldraw[fill=black] (2,-3) circle (.06);
            \filldraw[fill=black] (2,-3.5) circle (.06);
        \end{tikzpicture}
    +
        \begin{tikzpicture}[baseline={([yshift=-.5ex](.5,-.5))}]
            %%%%%%%%%%%%%%%%%%%%%%
            \path [draw=black, worldlineStatic] (-0.8,0) -- (-0.2,0);
            \path [draw=black, worldlineStatic] (0.2,0) -- (0.8,0);
            \path [draw=black, zParticle2] (.3,0) -- (-.3,0);
            \path [draw=black, photon] (-0.3,-1.) -- (-0.3,0);
            \path [draw=black, photon] (0.3,-1.) -- (0.3,0);
            \filldraw[fill=black] (-0.3,0) circle (.06);
            \filldraw[fill=black] (0.3,0) circle (.06);
        \end{tikzpicture}\,.
\end{align}
\end{subequations}
These can be identified as the (tree-level) contributions to the connected two-point function $\langle h_{\mu_1\nu_1}(k_1) h_{\mu_2\nu_2}(k_2)\rangle$.
The arrows then indicate the flow of causality, which in the in-in theory is denoted by $\langle h_{(+)\mu_1\nu_1}(k_1) h_{(-)\mu_2\nu_2}(k_2)\rangle$ or $\langle h_{(-)\mu_1\nu_1}(k_1) h_{(+)\mu_2\nu_2}(k_2)\rangle$ in the $+/-$ basis.
In these diagrams we leave the indices of the external graviton legs implicit.
The weightings of $1/2$ in \cref{eq:N2} are determined from the underlying symmetry of each graph
when one ignores $\iO$ prescriptions on the worldline propagators.

We can check that these expressions give rise to the expected observables using \cref{expObservables}.
Taking the momentum impulse $\Delta p_1^\mu$ as an example, at 1PM order we simply have
\begin{align}
    \Delta p_1^{(1)\mu}=\{N^{(1)},p_1^\mu\}=-\i\left\{\,\begin{tikzpicture}[baseline={([yshift=-.5ex](.5,-.5))}]
        %%%%%%%%%%%%%%%%%%%%%%
        \path [draw=black, worldlineStatic] (-0.6,0) -- (0,0);
        \path [draw=black, worldlineStatic] (0,0) -- (0.6,0);
        \path [draw=black, worldlineStatic] (-0.6,-1.) -- (0.6,-1.);
        \path [draw=black, photon] (0,-1.) -- (0,0);
        \filldraw[fill=black] (0,0) circle (.06);
        \filldraw[fill=black] (0,-1.) circle (.06);
    \end{tikzpicture},\,p_1^\mu\,\right\}\,
    =\i\,\begin{tikzpicture}[baseline={([yshift=-.5ex](.5,-.5))}]
        %%%%%%%%%%%%%%%%%%%%%%
        \path [draw=black, worldlineStatic] (-0.6,0) -- (0,0);
        \path [draw=black, zParticle] (0,0) -- (0.6,0);
        \path [draw=black, worldlineStatic] (-0.6,-1.) -- (0.6,-1.);
        \path [draw=black, photon] (0,-1.) -- (0,0);
        \filldraw[fill=black] (0,0) circle (.06);
        \filldraw[fill=black] (0,-1.) circle (.06);
    \end{tikzpicture}\,.
\end{align}
The one diagram on the right-hand side represents the only term in the 1PM momentum impulse.
At 2PM order there are two contributions:
\begin{align}
    \Delta p_1^{(2)\mu}=\{N^{(2)},p_1^\mu\}+\sfrac12\{N^{(1)},\{N^{(1)},p_1^\mu\}\}\,.
\end{align}
The first contribution, upon acting with the bracket on $p_1^\mu$, is
\begin{align}\label{bracketN2}
\begin{aligned}
    \{N^{(2)},p_1^\mu\}=\,
    &\i\left(\frac12\,
    \begin{tikzpicture}[baseline={([yshift=-.5ex](.5,-.5))}]
        %%%%%%%%%%%%%%%%%%%%%%
        \path [draw=black, worldlineStatic] (-1.,0) -- (-0.4,0);
        \path [draw=black, zParticle] (0.4,0) -- (1.,0);
        \path [draw=black, worldlineStatic] (-1.,-1.) -- (1.,-1.);
        \path [draw=black, zParticle] (-.4,0) -- (.4,0);
        \path [draw=black, photon] (-0.4,-1.) -- (-0.4,0);
        \path [draw=black, photon] (0.4,-1.) -- (0.4,0);
        \filldraw[fill=black] (-0.4,0) circle (.06);
        \filldraw[fill=black] (-0.4,-1.) circle (.06);
        \filldraw[fill=black] (0.4,0) circle (.06);
        \filldraw[fill=black] (0.4,-1.) circle (.06);
    \end{tikzpicture}
    +
    \frac12\,
    \begin{tikzpicture}[baseline={([yshift=-.5ex](.5,-.5))}]
        %%%%%%%%%%%%%%%%%%%%%%
        \path [draw=black, worldlineStatic] (-1.,0) -- (-0.4,0);
        \path [draw=black, zParticle] (0.4,0) -- (1.,0);
        \path [draw=black, worldlineStatic] (-1.,-1.) -- (1.,-1.);
        \path [draw=black, zParticle] (.4,0) -- (-.4,0);
        \path [draw=black, photon] (-0.4,-1.) -- (-0.4,0);
        \path [draw=black, photon] (0.4,-1.) -- (0.4,0);
        \filldraw[fill=black] (-0.4,0) circle (.06);
        \filldraw[fill=black] (-0.4,-1.) circle (.06);
        \filldraw[fill=black] (0.4,0) circle (.06);
        \filldraw[fill=black] (0.4,-1.) circle (.06);
    \end{tikzpicture}
    +
    \frac12\,
    \begin{tikzpicture}[baseline={([yshift=-.5ex](.5,-3.5))}]
        %%%%%%%%%%%%%%%%%%%%%%
        \path [draw=black, worldlineStatic] (-3.,-3) -- (-2.,-3);
        \path [draw=black, zParticle] (-2,-3) -- (-1.,-3);
        \path [draw=black, worldlineStatic] (-3.,-4.) -- (-1.,-4.);
        \path [draw=black, photon] (-2.4,-4.) -- (-2,-3);
        \path [draw=black, photon] (-1.6,-4.) -- (-2,-3);
        \filldraw[fill=black] (-2,-3) circle (.06);
        \filldraw[fill=black] (-2.4,-4.) circle (.06);
        \filldraw[fill=black] (-1.6,-4.) circle (.06);
    \end{tikzpicture}
    +
    \frac12\,
    \begin{tikzpicture}[baseline={([yshift=-.5ex](.5,-3.5))}]
        %%%%%%%%%%%%%%%%%%%%%%
        \path [draw=black, worldlineStatic] (1.,-3) -- (2.,-3);
        \path [draw=black, zParticle] (2,-3) -- (3.,-3);
        \path [draw=black, worldlineStatic] (1.,-4.) -- (3.,-4.);
        \path [draw=black, photon] (2,-3.5) -- (2,-3);
        \path [draw=black, photon] (1.6,-4.) -- (2,-3.5);
        \path [draw=black, photon] (2.4,-4.) -- (2,-3.5);
        \filldraw[fill=black] (2,-3) circle (.06);
        \filldraw[fill=black] (2,-3.5) circle (.06);
        \filldraw[fill=black] (1.6,-4.) circle (.06);
        \filldraw[fill=black] (2.4,-4.) circle (.06);
    \end{tikzpicture}\right.\\
    +
    &\quad\left.\frac12\,
    \begin{tikzpicture}[baseline={([yshift=-.5ex](.5,-.5))}]
        %%%%%%%%%%%%%%%%%%%%%%
        \path [draw=black, worldlineStatic] (-1.,0) -- (0.4,0);
        \path [draw=black, zParticle] (0.4,0) -- (1.,0);
        \path [draw=black, worldlineStatic] (-1.,-1.) -- (-.4,-1.);
        \path [draw=black, worldlineStatic] (.4,-1.) -- (1,-1.);
        \path [draw=black, zParticle] (-.4,-1) -- (.4,-1);
        \path [draw=black, photon] (-0.4,-1.) -- (-0.4,0);
        \path [draw=black, photon] (0.4,-1.) -- (0.4,0);
        \filldraw[fill=black] (-0.4,0) circle (.06);
        \filldraw[fill=black] (-0.4,-1.) circle (.06);
        \filldraw[fill=black] (0.4,0) circle (.06);
        \filldraw[fill=black] (0.4,-1.) circle (.06);
    \end{tikzpicture}
    +
    \frac12\,
    \begin{tikzpicture}[baseline={([yshift=-.5ex](.5,-.5))}]
        %%%%%%%%%%%%%%%%%%%%%%
        \path [draw=black, worldlineStatic] (-1.,0) -- (0.4,0);
        \path [draw=black, zParticle] (0.4,0) -- (1.,0);
        \path [draw=black, worldlineStatic] (-1.,-1.) -- (-.4,-1.);
        \path [draw=black, worldlineStatic] (.4,-1.) -- (1,-1.);
        \path [draw=black, zParticle] (.4,-1) -- (-.4,-1);
        \path [draw=black, photon] (-0.4,-1.) -- (-0.4,0);
        \path [draw=black, photon] (0.4,-1.) -- (0.4,0);
        \filldraw[fill=black] (-0.4,0) circle (.06);
        \filldraw[fill=black] (-0.4,-1.) circle (.06);
        \filldraw[fill=black] (0.4,0) circle (.06);
        \filldraw[fill=black] (0.4,-1.) circle (.06);
    \end{tikzpicture}
    +\quad\!
    \begin{tikzpicture}[baseline={([yshift=-.5ex](.5,-3.5))}]
        %%%%%%%%%%%%%%%%%%%%%%
        \path [draw=black, worldlineStatic] (-3.,-4) -- (-1.,-4);
        \path [draw=black, zParticle] (-1.6,-3) -- (-1.,-3);
        \path [draw=black, worldlineStatic] (-3.,-3) -- (-1.,-3);
        \path [draw=black, photon] (-2.4,-3) -- (-2,-4);
        \path [draw=black, photon] (-1.6,-3) -- (-2,-4);
        \filldraw[fill=black] (-2,-4) circle (.06);
        \filldraw[fill=black] (-2.4,-3) circle (.06);
        \filldraw[fill=black] (-1.6,-3) circle (.06);
    \end{tikzpicture}
   +\quad\!
    \begin{tikzpicture}[baseline={([yshift=-.5ex](.5,-3.5))}]
        %%%%%%%%%%%%%%%%%%%%%%
        \path [draw=black, worldlineStatic] (1.,-4) -- (3.,-4);
        \path [draw=black, zParticle] (2.4,-3) -- (3.,-3);
        \path [draw=black, worldlineStatic] (1.,-3) -- (3.,-3);
        \path [draw=black, photon] (2,-3.5) -- (2,-4);
        \path [draw=black, photon] (1.6,-3) -- (2,-3.5);
        \path [draw=black, photon] (2.4,-3) -- (2,-3.5);
        \filldraw[fill=black] (2,-4) circle (.06);
        \filldraw[fill=black] (2,-3.5) circle (.06);
        \filldraw[fill=black] (1.6,-3) circle (.06);
        \filldraw[fill=black] (2.4,-3) circle (.06);
    \end{tikzpicture}\,\right)\,.
\end{aligned}
\end{align}
While this nearly gives the expression for 2PM impulse,
the $\iO$ prescription on the worldline is incorrect ---
we require causality flow towards the outgoing line, which necessitates only retarded propagators.
This is cured by adding the other, nested contribution:
\begin{align}
    \{N^{(1)},\{N^{(1)},p_1^\mu\}\}=
    \left\{\,\begin{tikzpicture}[baseline={([yshift=-.5ex](.5,-.5))}]
        %%%%%%%%%%%%%%%%%%%%%%
        \path [draw=black, worldlineStatic] (-0.6,0) -- (0,0);
        \path [draw=black, worldlineStatic] (0,0) -- (0.6,0);
        \path [draw=black, worldlineStatic] (-0.6,-1.) -- (0.6,-1.);
        \path [draw=black, photon] (0,-1.) -- (0,0);
        \filldraw[fill=black] (0,0) circle (.06);
        \filldraw[fill=black] (0,-1.) circle (.06);
    \end{tikzpicture}\,,\,\begin{tikzpicture}[baseline={([yshift=-.5ex](.5,-.5))}]
        %%%%%%%%%%%%%%%%%%%%%%
        \path [draw=black, worldlineStatic] (-0.6,0) -- (0,0);
        \path [draw=black, zParticle] (0,0) -- (0.6,0);
        \path [draw=black, worldlineStatic] (-0.6,-1.) -- (0.6,-1.);
        \path [draw=black, photon] (0,-1.) -- (0,0);
        \filldraw[fill=black] (0,0) circle (.06);
        \filldraw[fill=black] (0,-1.) circle (.06);
    \end{tikzpicture}\,\right\}=\i\left(
    \begin{tikzpicture}[baseline={([yshift=-.5ex](.5,-.5))}]
        %%%%%%%%%%%%%%%%%%%%%%
        \path [draw=black, worldlineStatic] (-1.,0) -- (-0.4,0);
        \path [draw=black, zParticle] (0.4,0) -- (1.,0);
        \path [draw=black, worldlineStatic] (-1.,-1.) -- (1.,-1.);
        \path [draw=black, worldlineCut2] (-.4,0) -- (.4,0);
        \path [draw=black, photon] (-0.4,-1.) -- (-0.4,0);
        \path [draw=black, photon] (0.4,-1.) -- (0.4,0);
        \filldraw[fill=black] (-0.4,0) circle (.06);
        \filldraw[fill=black] (-0.4,-1.) circle (.06);
        \filldraw[fill=black] (0.4,0) circle (.06);
        \filldraw[fill=black] (0.4,-1.) circle (.06);
    \end{tikzpicture}+\begin{tikzpicture}[baseline={([yshift=-.5ex](.5,-.5))}]
        %%%%%%%%%%%%%%%%%%%%%%
        \path [draw=black, worldlineStatic] (-1.,-1) -- (-0.4,-1);
        \path [draw=black, worldlineStatic] (0.4,-1) -- (1,-1);
        \path [draw=black, zParticle] (0.4,0) -- (1.,0);
        \path [draw=black, worldlineStatic] (-1.,0) -- (1.,0);
        \path [draw=black, worldlineCut2] (-0.4,-1) -- (0.4,-1);
        \path [draw=black, photon] (-0.4,-1.) -- (-0.4,0);
        \path [draw=black, photon] (0.4,-1.) -- (0.4,0);
        \filldraw[fill=black] (-0.4,0) circle (.06);
        \filldraw[fill=black] (-0.4,-1.) circle (.06);
        \filldraw[fill=black] (0.4,0) circle (.06);
        \filldraw[fill=black] (0.4,-1.) circle (.06);
    \end{tikzpicture}\right)\,,
\end{align}
evaluated using the fundamental relationship \cref{diagBracket}.
These extra contributions with the directed cut worldlines serve to
``correct'' the $\iO$ prescriptions in \cref{bracketN2}.
Thus, we find that
\begin{align}
\begin{aligned}
    \Delta p_1^{(2)\mu}=\quad\!
    &\i\left(\,
    \begin{tikzpicture}[baseline={([yshift=-.5ex](.5,-.5))}]
        %%%%%%%%%%%%%%%%%%%%%%
        \path [draw=black, worldlineStatic] (-1.,0) -- (-0.4,0);
        \path [draw=black, zParticle] (0.4,0) -- (1.,0);
        \path [draw=black, worldlineStatic] (-1.,-1.) -- (1.,-1.);
        \path [draw=black, zParticle] (-.4,0) -- (.4,0);
        \path [draw=black, photon] (-0.4,-1.) -- (-0.4,0);
        \path [draw=black, photon] (0.4,-1.) -- (0.4,0);
        \filldraw[fill=black] (-0.4,0) circle (.06);
        \filldraw[fill=black] (-0.4,-1.) circle (.06);
        \filldraw[fill=black] (0.4,0) circle (.06);
        \filldraw[fill=black] (0.4,-1.) circle (.06);
    \end{tikzpicture}
    +
    \frac12\,
    \begin{tikzpicture}[baseline={([yshift=-.5ex](.5,-3.5))}]
        %%%%%%%%%%%%%%%%%%%%%%
        \path [draw=black, worldlineStatic] (-3.,-3) -- (-2.,-3);
        \path [draw=black, zParticle] (-2,-3) -- (-1.,-3);
        \path [draw=black, worldlineStatic] (-3.,-4.) -- (-1.,-4.);
        \path [draw=black, photon] (-2.4,-4.) -- (-2,-3);
        \path [draw=black, photon] (-1.6,-4.) -- (-2,-3);
        \filldraw[fill=black] (-2,-3) circle (.06);
        \filldraw[fill=black] (-2.4,-4.) circle (.06);
        \filldraw[fill=black] (-1.6,-4.) circle (.06);
    \end{tikzpicture}
    +
    \frac12\,
    \begin{tikzpicture}[baseline={([yshift=-.5ex](.5,-3.5))}]
        %%%%%%%%%%%%%%%%%%%%%%
        \path [draw=black, worldlineStatic] (1.,-3) -- (2.,-3);
        \path [draw=black, zParticle] (2,-3) -- (3.,-3);
        \path [draw=black, worldlineStatic] (1.,-4.) -- (3.,-4.);
        \path [draw=black, photon] (2,-3.5) -- (2,-3);
        \path [draw=black, photon] (1.6,-4.) -- (2,-3.5);
        \path [draw=black, photon] (2.4,-4.) -- (2,-3.5);
        \filldraw[fill=black] (2,-3) circle (.06);
        \filldraw[fill=black] (2,-3.5) circle (.06);
        \filldraw[fill=black] (1.6,-4.) circle (.06);
        \filldraw[fill=black] (2.4,-4.) circle (.06);
    \end{tikzpicture}\right.\\
    +
    &\quad\left.\,
    \begin{tikzpicture}[baseline={([yshift=-.5ex](.5,-.5))}]
        %%%%%%%%%%%%%%%%%%%%%%
        \path [draw=black, worldlineStatic] (-1.,0) -- (0.4,0);
        \path [draw=black, zParticle] (0.4,0) -- (1.,0);
        \path [draw=black, worldlineStatic] (-1.,-1.) -- (-.4,-1.);
        \path [draw=black, worldlineStatic] (.4,-1.) -- (1,-1.);
        \path [draw=black, zParticle] (-.4,-1) -- (.4,-1);
        \path [draw=black, photon] (-0.4,-1.) -- (-0.4,0);
        \path [draw=black, photon] (0.4,-1.) -- (0.4,0);
        \filldraw[fill=black] (-0.4,0) circle (.06);
        \filldraw[fill=black] (-0.4,-1.) circle (.06);
        \filldraw[fill=black] (0.4,0) circle (.06);
        \filldraw[fill=black] (0.4,-1.) circle (.06);
    \end{tikzpicture}
    +\quad\!
    \begin{tikzpicture}[baseline={([yshift=-.5ex](.5,-3.5))}]
        %%%%%%%%%%%%%%%%%%%%%%
        \path [draw=black, worldlineStatic] (-3.,-4) -- (-1.,-4);
        \path [draw=black, zParticle] (-1.6,-3) -- (-1.,-3);
        \path [draw=black, worldlineStatic] (-3.,-3) -- (-1.,-3);
        \path [draw=black, photon] (-2.4,-3) -- (-2,-4);
        \path [draw=black, photon] (-1.6,-3) -- (-2,-4);
        \filldraw[fill=black] (-2,-4) circle (.06);
        \filldraw[fill=black] (-2.4,-3) circle (.06);
        \filldraw[fill=black] (-1.6,-3) circle (.06);
    \end{tikzpicture}
   +\quad\!
    \begin{tikzpicture}[baseline={([yshift=-.5ex](.5,-3.5))}]
        %%%%%%%%%%%%%%%%%%%%%%
        \path [draw=black, worldlineStatic] (1.,-4) -- (3.,-4);
        \path [draw=black, zParticle] (2.4,-3) -- (3.,-3);
        \path [draw=black, worldlineStatic] (1.,-3) -- (3.,-3);
        \path [draw=black, photon] (2,-3.5) -- (2,-4);
        \path [draw=black, photon] (1.6,-3) -- (2,-3.5);
        \path [draw=black, photon] (2.4,-3) -- (2,-3.5);
        \filldraw[fill=black] (2,-4) circle (.06);
        \filldraw[fill=black] (2,-3.5) circle (.06);
        \filldraw[fill=black] (1.6,-3) circle (.06);
        \filldraw[fill=black] (2.4,-3) circle (.06);
    \end{tikzpicture}\,\right)\,,
\end{aligned}
\end{align}
as expected.

At 3PM order it is \emph{a priori} unclear what $\iO$ prescription we should use for the worldline propagators.
To determine this we use our knowledge of the scattering observables,
where causality always flows towards the outgoing line,
and infer the $\iO$ prescriptions on $N^{(3)}$ diagrammatically.
Using \cref{expObservables} the 3PM momentum impulse is given by
\begin{align}\label{3PMimpulse}
\begin{aligned}
    \Delta p_1^{(3)\mu}=\{N^{(3)},p_1^\mu\}&+\sfrac12\{N^{(1)},\{N^{(2)},p_1^\mu\}\}
    +\sfrac12\{N^{(2)},\{N^{(1)},p_1^\mu\}\}\\
    &+\sfrac16\{N^{(1)},\{N^{(1)},\{N^{(1)},p_1^\mu\}\}\}\,.
\end{aligned}
\end{align}
Accordingly, we determine that
\begin{align}\label{eq:N3}
    &\i N^{(3)}=
    \frac1{3!}\left(\,\,\frac23\,\,
    \begin{tikzpicture}[baseline={([yshift=-.5ex](.5,-3.5))}]
        %%%%%%%%%%%%%%%%%%%%%%
        \path [draw=black, worldlineStatic] (-3.,-3) -- (-1.,-3);
        \path [draw=black, worldlineStatic] (-3.,-4.) -- (-1.,-4.);
        \path [draw=black, photon] (-2.6,-4.) -- (-2.2,-3);
        \path [draw=black, photon] (-1.4,-4.) -- (-1.8,-3);
        \path [draw=black, photon] (-2,-4.) -- (-2,-3);
        \filldraw[fill=white, draw=black,thick] (-2.45,-3.2) rectangle (-1.55,-2.8);
        \filldraw[fill=black] (-2.6,-4.) circle (.06);
        \filldraw[fill=black] (-2,-4.) circle (.06);
        \filldraw[fill=black] (-1.4,-4.) circle (.06);
        \node (A) at (-2,-3.02) {$\rightarrow\rightarrow$};
    \end{tikzpicture}+
    \,\,
    \frac13\,\,
    \begin{tikzpicture}[baseline={([yshift=-.5ex](.5,-3.5))}]
        %%%%%%%%%%%%%%%%%%%%%%
        \path [draw=black, worldlineStatic] (-3.,-3) -- (-1.,-3);
        \path [draw=black, worldlineStatic] (-3.,-4.) -- (-1.,-4.);
        \path [draw=black, photon] (-2.6,-4.) -- (-2.2,-3);
        \path [draw=black, photon] (-1.4,-4.) -- (-1.8,-3);
        \path [draw=black, photon] (-2,-4.) -- (-2,-3);
        \filldraw[fill=white, draw=black,thick] (-2.45,-3.2) rectangle (-1.55,-2.8);
        \filldraw[fill=black] (-2.6,-4.) circle (.06);
        \filldraw[fill=black] (-2,-4.) circle (.06);
        \filldraw[fill=black] (-1.4,-4.) circle (.06);
        \node (A) at (-2,-3.02) {$\rightarrow\leftarrow$};
    \end{tikzpicture}
    +
    \frac23\,\,
    \begin{tikzpicture}[baseline={([yshift=-.5ex](.5,-3.5))}]
        %%%%%%%%%%%%%%%%%%%%%%
        \path [draw=black, worldlineStatic] (-3.,-4) -- (-1.,-4);
        \path [draw=black, worldlineStatic] (-3.,-3.) -- (-1.,-3.);
        \path [draw=black, photon] (-2.6,-3.) -- (-2.2,-4);
        \path [draw=black, photon] (-1.4,-3.) -- (-1.8,-4);
        \path [draw=black, photon] (-2,-3.) -- (-2,-4);
        \filldraw[fill=white, draw=black,thick] (-2.45,-4.2) rectangle (-1.55,-3.8);
        \filldraw[fill=black] (-2.6,-3.) circle (.06);
        \filldraw[fill=black] (-2,-3.) circle (.06);
        \filldraw[fill=black] (-1.4,-3.) circle (.06);
        \node (A) at (-2,-4.02) {$\rightarrow\rightarrow$};
    \end{tikzpicture}
    +
    \,\,
    \frac13\,\,
    \begin{tikzpicture}[baseline={([yshift=-.5ex](.5,-3.5))}]
        %%%%%%%%%%%%%%%%%%%%%%
        \path [draw=black, worldlineStatic] (-3.,-4) -- (-1.,-4);
        \path [draw=black, worldlineStatic] (-3.,-3.) -- (-1.,-3.);
        \path [draw=black, photon] (-2.6,-3.) -- (-2.2,-4);
        \path [draw=black, photon] (-1.4,-3.) -- (-1.8,-4);
        \path [draw=black, photon] (-2,-3.) -- (-2,-4);
        \filldraw[fill=white, draw=black,thick] (-2.45,-4.2) rectangle (-1.55,-3.8);
        \filldraw[fill=black] (-2.6,-3.) circle (.06);
        \filldraw[fill=black] (-2,-3.) circle (.06);
        \filldraw[fill=black] (-1.4,-3.) circle (.06);
        \node (A) at (-2,-4.02) {$\rightarrow\leftarrow$};
    \end{tikzpicture}
    \,\,\right)\nn \\
    &+
    \frac12\left[
        \left(\left(\,\,\frac23\,\,
    \begin{tikzpicture}[baseline={([yshift=-.5ex](.5,-3.5))}]
        %%%%%%%%%%%%%%%%%%%%%%
        \path [draw=black, worldlineStatic] (-3.,-3) -- (-1.,-3);
        \path [draw=black, worldlineStatic] (-3.,-4.) -- (-1.,-4.);
        \path [draw=black, photon] (-2.4,-4.) -- (-2.4,-3);
        \path [draw=black, photon] (-1.6,-4.) -- (-2.4,-3);
        \path [draw=black, photon] (-1.6,-4.) -- (-1.6,-3);
        \filldraw[fill=white, draw=black,thick] (-2.65,-3.2) rectangle (-2.15,-2.8);
        \filldraw[fill=white, draw=black,thick] (-1.85,-4.2) rectangle (-1.35,-3.8);
        \filldraw[fill=black] (-2.4,-4.) circle (.06);
        \filldraw[fill=black] (-1.6,-3.) circle (.06);
	    \node (A) at (-2.4,-3.03) {$\rightarrow$};
        \node (A) at (-1.6,-4.03) {$\rightarrow$};
    \end{tikzpicture}+
    \,\,\frac13\,\,
    \begin{tikzpicture}[baseline={([yshift=-.5ex](.5,-3.5))}]
        %%%%%%%%%%%%%%%%%%%%%%
        \path [draw=black, worldlineStatic] (-3.,-3) -- (-1.,-3);
        \path [draw=black, worldlineStatic] (-3.,-4.) -- (-1.,-4.);
        \path [draw=black, photon] (-2.4,-4.) -- (-2.4,-3);
        \path [draw=black, photon] (-1.6,-4.) -- (-2.4,-3);
        \path [draw=black, photon] (-1.6,-4.) -- (-1.6,-3);
        \filldraw[fill=white, draw=black,thick] (-2.65,-3.2) rectangle (-2.15,-2.8);
        \filldraw[fill=white, draw=black,thick] (-1.85,-4.2) rectangle (-1.35,-3.8);
        \filldraw[fill=black] (-2.4,-4.) circle (.06);
        \filldraw[fill=black] (-1.6,-3.) circle (.06);
        \node (A) at (-2.4,-3.03) {$\rightarrow$};
        \node (A) at (-1.6,-4.03) {$\leftarrow$};
    \end{tikzpicture}
    \,\,\right)
    % \cup{\rm reflected}
    % \cup
    % \left(\,\,
    \,\,\cup\,\,
    \begin{tikzpicture}[baseline={([yshift=-.5ex](.5,-3.5))}]
        %%%%%%%%%%%%%%%%%%%%%%
        \path [draw=black, worldlineStatic] (-3.,-3) -- (-1.,-3);
        \path [draw=black, worldlineStatic] (-3.,-4.) -- (-1.,-4.);
        \path [draw=black, photon] (-2.4,-4.) -- (-2.4,-3);
        \path [draw=black, photon] (-2.4,-3.) to[out=-60,in=-120] (-1.6,-3);
        \path [draw=black, photon] (-1.6,-4.) -- (-1.6,-3);
        \filldraw[fill=white, draw=black,thick] (-2.65,-3.2) rectangle (-2.15,-2.8);
        \filldraw[fill=white, draw=black,thick] (-1.85,-3.2) rectangle (-1.35,-2.8);
        \filldraw[fill=black] (-2.4,-4.) circle (.06);
        \filldraw[fill=black] (-1.6,-4.) circle (.06);
        \node (A) at (-2.4,-3.03) {$\rightarrow$};
        \node (A) at (-1.6,-3.03) {$\rightarrow$};
    \end{tikzpicture}
    \,\,\right)\right. \\
    &\,\,\quad\left.\cup\,\,
    \left(\left(\,\,\frac23\,\,
    \begin{tikzpicture}[baseline={([yshift=-.5ex](.5,-3.5))}]
        %%%%%%%%%%%%%%%%%%%%%%
        \path [draw=black, worldlineStatic] (-3.,-4) -- (-1.,-4);
        \path [draw=black, worldlineStatic] (-3.,-3.) -- (-1.,-3.);
        \path [draw=black, photon] (-2.4,-3.) -- (-2.4,-4);
        \path [draw=black, photon] (-1.6,-3.) -- (-2.4,-4);
        \path [draw=black, photon] (-1.6,-3.) -- (-1.6,-4);
        \filldraw[fill=white, draw=black,thick] (-2.65,-4.2) rectangle (-2.15,-3.8);
        \filldraw[fill=white, draw=black,thick] (-1.9,-3.2) rectangle (-1.4,-2.8);
        \filldraw[fill=black] (-2.4,-3.) circle (.06);
        \filldraw[fill=black] (-1.6,-4.) circle (.06);
	    \node (A) at (-2.4,-4.03) {$\rightarrow$};
        \node (A) at (-1.65,-3.03) {$\rightarrow$};
    \end{tikzpicture}+
    \,\,\frac13\,\,
    \begin{tikzpicture}[baseline={([yshift=-.5ex](.5,-3.5))}]
        %%%%%%%%%%%%%%%%%%%%%%
        \path [draw=black, worldlineStatic] (-3.,-4) -- (-1.,-4);
        \path [draw=black, worldlineStatic] (-3.,-3.) -- (-1.,-3.);
        \path [draw=black, photon] (-2.4,-3.) -- (-2.4,-4);
        \path [draw=black, photon] (-1.6,-3.) -- (-2.4,-4);
        \path [draw=black, photon] (-1.6,-3.) -- (-1.6,-4);
        \filldraw[fill=white, draw=black,thick] (-2.65,-4.2) rectangle (-2.15,-3.8);
        \filldraw[fill=white, draw=black,thick] (-1.9,-3.2) rectangle (-1.4,-2.8);
        \filldraw[fill=black] (-2.4,-3.) circle (.06);
        \filldraw[fill=black] (-1.6,-4.) circle (.06);
        \node (A) at (-2.4,-4.03) {$\rightarrow$};
        \node (A) at (-1.65,-3.03) {$\leftarrow$};
    \end{tikzpicture}
    \,\,\right)
    % \cup{\rm reflected}
    % \cup
    % \left(\,\,
    \,\,\cup\,\,
    \begin{tikzpicture}[baseline={([yshift=-.5ex](.5,-3.5))}]
        %%%%%%%%%%%%%%%%%%%%%%
        \path [draw=black, worldlineStatic] (-3.,-4) -- (-1.,-4);
        \path [draw=black, worldlineStatic] (-3.,-3.) -- (-1.,-3.);
        \path [draw=black, photon] (-2.4,-3.) -- (-2.4,-4);
        \path [draw=black, photon] (-2.4,-4.) to[out=60,in=120] (-1.6,-4);
        \path [draw=black, photon] (-1.6,-3.) -- (-1.6,-4);
        \filldraw[fill=white, draw=black,thick] (-2.65,-4.2) rectangle (-2.15,-3.8);
        \filldraw[fill=white, draw=black,thick] (-1.85,-4.2) rectangle (-1.35,-3.8);
        \filldraw[fill=black] (-2.4,-3.) circle (.06);
        \filldraw[fill=black] (-1.6,-3.) circle (.06);
        \node (A) at (-2.4,-4.03) {$\rightarrow$};
        \node (A) at (-1.6,-4.03) {$\rightarrow$};
    \end{tikzpicture}
    \,\,\right)\right]\!.\nn
\end{align}
Here we have ignored diagrams which vanish in four dimensions (see \cref{eq:PinchedN} and the surrounding discussion) and collected the rest into three-point functions, which now appear alongside the two-point functions already considered in \cref{eq:ComptonDiagrams}:
\begin{subequations}\label{eq:ThreeGravComptonDiagrams}
\begin{align}
    &\begin{tikzpicture}[baseline={([yshift=-.5ex](.5,-3.5))}]
        %%%%%%%%%%%%%%%%%%%%%%
        \path [draw=black, worldlineStatic] (-3.,-3) -- (-1.,-3);
        \path [draw=black, photon] (-2.6,-4.) -- (-2.2,-3);
        \path [draw=black, photon] (-1.4,-4.) -- (-1.8,-3);
        \path [draw=black, photon] (-2,-4.) -- (-2,-3);
        \filldraw[fill=white, draw=black,thick] (-2.45,-3.2) rectangle (-1.55,-2.8);
        \node (A) at (-2,-3.0) {$\rightarrow\rightarrow$};
    \end{tikzpicture}:=
	\frac{1}{3!}\,\begin{tikzpicture}[baseline={([yshift=-.5ex](.5,-3.5))}]
        %%%%%%%%%%%%%%%%%%%%%%
        \path [draw=black, worldlineStatic] (-2.8,-3) -- (-1.2,-3);
        \path [draw=black, photon] (-2.4,-4.) -- (-2.0,-3);
        \path [draw=black, photon] (-1.6,-4.) -- (-2.0,-3);
        \path [draw=black, photon] (-2,-4.) -- (-2,-3);
        \filldraw[fill=black] (-2,-3.) circle (.06);
    \end{tikzpicture}
	+
	\frac{1}{3!}\,\begin{tikzpicture}[baseline={([yshift=-.5ex](.5,-3.5))}]
        %%%%%%%%%%%%%%%%%%%%%%
        \path [draw=black, worldlineStatic] (-2.8,-3) -- (-1.2,-3);
        \path [draw=black, photon] (-2.4,-4.) -- (-2.0,-3.5);
        \path [draw=black, photon] (-1.6,-4.) -- (-2.0,-3.5);
        \path [draw=black, photon] (-2,-4.) -- (-2,-3);
        \filldraw[fill=black] (-2,-3) circle (.06);
        \filldraw[fill=black] (-2,-3.5) circle (.06);
    \end{tikzpicture}
	+
	\frac{1}{2}\,\begin{tikzpicture}[baseline={([yshift=-.5ex](.5,-3.5))}]
        %%%%%%%%%%%%%%%%%%%%%%
        \path [draw=black, worldlineStatic] (-2.8,-3) -- (-1.2,-3);
        \path [draw=black, photon] (-2.8,-4.) -- (-2.0,-3.3);
        \path [draw=black, photon] (-1.2,-4.) -- (-2.0,-3.3);
        \path [draw=black, photon] (-2,-3.) -- (-2,-3.3);
        \path [draw=black, photon] (-2,-4.) -- (-1.5,-3.6);
        \filldraw[fill=black] (-2,-3) circle (.06);
        \filldraw[fill=black] (-2,-3.3) circle (.06);
        \filldraw[fill=black] (-1.53,-3.67) circle (.06);
    \end{tikzpicture}
    +\frac{1}{2}\,\begin{tikzpicture}[baseline={([yshift=-.5ex](.5,-3.5))}]
        %%%%%%%%%%%%%%%%%%%%%%
        \path [draw=black, worldlineStatic] (-3.2,-3) -- (-2.8,-3);
        \path [draw=black, worldlineStatic] (-1.2,-3) -- (-0.8,-3);
        \path [draw=black, zParticle] (-2.8,-3) -- (-2.0,-3);
        \path [draw=black, zParticle] (-2.0,-3) -- (-1.2,-3);
        \path [draw=black, photon] (-2.0,-4.) -- (-2.0,-3.);
        \path [draw=black, photon] (-2.8,-4.) -- (-2.8,-3.);
        \path [draw=black, photon] (-1.2,-3.) -- (-1.2,-4.);
        \filldraw[fill=black] (-1.2,-3) circle (.06);
        \filldraw[fill=black] (-2.0,-3) circle (.06);
        \filldraw[fill=black] (-2.8,-3) circle (.06);
    \end{tikzpicture} \\
	&\quad\quad+\frac{1}{2}\,\begin{tikzpicture}[baseline={([yshift=-.5ex](.5,-3.5))}]
        %%%%%%%%%%%%%%%%%%%%%%
        \path [draw=black, worldlineStatic] (-2.8,-3) -- (-2.4,-3);
        \path [draw=black, worldlineStatic] (-1.6,-3) -- (-1.2,-3);
        \path [draw=black, zParticle] (-2.4,-3) -- (-1.6,-3);
        \path [draw=black, photon] (-2.0,-4.) -- (-1.6,-3.5);
        \path [draw=black, photon] (-1.2,-4.) -- (-1.6,-3.5);
        \path [draw=black, photon] (-1.6,-3.) -- (-1.6,-3.5);
        \path [draw=black, photon] (-2.4,-3.) -- (-2.4,-4.);
        \filldraw[fill=black] (-1.6,-3) circle (.06);
        \filldraw[fill=black] (-2.4,-3) circle (.06);
        \filldraw[fill=black] (-1.6,-3.5) circle (.06);
    \end{tikzpicture}
    +\frac{1}{2}\,\begin{tikzpicture}[baseline={([yshift=-.5ex](.5,-3.5))}]
        %%%%%%%%%%%%%%%%%%%%%%
        \path [draw=black, worldlineStatic] (-2.8,-3) -- (-2.4,-3);
        \path [draw=black, worldlineStatic] (-1.6,-3) -- (-1.2,-3);
        \path [draw=black, zParticle] (-2.4,-3) -- (-1.6,-3);
        \path [draw=black, photon] (-2.0,-4.) -- (-1.6,-3.);
        \path [draw=black, photon] (-1.2,-4.) -- (-1.6,-3.);
        \path [draw=black, photon] (-2.4,-3.) -- (-2.4,-4.);
        \filldraw[fill=black] (-1.6,-3) circle (.06);
        \filldraw[fill=black] (-2.4,-3) circle (.06);
    \end{tikzpicture}
    +\frac{1}{2}\,\begin{tikzpicture}[baseline={([yshift=-.5ex](.5,-3.5))}]
        %%%%%%%%%%%%%%%%%%%%%%
        \path [draw=black, worldlineStatic] (-2.8,-3) -- (-2.4,-3);
        \path [draw=black, worldlineStatic] (-1.6,-3) -- (-1.2,-3);
        \path [draw=black, zParticle] (-2.4,-3) -- (-1.6,-3);
        \path [draw=black, photon] (-2.0,-4.) -- (-2.4,-3.5);
        \path [draw=black, photon] (-2.8,-4.) -- (-2.4,-3.5);
        \path [draw=black, photon] (-2.4,-3.) -- (-2.4,-3.5);
        \path [draw=black, photon] (-1.6,-3.) -- (-1.6,-4.);
        \filldraw[fill=black] (-1.6,-3) circle (.06);
        \filldraw[fill=black] (-2.4,-3) circle (.06);
        \filldraw[fill=black] (-2.4,-3.5) circle (.06);
    \end{tikzpicture}
    +\frac{1}{2}\,\,\begin{tikzpicture}[baseline={([yshift=-.5ex](.5,-3.5))}]
        %%%%%%%%%%%%%%%%%%%%%%
        \path [draw=black, worldlineStatic] (-2.8,-3) -- (-2.4,-3);
        \path [draw=black, worldlineStatic] (-1.6,-3) -- (-1.2,-3);
        \path [draw=black, zParticle] (-2.4,-3) -- (-1.6,-3);
        \path [draw=black, photon] (-2.0,-4.) -- (-2.4,-3.);
        \path [draw=black, photon] (-2.8,-4.) -- (-2.4,-3.);
        \path [draw=black, photon] (-1.6,-3.) -- (-1.6,-4.);
        \filldraw[fill=black] (-1.6,-3) circle (.06);
        \filldraw[fill=black] (-2.4,-3) circle (.06);
    \end{tikzpicture}+{\rm perms.}\,,\nn \\
    %%%%%%%%%%%%%%%%%%%%%%%%%%%%%%%%%%%%%%%%%%%%%%%%%%%%%%%%%%%%%%%%%%%%%%%%%%%%%%%%%%%%%%%%%%%%%%%%%%%%%%%%%%%%%%%%
    &\begin{tikzpicture}[baseline={([yshift=-.5ex](.5,-3.5))}]
        %%%%%%%%%%%%%%%%%%%%%%
        \path [draw=black, worldlineStatic] (-3.,-3) -- (-1.,-3);
        \path [draw=black, photon] (-2.6,-4.) -- (-2.2,-3);
        \path [draw=black, photon] (-1.4,-4.) -- (-1.8,-3);
        \path [draw=black, photon] (-2,-4.) -- (-2,-3);
        \filldraw[fill=white, draw=black,thick] (-2.45,-3.2) rectangle (-1.55,-2.8);
        \node (A) at (-2,-3.0) {$\rightarrow\leftarrow$};
    \end{tikzpicture}:=
	\frac{1}{3!}\,\begin{tikzpicture}[baseline={([yshift=-.5ex](.5,-3.5))}]
        %%%%%%%%%%%%%%%%%%%%%%
        \path [draw=black, worldlineStatic] (-2.8,-3) -- (-1.2,-3);
        \path [draw=black, photon] (-2.4,-4.) -- (-2.0,-3);
        \path [draw=black, photon] (-1.6,-4.) -- (-2.0,-3);
        \path [draw=black, photon] (-2,-4.) -- (-2,-3);
        \filldraw[fill=black] (-2,-3.) circle (.06);
    \end{tikzpicture}
	+
	\frac{1}{3!}\,\begin{tikzpicture}[baseline={([yshift=-.5ex](.5,-3.5))}]
        %%%%%%%%%%%%%%%%%%%%%%
        \path [draw=black, worldlineStatic] (-2.8,-3) -- (-1.2,-3);
        \path [draw=black, photon] (-2.4,-4.) -- (-2.0,-3.5);
        \path [draw=black, photon] (-1.6,-4.) -- (-2.0,-3.5);
        \path [draw=black, photon] (-2,-4.) -- (-2,-3);
        \filldraw[fill=black] (-2,-3) circle (.06);
        \filldraw[fill=black] (-2,-3.5) circle (.06);
    \end{tikzpicture}
	+
	\frac{1}{2}\,\begin{tikzpicture}[baseline={([yshift=-.5ex](.5,-3.5))}]
        %%%%%%%%%%%%%%%%%%%%%%
        \path [draw=black, worldlineStatic] (-2.8,-3) -- (-1.2,-3);
        \path [draw=black, photon] (-2.8,-4.) -- (-2.0,-3.3);
        \path [draw=black, photon] (-1.2,-4.) -- (-2.0,-3.3);
        \path [draw=black, photon] (-2,-3.) -- (-2,-3.3);
        \path [draw=black, photon] (-2,-4.) -- (-1.5,-3.6);
        \filldraw[fill=black] (-2,-3) circle (.06);
        \filldraw[fill=black] (-2,-3.3) circle (.06);
        \filldraw[fill=black] (-1.53,-3.67) circle (.06);
    \end{tikzpicture}
    +\frac{1}{2}\,\begin{tikzpicture}[baseline={([yshift=-.5ex](.5,-3.5))}]
        %%%%%%%%%%%%%%%%%%%%%%
        \path [draw=black, worldlineStatic] (-3.2,-3) -- (-2.8,-3);
        \path [draw=black, worldlineStatic] (-1.2,-3) -- (-0.8,-3);
        \path [draw=black, zParticle] (-2.8,-3) -- (-2.0,-3);
        \path [draw=black, zParticle] (-1.2,-3) -- (-2.0,-3);
        \path [draw=black, photon] (-2.0,-4.) -- (-2.0,-3.);
        \path [draw=black, photon] (-2.8,-4.) -- (-2.8,-3.);
        \path [draw=black, photon] (-1.2,-3.) -- (-1.2,-4.);
        \filldraw[fill=black] (-1.2,-3) circle (.06);
        \filldraw[fill=black] (-2.0,-3) circle (.06);
        \filldraw[fill=black] (-2.8,-3) circle (.06);
    \end{tikzpicture} \\
	&\quad\quad+\frac{1}{2}\,\begin{tikzpicture}[baseline={([yshift=-.5ex](.5,-3.5))}]
        %%%%%%%%%%%%%%%%%%%%%%
        \path [draw=black, worldlineStatic] (-2.8,-3) -- (-2.4,-3);
        \path [draw=black, worldlineStatic] (-1.6,-3) -- (-1.2,-3);
        \path [draw=black, zParticle] (-2.4,-3) -- (-1.6,-3);
        \path [draw=black, photon] (-2.0,-4.) -- (-1.6,-3.5);
        \path [draw=black, photon] (-1.2,-4.) -- (-1.6,-3.5);
        \path [draw=black, photon] (-1.6,-3.) -- (-1.6,-3.5);
        \path [draw=black, photon] (-2.4,-3.) -- (-2.4,-4.);
        \filldraw[fill=black] (-1.6,-3) circle (.06);
        \filldraw[fill=black] (-2.4,-3) circle (.06);
        \filldraw[fill=black] (-1.6,-3.5) circle (.06);
    \end{tikzpicture}
    +\frac{1}{2}\,\begin{tikzpicture}[baseline={([yshift=-.5ex](.5,-3.5))}]
        %%%%%%%%%%%%%%%%%%%%%%
        \path [draw=black, worldlineStatic] (-2.8,-3) -- (-2.4,-3);
        \path [draw=black, worldlineStatic] (-1.6,-3) -- (-1.2,-3);
        \path [draw=black, zParticle] (-2.4,-3) -- (-1.6,-3);
        \path [draw=black, photon] (-2.0,-4.) -- (-1.6,-3.);
        \path [draw=black, photon] (-1.2,-4.) -- (-1.6,-3.);
        \path [draw=black, photon] (-2.4,-3.) -- (-2.4,-4.);
        \filldraw[fill=black] (-1.6,-3) circle (.06);
        \filldraw[fill=black] (-2.4,-3) circle (.06);
    \end{tikzpicture}
    +\frac{1}{2}\,\begin{tikzpicture}[baseline={([yshift=-.5ex](.5,-3.5))}]
        %%%%%%%%%%%%%%%%%%%%%%
        \path [draw=black, worldlineStatic] (-2.8,-3) -- (-2.4,-3);
        \path [draw=black, worldlineStatic] (-1.6,-3) -- (-1.2,-3);
        \path [draw=black, zParticle] (-1.6,-3) -- (-2.4,-3);
        \path [draw=black, photon] (-2.0,-4.) -- (-2.4,-3.5);
        \path [draw=black, photon] (-2.8,-4.) -- (-2.4,-3.5);
        \path [draw=black, photon] (-2.4,-3.) -- (-2.4,-3.5);
        \path [draw=black, photon] (-1.6,-3.) -- (-1.6,-4.);
        \filldraw[fill=black] (-1.6,-3) circle (.06);
        \filldraw[fill=black] (-2.4,-3) circle (.06);
        \filldraw[fill=black] (-2.4,-3.5) circle (.06);
    \end{tikzpicture}
    +\frac{1}{2}\,\,\begin{tikzpicture}[baseline={([yshift=-.5ex](.5,-3.5))}]
        %%%%%%%%%%%%%%%%%%%%%%
        \path [draw=black, worldlineStatic] (-2.8,-3) -- (-2.4,-3);
        \path [draw=black, worldlineStatic] (-1.6,-3) -- (-1.2,-3);
        \path [draw=black, zParticle] (-1.6,-3) -- (-2.4,-3);
        \path [draw=black, photon] (-2.0,-4.) -- (-2.4,-3.);
        \path [draw=black, photon] (-2.8,-4.) -- (-2.4,-3.);
        \path [draw=black, photon] (-1.6,-3.) -- (-1.6,-4.);
        \filldraw[fill=black] (-1.6,-3) circle (.06);
        \filldraw[fill=black] (-2.4,-3) circle (.06);
    \end{tikzpicture}+{\rm perms.}\,,\nn
\end{align}
\end{subequations}
including permutations of implicit graviton labels.

The ``merging" operation $\cup$ accounts for overcounting of diagrams.
In particular, the H diagram
\begin{align}\label{eq:HTopology}
	\vcenter{\hbox{\begin{tikzpicture}[baseline={([yshift=-.5ex](.5,-3.5))}]
            %%%%%%%%%%%%%%%%%%%%%%
	    \path [draw=black, worldlineStatic] (1.2,-3) -- (2.8,-3);
        \path [draw=black, worldlineStatic] (1.2,-4.5) -- (2.8,-4.5);
        \path [draw=black, photon] (1.6,-3) -- (1.6,-4.5);
        \path [draw=black, photon] (2.4,-3) -- (2.4,-4.5);
        \path [draw=black, photon] (1.6,-3.75) -- (2.4,-3.75);
        \filldraw[fill=black] (1.6,-3) circle (.06);
        \filldraw[fill=black] (2.4,-3) circle (.06);
	    \filldraw[fill=black] (1.6,-4.5) circle (.06);
        \filldraw[fill=black] (2.4,-4.5) circle (.06);
        \filldraw[fill=black] (1.6,-3.75) circle (.06);
        \filldraw[fill=black] (2.4,-3.75) circle (.06);
    \end{tikzpicture}}}
\end{align}
appears in all four merged collections of diagrams.
When reverse engineering the on-shell action from observables, one finds that this diagram comes with a factor of $1/2$, whereas simply adding all instances of \cref{eq:HTopology} in \cref{eq:N3} gives it a coefficient of $2$.
The merging operation, then, recovers the correct coefficient by normalising the diagram by the total number of instances it appears in the merged quantities: in this case four.
This meshes well with the merging of cuts in generalised unitarity;
we will come back to this in the next section.

At 3PM order we encounter an ``active graviton'',
i.e.~one which is allowed to go on-shell over the phase space of integration
and therefore where the $\i0^+$ prescription plays a role~\cite{Jakobsen:2022psy}.
In analogy to the worldline propagators,
the on-shell action will be a linear combination of different routings of
the active graviton(s)~\cite{Kim:2024svw,Kim:2025hpn} --- 
but, since the different graviton routings do not affect values of the ``b-type''~\cite{Jakobsen:2022psy}
loop integrals relevant to the on-shell action, this subtlety will not directly concern us.
Therefore, in this work we may use retarded graviton propagators exclusively.
We will capture all conservative effects and conservative-like dissipative effects,
namely all dissipative effects that do not involve loss of four-momentum or angular momentum.
For aligned spins, this means that we compute the full scattering angle including dissipative effects.
At higher PM orders with more active gravitons,
this subtlety will become non-trivial and will have to be worked out in detail.

Going back to \cref{eq:N3}, if one ignores the routings of worldline propagators (i.e.~the arrows inside the boxes),
then the two combined terms have symmetry factors of $1/6$ and $1/2$ respectively.
We thus observe that, at least up to 3PM order, when simply ignoring the $\iO$ prescriptions on the worldlines,
$N$ is given by the sum of all vacuum diagrams with appropriate symmetry factors \cite{Kim:2024svw}; \cref{eq:N3NoWL} highlights this for the two-loop case.
Moreover, the relative factors of $2$ between the pairs of terms in \cref{eq:N3}
differing by $\iO$ prescriptions on worldline propagators can also be argued for on simple symmetry grounds.
Taking the bracket of this expression with $p_1^\mu$
and plugging it into \cref{3PMimpulse} in combination with the $N^{(1)}$ and $N^{(2)}$
contributions, it is a straightforward exercise to check that this reproduces
the full 3PM momentum impulse $\Delta p_1^{(3)\mu}$ at the level of the constituent diagrams.

\section{Generalised unitarity on the worldline}\label{sec:Unitarity}

Our diagrammatic representation of the on-shell action in \cref{eq:N1,eq:N2,eq:N3}
highlights a convenient feature.
Namely, the diagrams can be combined such that tree-level-amplitude-like groupings can be recognised on the individual worldlines.
Although no propagators in that analysis were cut,
this is suggestive of the applicability of generalised unitarity methods to the construction of the worldline on-shell action.
In this section, we formalise this synthesis of WQFT and on-shell methods.

In analogy to tree-level scattering amplitudes,
the basic building blocks of the on-shell action through
unitarity cuts are single-worldline $n$-graviton functions,
defined diagrammatically as
\begin{align}\label{eq:nGravFunction}
    \i\mathcal{W}^{(n)}(q_{1}^{\lambda_{1}},\dots,q_{n}^{\lambda_{n}})
    :=
    \epsilon_{\mu_{1}\nu_{1}}^{\lambda_{1}}(q_{1})\dots\epsilon_{\mu_{n}\nu_{n}}^{\lambda_{n}}(q_{n})
    \left[
        \begin{tikzpicture}[baseline={([yshift=-.5ex](.5,-3.5))}]
        %%%%%%%%%%%%%%%%%%%%%%
        \path [draw=black, worldlineStatic] (-3.,-3) -- (-1.,-3);
        \path [draw=black, photon] (-2.6,-4.) -- (-2.2,-3);
        \path [draw=black, photon] (-1.4,-4.) -- (-1.8,-3);
        \filldraw[fill=white, draw=black,thick] (-2.45,-3.2) rectangle (-1.55,-2.8);
        \node (A) at (-1.95,-3.8) {$\boldsymbol{\dots}$};
        \node (A1) at (-3.,-3.6) {$q_{1}\,$\rotatebox[origin=c]{70}{$\leftarrow$}};
        \node (A2) at (-1.,-3.6) {\rotatebox[origin=c]{-70}{$\rightarrow$}$\,q_{n}$};
    \end{tikzpicture}
    \right]^{\mu_{1}\nu_{1}\dots\mu_{n}\nu_{n}}.
\end{align}
The empty white box represents the sum of all possible tree-level diagrams without specifying the $\iO$ prescription on worldline propagators, and we have explicitly indicated the Lorentz structure of this sum of diagrams.
Graviton $i$ carries helicity $\lambda_{i}$,
and the polarisation tensors $\eps^\lambda_{\mu\nu}(q)$ are symmetric, traceless,
and transverse to their corresponding graviton momentum.
These objects are connected, meaning they can be written as
\begin{align}\label{eq:nGravConn}
    \mathcal{W}^{(n)}=\dd(v\cdot q)e^{\i q\cdot b}\hat{\mathcal{W}}^{(n)},
\end{align}
where $q^{\mu}=\sum_{i=1}^{n}q_{i}^{\mu}$ and $\hat{\mathcal{W}}^{(n)}$ is free of Dirac delta functions and impact-parameter-dependent phases.
Diagrammatically, this means that any two graviton-worldline vertices in a constituent diagram are joined by one of the worldline propagators in \cref{WLprops}, as is illustrated in \cref{eq:ComptonDiagrams,eq:ThreeGravComptonDiagrams}.

Our key result in this section is that the integrand of the on-shell action, at least up to the formal 3PM (i.e. two-loop) order, is constructible by cutting the graviton lines in \cref{eq:N1,eq:N2,eq:N3} and gluing together the resulting $n$-graviton functions of \cref{eq:nGravFunction}, \textit{\`{a} la} generalised unitarity:
\begin{subequations}\label{eq:NUnitarity}
\begin{align}
	&\i N^{(1)}=\,\,\begin{tikzpicture}[baseline={([yshift=-.5ex](.5,-.5))}]
        %%%%%%%%%%%%%%%%%%%%%%
        \path [draw=black, worldlineStatic] (-0.6,0) -- (0.6,0);
        \path [draw=black, worldlineStatic] (-0.6,-1.) -- (0.6,-1.);
        \path [draw=black, photon] (0,-1.) -- (0,0);
		\path [draw=red, style=dashed, line width=0.40mm] (-0.4,-0.5) -- (0.4,-0.5);
        \filldraw[fill=black] (0,0) circle (.06);
        \filldraw[fill=black] (0,-1.) circle (.06);
    \end{tikzpicture}\,,\label{eq:N1Unitarity} \\
	&\i N^{(2)}=\,\,
	\frac12\,
        \begin{tikzpicture}[baseline={([yshift=-.5ex](.5,-3.5))}]
            %%%%%%%%%%%%%%%%%%%%%%
            \path [draw=black, worldlineStatic] (-3.,-3) -- (-1.,-3);
            \path [draw=black, worldlineStatic] (-3.,-4.) -- (-1.,-4.);
            \path [draw=black, photon] (-2.4,-4.) -- (-2,-3);
            \path [draw=black, photon] (-1.6,-4.) -- (-2,-3);
	    \path [draw=red, style=dashed, line width=0.40mm] (-2.6,-3.5) -- (-1.4,-3.5);
            \filldraw[fill=white, draw=black,thick] (-2.25,-3.2) rectangle (-1.75,-2.8);
            \node (A) at (-2,-3.02) {$\rightarrow$};
            \filldraw[fill=black] (-2.4,-4.) circle (.06);
            \filldraw[fill=black] (-1.6,-4.) circle (.06);
        \end{tikzpicture}
	+\frac12\,
        \begin{tikzpicture}[baseline={([yshift=-.5ex](.5,-3.5))}]
            %%%%%%%%%%%%%%%%%%%%%%
            \path [draw=black, worldlineStatic] (-3.,-3) -- (-1.,-3);
            \path [draw=black, worldlineStatic] (-3.,-4.) -- (-1.,-4.);
            \path [draw=black, photon] (-2.4,-3) -- (-2,-4);
            \path [draw=black, photon] (-1.6,-3) -- (-2,-4);
	    \path [draw=red, style=dashed, line width=0.40mm] (-2.6,-3.5) -- (-1.4,-3.5);
            \filldraw[fill=white, draw=black,thick] (-2.25,-4.2) rectangle (-1.75,-3.8);
            \node (A) at (-2,-4.02) {$\rightarrow$};
            \filldraw[fill=black] (-2.4,-3) circle (.06);
            \filldraw[fill=black] (-1.6,-3) circle (.06);
        \end{tikzpicture}\,,\label{eq:N2Unitarity} \\
	&\i N^{(3)}=\,\,
	\frac{1}{3!}\left(\,\,\frac{2}{3}
    \begin{tikzpicture}[baseline={([yshift=-.5ex](.5,-3.5))}]
        %%%%%%%%%%%%%%%%%%%%%%
        \path [draw=black, worldlineStatic] (-3.,-3) -- (-1.,-3);
        \path [draw=black, worldlineStatic] (-3.,-4.) -- (-1.,-4.);
        \path [draw=black, photon] (-2.6,-4.) -- (-2.2,-3);
        \path [draw=black, photon] (-1.4,-4.) -- (-1.8,-3);
        \path [draw=black, photon] (-2,-4.) -- (-2,-3);
	\path [draw=red, style=dashed, line width=0.40mm] (-2.6,-3.5) -- (-1.4,-3.5);
        \filldraw[fill=white, draw=black,thick] (-2.45,-3.2) rectangle (-1.55,-2.8);
        \filldraw[fill=black] (-2.6,-4.) circle (.06);
        \filldraw[fill=black] (-2,-4.) circle (.06);
        \filldraw[fill=black] (-1.4,-4.) circle (.06);
        \node (A) at (-2,-3.02) {$\rightarrow\rightarrow$};
    \end{tikzpicture}+\frac{1}{3}
    \,
    \begin{tikzpicture}[baseline={([yshift=-.5ex](.5,-3.5))}]
        %%%%%%%%%%%%%%%%%%%%%%
        \path [draw=black, worldlineStatic] (-3.,-3) -- (-1.,-3);
        \path [draw=black, worldlineStatic] (-3.,-4.) -- (-1.,-4.);
        \path [draw=black, photon] (-2.6,-4.) -- (-2.2,-3);
        \path [draw=black, photon] (-1.4,-4.) -- (-1.8,-3);
        \path [draw=black, photon] (-2,-4.) -- (-2,-3);
	\path [draw=red, style=dashed, line width=0.40mm] (-2.6,-3.5) -- (-1.4,-3.5);
        \filldraw[fill=white, draw=black,thick] (-2.45,-3.2) rectangle (-1.55,-2.8);
        \filldraw[fill=black] (-2.6,-4.) circle (.06);
        \filldraw[fill=black] (-2,-4.) circle (.06);
        \filldraw[fill=black] (-1.4,-4.) circle (.06);
        \node (A) at (-2,-3.02) {$\rightarrow\leftarrow$};
    \end{tikzpicture}
    +
    \text{reflected}
    \,\,\right)\label{eq:N3Unitarity} \\
    &+
    \frac12\left[\left(\left(\,\,\frac{2}{3}\,\,
    \begin{tikzpicture}[baseline={([yshift=-.5ex](.5,-3.5))}]
        %%%%%%%%%%%%%%%%%%%%%%
        \path [draw=black, worldlineStatic] (-3.,-3) -- (-1.,-3);
        \path [draw=black, worldlineStatic] (-3.,-4.) -- (-1.,-4.);
        \path [draw=black, photon] (-2.4,-4.) -- (-2.4,-3);
        \path [draw=black, photon] (-1.6,-4.) -- (-2.4,-3);
        \path [draw=black, photon] (-1.6,-4.) -- (-1.6,-3);
	\path [draw=red, style=dashed, line width=0.40mm] (-2.6,-3.5) -- (-1.4,-3.5);
        \filldraw[fill=white, draw=black,thick] (-2.65,-3.2) rectangle (-2.15,-2.8);
        \filldraw[fill=white, draw=black,thick] (-1.85,-4.2) rectangle (-1.35,-3.8);
        \filldraw[fill=black] (-2.4,-4.) circle (.06);
        \filldraw[fill=black] (-1.6,-3.) circle (.06);
            \node (A) at (-2.4,-3.03) {$\rightarrow$};
        \node (A) at (-1.6,-4.03) {$\rightarrow$};
    \end{tikzpicture}+
    \frac{1}{3}\begin{tikzpicture}[baseline={([yshift=-.5ex](.5,-3.5))}]
        %%%%%%%%%%%%%%%%%%%%%%
        \path [draw=black, worldlineStatic] (-3.,-3) -- (-1.,-3);
        \path [draw=black, worldlineStatic] (-3.,-4.) -- (-1.,-4.);
        \path [draw=black, photon] (-2.4,-4.) -- (-2.4,-3);
        \path [draw=black, photon] (-1.6,-4.) -- (-2.4,-3);
        \path [draw=black, photon] (-1.6,-4.) -- (-1.6,-3);
	\path [draw=red, style=dashed, line width=0.40mm] (-2.6,-3.5) -- (-1.4,-3.5);
        \filldraw[fill=white, draw=black,thick] (-2.65,-3.2) rectangle (-2.15,-2.8);
        \filldraw[fill=white, draw=black,thick] (-1.85,-4.2) rectangle (-1.35,-3.8);
        \filldraw[fill=black] (-2.4,-4.) circle (.06);
        \filldraw[fill=black] (-1.6,-3.) circle (.06);
        \node (A) at (-2.4,-3.03) {$\rightarrow$};
        \node (A) at (-1.6,-4.03) {$\leftarrow$};
    \end{tikzpicture}
    \,\,\right)
    \cup
    \,\,
    \begin{tikzpicture}[baseline={([yshift=-.5ex](.5,-3.5))}]
        %%%%%%%%%%%%%%%%%%%%%%
        \path [draw=black, worldlineStatic] (-3.,-3) -- (-1.,-3);
        \path [draw=black, worldlineStatic] (-3.,-4.) -- (-1.,-4.);
        \path [draw=black, photon] (-2.4,-4.) -- (-2.4,-3);
        \path [draw=black, photon] (-2.4,-3.) to[out=-60,in=-120] (-1.6,-3);
        \path [draw=black, photon] (-1.6,-4.) -- (-1.6,-3);
	\path [draw=red, style=dashed, line width=0.40mm] (-2.6,-3.5) -- (-2.2,-3.5);
	\path [draw=red, style=dashed, line width=0.40mm] (-1.8,-3.5) -- (-1.4,-3.5);
	\path [draw=red, style=dashed, line width=0.40mm] (-2.,-3.1) -- (-2.,-3.5);
        \filldraw[fill=white, draw=black,thick] (-2.65,-3.2) rectangle (-2.15,-2.8);
        \filldraw[fill=white, draw=black,thick] (-1.85,-3.2) rectangle (-1.35,-2.8);
        \filldraw[fill=black] (-2.4,-4.) circle (.06);
        \filldraw[fill=black] (-1.6,-4.) circle (.06);
        \node (A) at (-2.4,-3.03) {$\rightarrow$};
        \node (A) at (-1.6,-3.03) {$\rightarrow$};
    \end{tikzpicture}\right)
        \cup
    \text{reflected}
    \,\,\right]\!.\nn
\end{align}
\end{subequations}
We adopt the dashed red line notation of ref.~\cite{Brandhuber:2021eyq} to indicate the cut construction of the integrand.
For example, 
\begin{subequations}\label{eq:CutIntegrands}
\begin{align}
    \begin{tikzpicture}[baseline={([yshift=-.5ex](.5,-3.5))}]
            %%%%%%%%%%%%%%%%%%%%%%
            \path [draw=black, worldlineStatic] (-3.,-3) -- (-1.,-3);
            \path [draw=black, worldlineStatic] (-3.,-4.) -- (-1.,-4.);
            \path [draw=black, photon] (-2.4,-4.) -- (-2,-3);
            \path [draw=black, photon] (-1.6,-4.) -- (-2,-3);
	    \path [draw=red, style=dashed, line width=0.40mm] (-2.6,-3.5) -- (-1.4,-3.5);
            \filldraw[fill=white, draw=black,thick] (-2.25,-3.2) rectangle (-1.75,-2.8);
            \node (A) at (-2,-3.02) {$\rightarrow$};
            \filldraw[fill=black] (-2.4,-4.) circle (.06);
            \filldraw[fill=black] (-1.6,-4.) circle (.06);
        \end{tikzpicture}
        &=
        \i^{2}\int_{\ell_{1},\ell_{2}}\sum_{\lambda_{i}}\frac{\i\mathcal{W}_{1,\rightarrow}^{(2)}(\ell_{1}^{\lambda_{1}},\ell_{2}^{\lambda_{2}})\prod_{j=1}^{2}\i\mathcal{W}_{2}^{(1)}(-\ell_{j}^{-\lambda_{j}})}{\ell_{1}^2\ell_{2}^2} \\
        \begin{tikzpicture}[baseline={([yshift=-.5ex](.5,-3.5))}]
        %%%%%%%%%%%%%%%%%%%%%%
        \path [draw=black, worldlineStatic] (-3.,-3) -- (-1.,-3);
        \path [draw=black, worldlineStatic] (-3.,-4.) -- (-1.,-4.);
        \path [draw=black, photon] (-2.6,-4.) -- (-2.2,-3);
        \path [draw=black, photon] (-1.4,-4.) -- (-1.8,-3);
        \path [draw=black, photon] (-2,-4.) -- (-2,-3);
	\path [draw=red, style=dashed, line width=0.40mm] (-2.6,-3.5) -- (-1.4,-3.5);
        \filldraw[fill=white, draw=black,thick] (-2.45,-3.2) rectangle (-1.55,-2.8);
        \filldraw[fill=black] (-2.6,-4.) circle (.06);
        \filldraw[fill=black] (-2,-4.) circle (.06);
        \filldraw[fill=black] (-1.4,-4.) circle (.06);
        \node (A) at (-2,-3.02) {$\rightarrow\rightarrow$};
    \end{tikzpicture}%&=\i^{3}\int_{q\ell_{1}\ell_{2}}\sum_{\lambda_{i}}\frac{\i\mathcal{W}_{1,\rightarrow\rightarrow}^{(3)}(\ell_{1}^{\lambda_{1}},\ell_{2}^{\lambda_{2}},\ell_{3}^{\lambda_{3}})\i\mathcal{W}_{2}^{(1)}(-\ell_{1}^{-\lambda_{1}})\i\mathcal{W}_{2}^{(1)}(-\ell_{2}^{-\lambda_{2}})\i\mathcal{W}_{2}^{(1)}(-\ell_{3}^{-\lambda_{3}})}{\ell_{1}^2\ell_{2}^2\ell_{3}^2},
    &=\i^{3}\int_{\ell_{1},\ell_{2},\ell_{3}}\sum_{\lambda_{i}}\frac{\i\mathcal{W}_{1,\rightarrow\rightarrow}^{(3)}(\ell_{1}^{\lambda_{1}},\ell_{2}^{\lambda_{2}},\ell_{3}^{\lambda_{3}})\prod_{j=1}^{3}\i\mathcal{W}_{2}^{(1)}(-\ell_{j}^{-\lambda_{j}})}{\ell_{1}^2\ell_{2}^2\ell_{3}^2},
\end{align}
\end{subequations}
with other cuts being evaluated similarly.
Here, one of the loop integrations can be cast as a Fourier transform over the total transferred momentum $q^\mu=-\sum_{i=1}^{n}\ell_{i}^\mu$, the overall powers of $\i$ correspond to how many propagators are taken on shell in the numerator, and the arrow subscripts on the $\mathcal{W}^{(n)}$ correspond to the direction of causality flow in the graph.
Each $\mathcal{W}^{(n)}$ has also been given a worldline label.

Our integrals are performed in $D=4-2\eps$ dimensions, with measure
\begin{align}\label{eq:LoopMeasure}
    \int_{\ell}\,:=\int\!\frac{{\rm d}^{D}\ell}{(2\pi)^{D}}\,.
\end{align}
In line with previous calculations \cite{Bern:2019nnu,Bern:2019crd,Brandhuber:2021eyq},
we find that it is nevertheless sufficient to construct integrands in four dimensions.
Intuitively this is because  the on-shell action in four dimensions is finite and
there is no need for a specific dimensional regularisation scheme ---
any discrepancies caused by use of a four vs.~$D$-dimensional integrand will ultimately
disappear in the $\eps\to0$ limit. On a technical level this is found to be true as
divergences in $\epsilon$ only arise from master integrals with worldline propagators.
On the other hand, the upshot of summing identical cuts with different $\iO$ prescriptions on worldline propagators is precisely the cancellation of these master integrals, as can be easily seen from their values in ref.~\cite{Jakobsen:2022fcj}.\footnote{The master integral in eq.~(A8b) of ref.~\cite{Jakobsen:2022fcj} does not contribute to the conservative-like on-shell action.}
Thus, $\epsilon/\epsilon$ effects never become finite at the orders considered.
The polarisation sum is accordingly evaluated in four dimensions:
\begin{align}\label{eq:PolSum}
    \sum_{\lambda}\epsilon_{\mu\nu}^{\lambda}(q)\epsilon_{\alpha\beta}^{-\lambda}(-q)&=\Pi_{\mu(\alpha}\Pi_{\beta)\nu}-\frac{1}{2}\Pi_{\mu\nu}\Pi_{\alpha\beta}\,,
\end{align}
with $\Pi^{\mu\nu}(q,k)=\eta^{\mu\nu}-(q^{\mu}k^{\nu}+q^{\nu}k^{\mu})/(q\cdot k)$ for some reference vector $k^{\mu}$.
Apart from the requirement that $k^{\mu}$ be null, the reference vector may be freely chosen for each cut graviton; the final answer will not depend on the vectors chosen.
The right-hand side of the sum resembles $P_{\mu\nu;\alpha\beta}$ from the graviton
propagator in \cref{eq:gravProp}, only with $\eta^{\mu\nu}\rightarrow\Pi^{\mu\nu}$.
Such an augmentation is necessary to ensure that only on-shell graviton modes cross the cut, and can be thought of as compensating for terms in the graphs of \cref{eq:nGravFunction} that are killed by transversality and tracelessness of the polarizations.
Up to quadratic order in spin, all of the $\mathcal{W}^{(n)}$ entering the cuts in \cref{eq:NUnitarity} satisfied generalised Ward identities, such that $P_{\mu\nu;\alpha\beta}$ could be used in place of the right-hand side of \cref{eq:PolSum} \cite{Kosmopoulos:2020pcd}.
We could not use such a simplification at cubic and quartic orders in spin.

Our justification for the cut decomposition \cref{eq:NUnitarity} relies on the fact that,
at all orders in perturbation theory considered,
the on-shell action in momentum space must be non-analytic in the transferred momentum $q^\mu$.
This is required in order to produce long-range effects in impact-parameter ($b^\mu$) space.
Up to 2PM, non-analyticity in momentum space simply requires that no
graviton propagators in \cref{eq:N1,eq:N2} be pinched.
This will be achieved if the integrand is constructed on the mass shell of the transferred gravitons: said otherwise, the relevant part of the integrand is captured by the cut of the transferred gravitons.
The same is true at 3PM in the first line of \cref{eq:N3}.
On the other hand, as they can go on shell --- and thereby give integrals with non-vanishing radiative regions --- the middle gravitons in each diagram of the second line (dubbed the N topology for the two on the left, and the mushroom topology for the one on the right) of \cref{eq:N3} can be pinched without spoiling non-analyticity of the impact-parameter-space on-shell action.

The topologies resulting from such a pinch in the N topology are
\begin{align}\label{eq:PinchedN}
    \vcenter{\hbox{\begin{tikzpicture}[baseline={([yshift=-.5ex](.5,-3.5))}]
            %%%%%%%%%%%%%%%%%%%%%%
            \path [draw=black, worldlineStatic] (1.2,-3) -- (2.8,-3);
            \path [draw=black, photon] (1.6,-3) -- (2,-3.5);
            \path [draw=black, photon] (2.4,-3) -- (2,-3.5);
            \path [draw=black, photon] (2,-4) -- (2,-3.5);
            \path [draw=black, photon] (1.6,-4.5) -- (2,-4);
            \path [draw=black, photon] (2.4,-4.5) -- (2,-4);
            \filldraw[fill=black] (1.6,-3) circle (.06);
            \filldraw[fill=black] (2.4,-3) circle (.06);
            \filldraw[fill=black] (1.6,-4.5) circle (.06);
            \filldraw[fill=black] (2.4,-4.5) circle (.06);
            \filldraw[fill=black] (2,-3.5) circle (.06);
            \filldraw[fill=black] (2,-4) circle (.06);
            \path [draw=black, worldlineStatic] (1.2,-4.5) -- (2.8,-4.5);
    \end{tikzpicture}}}\,,\quad
    \vcenter{\hbox{\begin{tikzpicture}[baseline={([yshift=-.5ex](.5,-3.5))}]
            %%%%%%%%%%%%%%%%%%%%%%
            \path [draw=black, worldlineStatic] (1.2,-3) -- (2.8,-3);
            \path [draw=black, photon] (1.6,-3) -- (2,-3.75);
            \path [draw=black, photon] (2.4,-3) -- (2,-3.75);
            \path [draw=black, photon] (1.6,-4.5) -- (2,-3.75);
            \path [draw=black, photon] (2.4,-4.5) -- (2,-3.75);
            \filldraw[fill=black] (1.6,-3) circle (.06);
            \filldraw[fill=black] (2.4,-3) circle (.06);
            \filldraw[fill=black] (1.6,-4.5) circle (.06);
            \filldraw[fill=black] (2.4,-4.5) circle (.06);
            \filldraw[fill=black] (2,-3.75) circle (.06);
            \path [draw=black, worldlineStatic] (1.2,-4.5) -- (2.8,-4.5);
    \end{tikzpicture}}}\,.
\end{align}
Contributions to classical observables from these topologies vanish in four dimensions.
To see this, we note that they involve a product of two one-loop integrals, which are finite in dimensional regularisation.
Similarly, the tree-level building blocks entering the cuts are naturally also finite in the limit $D\rightarrow4$.
On the other hand, the Fourier transform into impact parameter space will enter at $\cO(D-4)$,
such that, when all pieces are assembled, these topologies vanish in the $D\rightarrow4$ limit.
It is even simpler to see that pinching the middle graviton in the mushroom topologies gives zero, since the resulting integrals are scaleless in one of the loop momenta.
We conclude that it is also safe at 3PM to construct the integrand by cutting the gravitons in \cref{eq:N3}. \Cref{eq:NUnitarity} summarises our findings.

As we took care not to overcount diagrams in \cref{eq:N3}, so too must we merge the cuts in the 3PM-1SF sector, i.e. the second line of \cref{eq:N3Unitarity}.
This is again indicated by the $\cup$ symbol.
Our method for merging the cuts (rather than the diagrams) is as follows (see also ref.~\cite{Brandhuber:2021eyq}):
\begin{enumerate}
\item Having constructed our integrand using generalised unitarity,
we reduce the integrals using integration-by-parts (IBP) algorithms to a set of master integrals.
We choose the sets adopted in refs.~\cite{Jakobsen:2022fcj,Jakobsen:2022psy} respectively for the 1SF and 0SF integrals.
\item We identify the master integrals common to all 1SF cuts,
and observe that these are only the integrals with an H topology, \cref{eq:HTopology}.
In addition to being present in all four 1SF cuts, we find that the coefficients of these master integrals are identical in all four cut sectors.
Since this contribution is repeated four times, avoiding overcounting means that we simply scale these master integrals by $1/4$ before summing the four 1SF cuts.
\end{enumerate}
In contrast with comparable scattering amplitudes approaches,
the 1SF cuts need not be merged with the 0SF ones~\cite{Bern:2019nnu,Bern:2019crd}.
With the topology in \cref{eq:HTopology} captured by the 1SF cuts in \cref{eq:N3Unitarity}, there is no need to separately include an H cut,
\begin{align}\label{eq:HCut}
        \vcenter{\hbox{\begin{tikzpicture}[baseline={([yshift=-.5ex](.5,-3.5))}]
            %%%%%%%%%%%%%%%%%%%%%%
            \draw[pattern=north east lines] (2.,-3.75) ellipse (.6cm and 0.25cm);
            \path [draw=black, worldlineStatic] (1.2,-3) -- (2.8,-3);
            \path [draw=black, worldlineStatic] (1.2,-4.5) -- (2.8,-4.5);
            \path [draw=black, photon] (1.6,-3) -- (1.6,-3.57);
            \path [draw=black, photon] (1.6,-3.93) -- (1.6,-4.5);
            \path [draw=black, photon] (2.4,-3) -- (2.4,-3.57);
            \path [draw=black, photon] (2.4,-3.93) -- (2.4,-4.5);
            \filldraw[fill=black] (1.6,-3) circle (.06);
            \filldraw[fill=black] (2.4,-3) circle (.06);
            \filldraw[fill=black] (1.6,-4.5) circle (.06);
            \filldraw[fill=black] (2.4,-4.5) circle (.06);
            \path [draw=red, style=dashed, line width=0.40mm] (1.42,-3.25) -- (1.82,-3.25);
            \path [draw=red, style=dashed, line width=0.40mm] (2.22,-3.25) -- (2.62,-3.25);
            \path [draw=red, style=dashed, line width=0.40mm] (1.42,-4.25) -- (1.82,-4.25);
            \path [draw=red, style=dashed, line width=0.40mm] (2.22,-4.25) -- (2.62,-4.25);
    \end{tikzpicture}}}\,,
\end{align}
as was noted in ref.~\cite{Brandhuber:2021eyq}.
This is because the only unique information in this cut corresponds to the topologies (\ref{eq:PinchedN}), which vanish in four dimensions.

A word is in order about IBP reducing the integrals appearing in the cuts.
Taking as an example the N cut, once an integral in this sector is reduced it may depend on a master integral with a reflected-N topology.
Given the construction of this piece of the integrand specifically on the residues of the N cut, we would expect that with this cut we have not faithfully captured all information about contributions from such master integrals.
On the other hand, we are guaranteed by construction that the reflected-N cut will capture all this information (while in turn missing pieces with an N topology).
Therefore, when reducing the N cut, we discard any masters with vanishing residues on any of the three cut graviton momenta.
Practically, for an integral in the N sector,
\begin{align}
    \int_{\ell_{1}\ell_{2}}\frac{\dd(v_{1}\cdot\ell_{2})\dd(v_{2}\cdot\ell_{1})}{\mathcal{D}}\frac{1}{\ell_{1}^{2}\ell_{2}^{2}\ell_{3}^{2}},
\end{align}
with some topology-appropriate denominator $\mathcal{D}$, we perform the IBP reduction for the integral with $1/\ell_{i}^{2}\rightarrow\dd(\ell_{i}^{2})$ instead.
When we plug in expressions for the master integrals we once again treat all graviton propagators as off-shell.
Whatever information is lost by this procedure is recovered by the reflected-N cut.

%%%%%%%%%%%%%%%%%%%%%%%%%%%%%%%%%%%%%%%%%%%%%%%%%

\section{Connection to scattering amplitudes}\label{sec:AmplitudesConnection}

Having now imported generalised unitarity into the WQFT,
as exemplified by \cref{eq:NUnitarity},
in this section we use this to argue for the equivalence of WQFT and
scattering amplitude extraction of classical observables.
To begin, let us focus on $\hat{\mathcal{W}}^{(n)}$, defined implicitly in \cref{eq:nGravConn}.
These objects correspond directly to the classical amplitudes for the scattering of $n$ gravitons off a massive spinning body.
Computing the one- and two-graviton functions at leading order Feynman-diagrammatically from the action in \cref{eq:sWQFT2}, we find\footnote{We have also produced $\hat{\mathcal{W}}^{(3)}$ up to cubic order in spin, and find that \cref{eq:12GravAmps} extends to the spinless three-graviton case when comparing with ref.~\cite{Brandhuber:2023hhy}.
We are not aware of results in the literature for $\cM^{(3)}$ with spin.}
\begin{align}\label{eq:12GravAmps}
    \hat{\mathcal{W}}^{(1,2)}(Z)&=-\frac{1}{2m}\left[1+\frac{\i (q\cdot Z)}{m}\right]\cM^{(1,2)},
\end{align}
where we have highlighted the dependence of $\hat{\mathcal{W}}^{(n)}$ on the SSC vector.
Here, $\cM^{(1)}$ is precisely the tree-level three-point amplitude describing a graviton emitted from a Kerr black hole to all orders in the spin tensor \cite{Vines:2017hyw,Levi:2015msa,Guevara:2018wpp,Chung:2018kqs,Guevara:2019fsj}.
Adding a graviton, $\cM^{(2)}$ is the tree-level classical Kerr Compton amplitude up to fourth order in the spin tensor \cite{Arkani-Hamed:2017jhn,Guevara:2018wpp,Aoude:2020onz,Bautista:2022wjf,Bautista:2023sdf}.\footnote{Both $\cM^{(1,2)}$ are in the normalisation of ref.~\cite{Aoude:2023dui}.}
Deviating from Kerr by means of the action of ref.~\cite{Haddad:2024ebn}, $\hat{\mathcal{W}}^{(2)}$ also agrees with the neutron-star Compton amplitude computed in refs.~\cite{Saketh:2022wap,Haddad:2023ylx}.
The factor of $1/2m$ is easily understood as linking $\dd(v\cdot q)$ in \cref{eq:nGravConn} to the $\dd(2p\cdot q)$ imposing four-momentum conservation in the classical limit \cite{Kosower:2018adc}.
By way of \cref{eq:12GravAmps}, we have thus established a connection between
the WQFT on-shell building blocks and scattering amplitudes.

Two important differences between the WQFT on-shell building blocks~\eqref{eq:nGravConn}, \eqref{eq:12GravAmps}
and classical scattering amplitudes are apparent.
First is the overall energy-conserving delta function, whose influence becomes significant in loop-corrected one-massive-body processes.
In particular, its presence takes care of iterative contributions to loop calculations from worldline propagators, which are not present in the WQFT approach but which do emerge in the scattering amplitude.
At the same time, bulk fields may still introduce iterations into loops, placing \cref{eq:nGravFunction} somewhere between the exponential~\eqref{eq:exponential} and the $1+i\hat{T}$ representation of the S-matrix.
We consequently might expect that \cref{eq:12GravAmps} and its analogs with more gravitons are strictly leading-order statements.

Second is the SSC-dependent factor.
In appendix~\ref{app:SSC} we show that the factorisation
\begin{align}\label{qdotZ}
    \hat{\mathcal{W}}^{(n)}(Z)=\left[1+\frac{\i(q\cdot Z)}{m}\right]\hat{\mathcal{W}}^{(n)}(0)
\end{align}
results in the on-shell action depending on the SSC vector only through the
invariant impact parameter~\eqref{eq:GeneralisedIP}.
As such, we may construct the on-shell action instead with $\hat{\mathcal{W}}^{(n)}(0)$, i.e.~\textit{exactly the classical tree-level one-body amplitude}, per \cref{eq:12GravAmps}, and reintroduce the SSC vector at the end according to the shift $b_{i}^{\mu}\rightarrow b_{i}^{\mu}+Z_{i}^{\mu}$ for each worldline.
This also provides rigorous justification for \cref{Free2Gauge}.

We note the parallel of \eqn{qdotZ} with the covariant SSC-conserving Hamiltonian construction of WQFT in
\rcite{Haddad:2024ebn},
where it was observed that the linear-in-curvature Hamiltonian carries an overall factor of
$(1+|\pi|^{-1}\nabla_Z)$ for $\pi^\mu=m\dot{x}^\mu+\cdots$,
which ensures that the covariant SSC, $Z^{\mu}=0$, is preserved.
The same prefactor also occurs in our non-minimal Lagrangian $\cL_{\rm nm}$~\eqref{eq:Lnm}.
This connection is more than superficial: in \cref{app:SSC} we also show that the above factorisation of the SSC-vector dependence is a necessary and sufficient condition for conservation of the covariant SSC after a two-body encounter.

\subsection{Equating WQFT and scattering amplitude on-shell actions}

We have shown that the objects entering the cuts in \cref{eq:NUnitarity} are equivalent to classical tree-level scattering amplitudes, up to the delta functions in \cref{eq:nGravConn} (and a convention-dependent phase).
From this, it is immediately obvious that at 1PM order,
corresponding with the diagram in \cref{eq:N1Unitarity},
the same on-shell action for WQFT is derived as using scattering amplitudes.

At 2PM (and beyond),
we must do slightly more work to argue that \cref{eq:N2Unitarity} is identical to the amplitudes result.
Equivalence follows from reverse unitarity:
\begin{align}\label{eq:ReverseUnitarity}
    \dd(v_{i}\cdot\ell)=\frac{\i}{v_{i}\cdot\ell+\iO}+\frac{-\i}{v_{i}\cdot\ell-\iO}\,.
\end{align}
Inserting this into the cuts in \cref{eq:N2Unitarity}, we produce a sum of triangle integrals familiar from the one-loop calculation in scattering amplitudes.
Since the value of the triangle integral in the classical limit is real (see e.g.~\rcite{Holstein:2008sw}), both terms coming from reverse unitarity are equal, thus cancelling the factors of $1/2$ in \cref{eq:N2Unitarity} and giving the familiar sum of triangle cuts which isolate the classical 2PM amplitude~\cite{Holstein:2008sx,Bjerrum-Bohr:2018xdl,Cheung:2018wkq}.
Without applying \cref{eq:ReverseUnitarity} to ``un-cut" the static worldlines (the dotted lines), we may understand the factors of $1/2$ in \cref{eq:N2Unitarity} relative to the triangles in the classical amplitude as symmetry factors since these diagrams are left-right symmetric.
This symmetry is spoiled by the matter propagator in the scattering amplitude, such that no symmetry factor is required in such a calculation.

Moving to two-loop order (3PM),
the mapping between amplitudes and WQFT must relate the $\iO$ prescriptions of one formulation to the other.
One facet of this is the inclusion of identical cuts with different causality flows in \cref{eq:N3Unitarity}, which results in the cancellation of master integrals with worldline propagators thanks to the relative factors between identical master integrals with different $\iO$ prescriptions \cite{Jakobsen:2022fcj}.
This is related to the cancellation of super-classical parts of the amplitude.
To make this connection, we notice that the contribution from masters with worldline propagators is isolated by the cuts\footnote{Propagating worldline states across a cut is a subtle affair since the propagating object, $\mathcal{Z}_{I}^{\mu}$, is a composite field. An alternative method for cutting the worldlines rather takes residues of $n$-graviton functions, analogously to, e.g., eq.~(3.9) of ref.~\cite{Chen:2021kxt}.}\textsuperscript{,}\footnote{We have found that masters with one worldline propagator do not contribute to the on-shell action.}
\begin{align}
    \begin{tikzpicture}[baseline={([yshift=-.5ex](.5,-3.5))}]
        %%%%%%%%%%%%%%%%%%%%%%
        \path [draw=black, worldlineStatic] (-3.2,-3) -- (-2.8,-3);
        \path [draw=black, worldlineStatic] (-1.2,-3) -- (-0.8,-3);
        \path [draw=black, zParticleF] (-2.8,-3) -- (-2.,-3);
        \path [draw=black, zParticleF] (-2.,-3) -- (-1.2,-3);
        \path [draw=black, worldlineStatic] (-3.2,-4.) -- (-0.8,-4.);
        \path [draw=black, photon] (-2.8,-4.) -- (-2.8,-3);
        \path [draw=black, photon] (-1.2,-4.) -- (-1.2,-3);
        \path [draw=black, photon] (-2,-4.) -- (-2,-3);
	    \path [draw=red, style=dashed, line width=0.40mm] (-3.0,-3.5) -- (-2.6,-3.5);
        \path [draw=red, style=dashed, line width=0.40mm] (-2.2,-3.5) -- (-1.8,-3.5);
        \path [draw=red, style=dashed, line width=0.40mm] (-1.4,-3.5) -- (-1.0,-3.5);
        \path [draw=red, style=dashed, line width=0.40mm] (-2.4,-2.8) -- (-2.4,-3.2);
        \path [draw=red, style=dashed, line width=0.40mm] (-1.6,-2.8) -- (-1.6,-3.2);
        \filldraw[fill=black] (-2.8,-4.) circle (.06);
        \filldraw[fill=black] (-2,-4.) circle (.06);
        \filldraw[fill=black] (-1.2,-4.) circle (.06);
        \filldraw[fill=black] (-2.8,-3.) circle (.06);
        \filldraw[fill=black] (-2,-3.) circle (.06);
        \filldraw[fill=black] (-1.2,-3.) circle (.06);
    \end{tikzpicture}\,\,,
    \quad
    \begin{tikzpicture}[baseline={([yshift=-.5ex](.5,-3.5))}]
        %%%%%%%%%%%%%%%%%%%%%%
        \path [draw=black, worldlineStatic] (-3.2,-3) -- (-2.8,-3);
        \path [draw=black, worldlineStatic] (-2.,-3) -- (-0.8,-3);
        \path [draw=black, zParticleF] (-2.8,-3) -- (-2.,-3);
        \path [draw=black, zParticleF] (-2.,-4) -- (-1.2,-4);
        \path [draw=black, worldlineStatic] (-3.2,-4.) -- (-2.,-4.);
        \path [draw=black, worldlineStatic] (-1.2,-4.) -- (-0.8,-4.);
        \path [draw=black, photon] (-2.8,-4.) -- (-2.8,-3);
        \path [draw=black, photon] (-1.2,-4.) -- (-1.2,-3);
        \path [draw=black, photon] (-2,-4.) -- (-2,-3);
	    \path [draw=red, style=dashed, line width=0.40mm] (-3.0,-3.5) -- (-2.6,-3.5);
        \path [draw=red, style=dashed, line width=0.40mm] (-2.2,-3.5) -- (-1.8,-3.5);
        \path [draw=red, style=dashed, line width=0.40mm] (-1.4,-3.5) -- (-1.0,-3.5);
        \path [draw=red, style=dashed, line width=0.40mm] (-2.4,-2.8) -- (-2.4,-3.2);
        \path [draw=red, style=dashed, line width=0.40mm] (-1.6,-3.8) -- (-1.6,-4.2);
        \filldraw[fill=black] (-2.8,-4.) circle (.06);
        \filldraw[fill=black] (-2,-4.) circle (.06);
        \filldraw[fill=black] (-1.2,-4.) circle (.06);
        \filldraw[fill=black] (-2.8,-3.) circle (.06);
        \filldraw[fill=black] (-2,-3.) circle (.06);
        \filldraw[fill=black] (-1.2,-3.) circle (.06);
    \end{tikzpicture}\,\,,
\end{align}
at 0SF and 1SF respectively, and their analogs with both worldlines swapped.
Expanding the delta functions on the static worldlines using reverse unitarity, we find exactly the double-(crossed-)box integrals which produce the super-classical contributions to the two-loop amplitude
\cite{Bern:2019nnu,Bern:2019crd,Bjerrum-Bohr:2021din}.
Thus, the cancellation of super-classical parts of the amplitude is linked to
the presence of identical cuts with different $\iO$ prescriptions in the WQFT on-shell action.

To complete the connection to amplitudes, it will be useful to introduce
a new representation of \cref{eq:N3Unitarity}:
\be
\begin{aligned}
    \i N^{(3)}=\,\,&
	\frac{1}{3!}\left(\,\,
    \begin{tikzpicture}[baseline={([yshift=-.5ex](.5,-3.5))}]
        %%%%%%%%%%%%%%%%%%%%%%
        \path [draw=black, worldlineStatic] (-3.,-3) -- (-1.,-3);
        \path [draw=black, worldlineStatic] (-3.,-4.) -- (-1.,-4.);
        \path [draw=black, photon] (-2.6,-4.) -- (-2.2,-3);
        \path [draw=black, photon] (-1.4,-4.) -- (-1.8,-3);
        \path [draw=black, photon] (-2,-4.) -- (-2,-3);
	\path [draw=red, style=dashed, line width=0.40mm] (-2.6,-3.5) -- (-1.4,-3.5);
        \filldraw[fill=white, draw=black,thick] (-2.45,-3.2) rectangle (-1.55,-2.8);
        \filldraw[fill=white, pattern=north east lines, draw=black,thick] (-2.45,-3.2) rectangle (-1.55,-2.8);
        \filldraw[fill=black] (-2.6,-4.) circle (.06);
        \filldraw[fill=black] (-2,-4.) circle (.06);
        \filldraw[fill=black] (-1.4,-4.) circle (.06);
    \end{tikzpicture}
    +
    \text{reflected}
    \,\,\right)\label{eq:N3NoWL} \\
    +\,\,&
    \frac1{2!}\left[\left(\,\,
    \begin{tikzpicture}[baseline={([yshift=-.5ex](.5,-3.5))}]
        %%%%%%%%%%%%%%%%%%%%%%
        \path [draw=black, worldlineStatic] (-3.,-3) -- (-1.,-3);
        \path [draw=black, worldlineStatic] (-3.,-4.) -- (-1.,-4.);
        \path [draw=black, photon] (-2.4,-4.) -- (-2.4,-3);
        \path [draw=black, photon] (-1.6,-4.) -- (-2.4,-3);
        \path [draw=black, photon] (-1.6,-4.) -- (-1.6,-3);
	\path [draw=red, style=dashed, line width=0.40mm] (-2.6,-3.5) -- (-1.4,-3.5);
        \filldraw[fill=white, draw=black,thick] (-2.65,-3.2) rectangle (-2.15,-2.8);
        \filldraw[fill=white, draw=black,thick] (-1.85,-4.2) rectangle (-1.35,-3.8);
        \filldraw[fill=white, pattern=north east lines, draw=black,thick] (-2.65,-3.2) rectangle (-2.15,-2.8);
        \filldraw[fill=white, pattern=north east lines, draw=black,thick] (-1.85,-4.2) rectangle (-1.35,-3.8);
        \filldraw[fill=black] (-2.4,-4.) circle (.06);
        \filldraw[fill=black] (-1.6,-3.) circle (.06);
    \end{tikzpicture}
    \cup
    \begin{tikzpicture}[baseline={([yshift=-.5ex](.5,-3.5))}]
        %%%%%%%%%%%%%%%%%%%%%%
        \path [draw=black, worldlineStatic] (-3.,-3) -- (-1.,-3);
        \path [draw=black, worldlineStatic] (-3.,-4.) -- (-1.,-4.);
        \path [draw=black, photon] (-2.4,-4.) -- (-2.4,-3);
        \path [draw=black, photon] (-2.4,-3.) to[out=-60,in=-120] (-1.6,-3);
        \path [draw=black, photon] (-1.6,-4.) -- (-1.6,-3);
	\path [draw=red, style=dashed, line width=0.40mm] (-2.6,-3.5) -- (-2.2,-3.5);
	\path [draw=red, style=dashed, line width=0.40mm] (-1.8,-3.5) -- (-1.4,-3.5);
	\path [draw=red, style=dashed, line width=0.40mm] (-2.,-3.1) -- (-2.,-3.5);
        \filldraw[fill=white, draw=black,thick] (-2.65,-3.2) rectangle (-2.15,-2.8);
        \filldraw[fill=white, draw=black,thick] (-1.85,-3.2) rectangle (-1.35,-2.8);
        \filldraw[fill=white, pattern=north east lines, draw=black,thick] (-2.65,-3.2) rectangle (-2.15,-2.8);
        \filldraw[fill=white, pattern=north east lines, draw=black,thick] (-1.85,-3.2) rectangle (-1.35,-2.8);
        \filldraw[fill=black] (-2.4,-4.) circle (.06);
        \filldraw[fill=black] (-1.6,-4.) circle (.06);
    \end{tikzpicture}\right)
        \cup
    \text{reflected}
    \,\,\right]\!,
\end{aligned}
\ee
where shaded boxes indicate that all master integrals with worldline propagators should be set to zero; that is, super-classical contributions are ignored.
Having handled the flow of causality in this way, the relation of the remaining factors to symmetry factors of the diagrams mentioned at the end of \cref{sec:diagObservables} is apparent.
As for lower loops, matter propagators in the analogous scattering-amplitude diagrams spoil these symmetries.
Reverse unitarity reinstates these propagators in \cref{eq:N3NoWL}, and thus, as we will now show, accounts for the remaining prefactors, and also extracts all the cuts of the amplitudes calculation with the correct $\iO$ prescriptions. 

Written as in \cref{eq:N3NoWL}, the 0SF contribution (i.e. the first term in the first line of that equation) takes the schematic form
\begin{align}
    \frac{1}{3!}\mathcal{I}:=\frac{1}{3!}\int_{q\, \ell_{1}\ell_{2}}\frac{e^{-\i q\cdot b}\dd(v_{1}\cdot q)\dd(v_{2}\cdot q)\dd(v_{2}\cdot\ell_{1})\dd(v_{2}\cdot\ell_{2})}{\ell_{1}^{2}\ell_{2}^{2}(q-\ell_{1}-\ell_{2})^{2}}I_{\rm 0SF},
\end{align}
where $I_{\rm 0SF}$ is the cut constructed integrand with delta functions and phase factors extracted.
We may assume $I_{\rm 0SF}$ to be even under the reflections $\mathcal{R}=\{\ell_{1,2}\rightarrow-\ell_{1,2},q\rightarrow-q\}$,\footnote{Swapping the sign of the transfer momentum makes use of the fact that the Fourier transform only depends on the impact parameter through its modulus.} since any odd piece must vanish once integrated over the symmetric domain.
Expanding two of the delta functions using reverse unitarity gives
\begin{align}
    \mathcal{I}=-2\left(\mathcal{I}^{+}+\mathcal{I}^{-}\right),
\end{align}
where
\begin{align}\label{eq:0SFAmpInt}
    \mathcal{I}^{\sigma}:=\int_{q\, \ell_{1}\ell_{2}}\frac{e^{-\i q\cdot b}\dd(v_{1}\cdot q)\dd(v_{2}\cdot q)}{\ell_{1}^{2}\ell_{2}^{2}(q-\ell_{1}-\ell_{2})^{2}}\frac{I_{\rm 0SF}}{(v_{2}\cdot\ell_{1}+\iO)(\sigma v_{2}\cdot\ell_{2}+\iO)},
\end{align}
and we have used the reflection $\mathcal{R}$ to combine terms.
Partial fractioning $\mathcal{I}^{+}$ as in ref.~\cite{Driesse:2024xad} and relabelling loop momenta yields $\mathcal{I}^{+}=2\mathcal{I}^{-}$.\footnote{Note that this is the same relation as in eq.~(A8c) of ref.~\cite{Jakobsen:2022fcj}, except now the integrand does not contain factors of $\dd(v_{2}\cdot\ell_{i})$.} This means that the 0SF contribution is given by
\begin{align}
    \frac{1}{3!}\mathcal{I}=-\mathcal{I}^{-}.
\end{align}
Given the equivalence \cref{eq:12GravAmps} and its expected analog for three gravitons, the cut constructed integrand $I_{\rm 0SF}$ must be identical when computed in WQFT or scattering amplitudes.
With this in mind, $-\mathcal{I}^{-}$ is exactly the contribution from the fan cut (see, e.g., figure 12(a) of ref.~\cite{Bern:2019crd}) in the amplitudes calculation.\footnote{The additional sign arises because the amplitudes approach cuts two additional propagators (the matter propagators) relative to the WQFT approach, each of which introduces a factor of $\i$.}
As at one-loop order, including a causal direction on the static worldline through reverse unitarity cancels the $1/3!$ symmetry factor.

The 1SF case involves another consideration, as cuts in the second line of \cref{eq:N3NoWL} comprise exclusively planar topologies, in contrast to analogs in the amplitudes computation (e.g. figure 12(b) of ref.~\cite{Bern:2019crd}).
Un-cutting the static worldlines through reverse unitarity again yields the resolution.
The contribution from the N-cut is
\begin{align}
    \frac{1}{2!}\mathcal{J}:=\frac{1}{2!}\int_{q\,\ell_{1}\ell_{2}}\frac{e^{-\i q\cdot b}\dd(v_{1}\cdot q)\dd(v_{2}\cdot q)\dd(v_{1}\cdot\ell_{2})\dd(v_{2}\cdot\ell_{1})}{\ell_{1}^{2}\ell_{2}^{2}(q-\ell_{1}-\ell_{2})^{2}}I_{\rm N}=-\left(\mathcal{J}^{+}+\mathcal{J}^{-}\right),
\end{align}
where $\mathcal{J}^{\sigma}$ is defined analogously to \cref{eq:0SFAmpInt}, with the sign of the $v_{1}\cdot\ell_{2}$ propagator being controlled by $\sigma$.
These correspond to the topologies
\begin{alignat}{2}
    &\qquad\mathcal{J}^{+}\qquad\qquad &&\qquad\mathcal{J}^{-}\nn \\
    &\qquad\downarrow &&\qquad\downarrow \\
    &\begin{tikzpicture}[baseline={([yshift=-.5ex](.5,-3.5)
        )}]
        %%%%%%%%%%%%%%%%%%%%%%
        \path [draw=black, worldlineStatic] (-3.,-3) -- (-2.2,-3);
        \path [draw=black, worldlineStatic] (-1.6,-3) -- (-1.,-3);
        \path [draw=black, zParticle] (-2.2,-3) -- (-1.6,-3);
        \path [draw=black, worldlineStatic] (-3.,-4.) -- (-2.4,-4.);
        \path [draw=black, worldlineStatic] (-1.6,-4.) -- (-1.,-4.);
        \path [draw=black, zParticle] (-2.4,-4) -- (-1.6,-4);
        \path [draw=black, photon] (-2.4,-4.) -- (-2.4,-3);
        \path [draw=black, photon] (-1.6,-4.) -- (-2.4,-3);
        \path [draw=black, photon] (-1.6,-4.) -- (-1.6,-3);
	%\path [draw=red, style=dashed, line width=0.40mm] (-2.6,-3.5) -- (-1.4,-3.5);
        \filldraw[fill=white, draw=black,thick] (-2.65,-3.2) rectangle (-2.15,-2.8);
        \filldraw[fill=white, draw=black,thick] (-1.85,-4.2) rectangle (-1.35,-3.8);
        \filldraw[pattern=north east lines, draw=black,thick] (-2.65,-3.2) rectangle (-2.15,-2.8);
        \filldraw[pattern=north east lines, draw=black,thick] (-1.85,-4.2) rectangle (-1.35,-3.8);
        \filldraw[fill=black] (-2.4,-4.) circle (.06);
        \filldraw[fill=black] (-1.6,-3.) circle (.06);
    \end{tikzpicture}\qquad
    &&
    \begin{tikzpicture}[baseline={([yshift=-.5ex](.5,-3.5))}]
        %%%%%%%%%%%%%%%%%%%%%%
        \path [draw=black, worldlineStatic] (-3.,-3) -- (-2.2,-3);
        \path [draw=black, worldlineStatic] (-1.6,-3) -- (-1.,-3);
        \path [draw=black, zParticle] (-1.6,-3) -- (-2.4,-3);
        \path [draw=black, worldlineStatic] (-3.,-4.) -- (-2.4,-4.);
        \path [draw=black, worldlineStatic] (-1.6,-4.) -- (-1.,-4.);
        \path [draw=black, zParticle] (-2.4,-4) -- (-1.6,-4);
        \path [draw=black, photon] (-2.4,-4.) -- (-2.4,-3);
        \path [draw=black, photon] (-1.6,-4.) -- (-2.4,-3);
        \path [draw=black, photon] (-1.6,-4.) -- (-1.6,-3);
	%\path [draw=red, style=dashed, line width=0.40mm] (-2.6,-3.5) -- (-1.4,-3.5);
        \filldraw[fill=white, draw=black,thick] (-2.65,-3.2) rectangle (-2.15,-2.8);
        \filldraw[fill=white, draw=black,thick] (-1.85,-4.2) rectangle (-1.35,-3.8);
        \filldraw[pattern=north east lines, draw=black,thick] (-2.65,-3.2) rectangle (-2.15,-2.8);
        \filldraw[pattern=north east lines, draw=black,thick] (-1.85,-4.2) rectangle (-1.35,-3.8);
        \filldraw[fill=black] (-2.4,-4.) circle (.06);
        \filldraw[fill=black] (-1.6,-3.) circle (.06);
    \end{tikzpicture}\,\,,\nn
\end{alignat}
mapping on to the N and crossed-N cuts of the amplitudes calculation \cite{Bern:2019crd,Akpinar:2024meg}.
An identical manipulation decomposes the mushroom cut on the worldline to the mushroom and crossed-mushroom that contribute to the dissipative piece of the amplitude \cite{Bjerrum-Bohr:2021din}.
Notice that, again, all symmetry factors have canceled.

In this way, we have seen that we should expect complete equivalence between the WQFT and scattering amplitude on-shell action, at least up to two-loop order.
This is inherited via generalised unitarity from the leading-order equivalence $\hat{\mathcal{W}}^{(n)}(0)\propto\cM^{(n)}$ between the $n$-graviton function and the analogous scattering amplitude for one massive body.
In fact, we have not made reference to a specific theory here, and thus conclude that WQFT and scattering amplitudes will produce the same observables in any theory where this one-body equivalence holds.

Evaluating \cref{eq:NUnitarity}, we have produced the two-spinning-body on-shell action up to two loops and for spin orders $a_{1}^{n_{1}}a_{2}^{n_{2}}$ with $n_{1}+n_{2}\leq4$, 
with the exception of the $\cO(G^{3}a_{1}^{4}a_{2}^{0})$ contribution in the probe limit $m_{1}\ll m_{2}$, which is known from \cite{Akpinar:2025bkt,Hoogeveen:2025tew} and included in the
ancillary file.
In line with expectations, we obtain precisely the scattering-amplitude on-shell action where there is overlap \cite{Bern:2019nnu,Bjerrum-Bohr:2021din,Chen:2021kxt,Aoude:2022trd} and extend existing two-loop results which have considered only one spinning body \cite{Akpinar:2024meg,Akpinar:2025bkt}.
These results will be reported in the next section.

Our analysis here also implies that the WQFT approach will produce the classical two-body scattering amplitude itself as an intermediate quantity.
Specifically, the amplitude at $n^{\rm th}$ order in perturbation theory is the object whose Fourier transform is the on-shell action:
\begin{align}
    N^{(n)}=\int_{q}\dd(v_{1}\cdot q)\dd(v_{2}\cdot q)e^{-\i q\cdot b}\mathcal{M}^{(n)}_{\rm cl.}.
\end{align}
We have indeed verified that we reproduce known amplitudes everywhere we overlap with the literature \cite{Bern:2019nnu,Bjerrum-Bohr:2021din,Chen:2021kxt,Aoude:2022trd,Akpinar:2024meg,Akpinar:2025bkt}.
It was observed in ref.~\cite{Akpinar:2025bkt} that the spin-shift symmetry of refs.~\cite{Bern:2022kto,Aoude:2022trd} is exhibited by the spinning-spinless amplitude at 3PM in either probe limit.
We report the presence of this symmetry when either $m_{1}\ll m_{2}$ (0SF) or $m_{2}\ll m_{1}$ ($\bar 0$SF) also when both worldlines are spinning for all new sectors produced here: $\cO(G^{3}a_{1}^{2}a_{2}^{1})$, $\cO(G^{3}a_{1}^{2}a_{2}^{2})$, and $\cO(G^{3}a_{1}^{3}a_{2}^{1})$.

\section{Results}\label{sec:Observables}

In this section we present the final results of this work,
being the on-shell action and scattering observables up to $\cO(G^3S^4)$.
The summary here is accompanied by full results provided in a {\zenodo} repository.

\subsection{On-shell action}

In this work, we provide the on-shell action $N$ to $\cO(G^3 S^4)$.
Its simple building blocks in a PM expansion are made clear by a schematic expansion in the black hole masses $m_i$ and spin vectors $a_i^\mu$ with power counting parameters $\tilde a_i=1$ as follows,
\begin{align}\label{eq:NExpansion0}
    \frac{N}{G m_1 m_2}
    &=
    \sum_{n=0}^2
    \sum_{k=0}^{n}
    \sum_{l_1=0}^4
    \sum_{l_2=0}^{4-l_1}
    \frac{
        G^{n} m_1^{n-k} m_2^{k}
        \tilde a_1^{l_1} \tilde a_2^{l_2}
    }{|b|^{(n+l_1+l_2)}}
    c_{(n;n-k,k;l_1,l_2)}
    +
    \dots
\end{align}
where the ellipsis denotes terms of order $G^4$ or $S^5$.
Here, $c_{(n;k_1,k_2;l_1,l_2)}$ are expansion coefficients of $N$ with the formal PM order $n+1$, the order in each black hole mass $k_i$ and the order in each black hole spin $l_i$.
The expansion coefficients $c_{(n;k_1,k_2;l_1,l_2)}$ depend only on $\gamma$ and on the spin vectors projected onto a dimensionless basis of vectors: $\{v_i^\mu, \hat b^\mu,\hat L^\mu\}$.
Here, $\hat L^\mu$ is the unit orbital angular momentum vector orthogonal to $v_i^\mu$ and $b^\mu$,
introduced below in \cref{eq:AngularMomentum}.
Finally, we defined
\begin{align}\label{eq:funnyPower}
    \frac1{|b|^{(m)}}
    =
    \begin{cases}
        \log\frac{|b|}{|b_0|}
        & m=0\,,
        \\
        \frac1{|b|^m}
        & \text{otherwise}\,,
    \end{cases}
\end{align}
to capture the logarithmic behaviour of the non-spinning 0PM contribution with an arbitrary IR scale $|b_0|$.

The on-shell action $N$ is invariant under the exchange of the two black hole labels, $1\leftrightarrow2$, which for the expansion coefficients implies that
$
    c_{(n;k_1,k_2;l_1,l_2)}
    =
    c_{(n;k_2,k_1;l_2,l_1)}
$.
Thus, for example, one may focus on mass sectors with $k_2\ge k_1$ and for $k_1=k_2$ one may focus on spin sectors with $l_2\ge l_1$.
The implications of this $1\leftrightarrow2$ symmetry can be made more explicit by using symmetric variables.
For the spin vectors and velocities we use
\begin{align}\label{aPM}
    a_\pm^\mu
    &=
    a_1^\mu \pm a_2^\mu
    \,, &
    v_+^\mu&=v_1^\mu+v_2^\mu\,.
\end{align}
We also introduce the total mass, reduced mass, the symmetric mass ratio and mass difference:
\begin{align}
    M=m_1+m_2
    \ ,\quad
    \mu =\frac{m_1 m_2}{M}
    \ ,\quad
    \nu = \frac{\mu}{M}
    \ ,\quad
    \delta = \frac{m_1-m_2}{M} = \sqrt{1-4 \nu}
    \ ,
\end{align}
where $m_1\ge m_2$ is assumed in the last equality.

In terms of these building blocks, we expand the on-shell action in $G$ as
\begin{align}\label{eq:NExpansion}
    \frac{N}{|L|}
    =
    \Gamma\bigg[
%    \Big(
        \frac{GM}{|b|}
%    \Big)^{1}
    N_{(1,0)}
    +
    \bigg(\frac{GM}{|b|}\bigg)^2
    N_{(2,0)}
    +
    \bigg(\frac{GM}{|b|}\bigg)^3
    \big[
        \Gamma^2 N_{(3,0)}    
        +
        \nu
        N_{(3,1)}
    \big] 
    \bigg]
    +
    \cO(G^4) 
    \ ,
\end{align}
with spin-dependent expansion coefficients $N_{(n,m)}$ at $n$th PM order and $m$th self-force order.
The reduced energy, $\Gamma$, is defined as
\begin{align}
    \Gamma= \frac{E}{M}
    =\sqrt{1+2\nu(\gamma-1)}
    \ .
\end{align}
The spin expansion of $N_{(n,m)}$ reads
\begin{align}
    \frac{N_{(n,m)}}{|b|^{n-1}}
    =
    \sum_{s=0}^4
    \sum_{i=0}^s
    \frac{
        \tilde a_+^{s-i}
        \tilde a_-^i
    }{|b|^{(s+n-1)}}
    \delta^{\sigma(i)}
    N_{(n,m;s-i,i)}
    +
    \cO(S^5)
    \ ,
\end{align}
where $\tilde a_\pm=1$ are formal power counting parameters of $a_\pm^\mu$ and
\begin{align}
    \sigma(i)
    =
    \begin{cases}
        0,\qquad i\text{ even,}
        \\
        1,\qquad i\text{ odd.}
    \end{cases}
\end{align}
The function $\sigma(i)$ ensures that odd orders of $a_-^\mu$ come with a factor $\delta$~\cite{Buonanno:2024vkx}.
Finally, the power function $|b|^{(s)}$ is defined in Eq.~\eqref{eq:funnyPower} and the overall power of $|b|^{n-1}$ is there to insert the logarithm only at the spinless, 1PM order.

The expansion coefficients $N_{(n,m;s_+,s_-)}$ are functions only of the Lorentz factor $\gamma$ and the spin vectors projected onto the basis $\{v_+,\hat b^\mu,\hat L^\mu\}$,
\begin{align}
    \frac{a_\pm^\mu \hat L_\mu}{|b|}
    \ ,
    \qquad
    \frac{a_\pm^\mu \hat b_\mu}{|b|}
    \ ,
    \qquad
    \frac{a_\pm^\mu v_{+\mu}}{|b|}
    \ ,
\end{align}
where $L^\mu$ is the initial orbital angular momentum:
\begin{align}\label{eq:AngularMomentum}
    L^\sigma
    &=
    |L| \hat L^\sigma
    =
    -\frac{\mu}{\Gamma}
    \eps^\sigma_{\ \nu\ab}
    b^\nu v_1^\alpha v_2^\beta
    \ ,
    \qquad
    |L| 
    =
    \sqrt{-L^2}
    =
    p_{\infty}|b|
    \ .
\end{align}
Written out in terms of this basis, the on-shell action may be considered a function of the vectors $a_\pm^\mu$, $L^\mu$, $b^\mu$ and $v_i^\mu$ and does not explicitly depend on $\eps^{\mn\ab}$.
Once again, the spinless 1PM contribution $N_{(1,0;0,0)}$ additionally depends on $\log |b|/|b_0|$.

We note the following general symmetries of $N$:
\begin{align}
    N\Big|_{v_i^\mu\to-v_i^\mu}
    =
    N
    \ ,\qquad
    N\Big|_{\substack{a_i^\mu\to-a_i^\mu \\ L^\mu\to -L^\mu}}
    =
    N
    \ ,\qquad
    N\Big|_{b^\mu\to-b^\mu}
    =
    N
    \ .
\end{align}
The first symmetry is a time-reversal, leaving space-like vectors invariant but reversing time-like vectors, and tells us that $N$ gets contributions only from even-parity integrals
(in the sense of \rcite{Driesse:2024feo}).
The second symmetry is parity invariance and the last one follows from the two first ones together with the obvious symmetry that $N$ is invariant when all vectors flip their sign.

The schematic expansion of $N$ in eq.~\eqref{eq:NExpansion} clearly shows the combined PM and SF expansion.
At the leading 1PM and 2PM orders, the on-shell action $N$ is 0SF exact, such that no self-force or dissipative effects are present.
This changes at 3PM where 1SF effects appear.
This piece, the 3PM-1SF contribution, $N_{(3,1)}$, is, analytically, the most non-trivial part computed in this work and its full dependence on both $a_1^\mu$ and $a_2^\mu$ is a novel result not previously reported.
Conservative and dissipative effects are easily distinguished by the two regions of integration, potential and radiative.
Correspondingly, the 3PM-1SF on-shell action has a conservative part and a dissipative part,
\begin{align}
    N_{(3,1)}
    =
    N_{(3,1)}^{\rm (cons,rat)}
    +
    N_{(3,1)}^{\rm (cons,rap)}
    \text{arccosh}\gamma
    +
    N_{(3,1)}^{(\rm rad)}
    \mathcal{I}(\gamma)
    \ ,
\end{align}
with corresponding expansions of the spin coefficients $N_{(3,1;s_+,s_-)}$.
Here, $N_{(3,1)}^{(\dots)}$ are all polynomial in $\gamma$ (up to poles in $\gamma^2-1$) with a rational conservative piece, a rapidity conservative piece (as $\text{arccosh}\gamma$ is the rapidity), and a radiative piece.
The function $\mathcal{I}(\gamma)$ is
\begin{align}
    \mathcal{I}(\gamma)
    =
    \frac{\left(6 \gamma ^3-9 \gamma \right) \text{arccosh}\gamma+\left(8-5 \gamma ^2\right) \sqrt{\gamma ^2-1}}{3 \left(\gamma ^2-1\right)^{3/2}}
    \ ,
\end{align}
and appears in the 2PM loss of angular momentum~\cite{Damour:2020tta,Jakobsen:2021smu,Jakobsen:2021lvp}.
It enters our result via linear response, to be discussed in \cref{linearResponse}.

The mass dependence of Eq.~\eqref{eq:NExpansion0} implies that dependence on $\delta$ cannot appear at 3PM-1SF, and thus odd orders of $a_-^\mu$ cannot appear.
The only non-zero spin coefficients $N_{(3,1;s_+,s_-)}$, then, are those for even values of $s_-$.
The novel results of this work may be identified with the orders where dependence on $a_-^\mu$ is present: $N_{(3,1;1,2)}$, $N_{(3,1;2,2)}$ and $N_{(3,1;0,4)}$.

\subsection{Observables}

Observables are easily derived from the on-shell action through the repeated use of Poisson brackets.
Using our two-loop results for $N$, we have computed the impulse, the spin kick, and the change in the relative impact parameter:
\begin{subequations}
\begin{align}
    &\Delta p^\mu_{i,N}
    =
    \{
        N,
        {p_i^\mu}
    \}
    \!+\!
    \sfrac1{2!} 
    \big\{N,
        \{N,{p_i^\mu}
    \}
    \big\}
    \!+\!
    \sfrac1{3!} 
    \big\{N,\big\{N,
        \{N,{p_i^\mu}
    \}
    \big\}\big\}
    +
    \cO(G^4)\,,
    \\
        &\Delta S^\mu_{i,N}
    =
    \{
        N,
        {S_i^\mu}
    \}
    \!+\!
    \sfrac1{2!} 
    \big\{N,
        \{N,{S_i^\mu}
    \}
    \big\}
    \!+\!
    \sfrac1{3!} 
    \big\{N,\big\{N,
        \{N,{S_i^\mu}
    \}
    \big\}\big\}
    +
    \cO(G^4)\,,
    \\
    &\Delta b^\mu_N
    =
    \{
        N,
        {b^\mu}
    \}
    \!+\!
    \sfrac1{2!} 
    \big\{N,
        \{N,{b^\mu}
    \}
    \big\}
    \!+\!
    \sfrac1{3!} 
    \big\{N,\big\{N,
        \{N,{b^\mu}
    \}
    \big\}\big\}
    +
    \cO(G^4)
    \ .
\end{align}
\end{subequations}
We have put a subscript $N$ on these observables to highlight that they include only contributions that can be described by $N$.
The missing contributions are dissipative and related to loss of linear and angular momentum through gravitational waves.
This dissipative addition starts at 3PM for the impulse and spin kick but at 2PM for $\Delta b^\mu$ (or equivalently $\Delta J^\mu$).

The observables $\Delta \lambda_N^\mu$ are conservative-like,
implying that total linear and angular momentum are conserved:
\begin{align}\label{eq:ConservativeLike}
    \Delta J^\mu_N &= 0
    \ , &
    \Delta P^\mu_N &= 0
    \ ,
\end{align}
with $P^\mu$ and $J^\mu$ given by~\cite{Jakobsen:2022zsx}
\begin{align}
    P^\mu 
    &=
    p_1^\mu + p_2^\mu
    =
    E \hat P^\mu
    \ ,
    &
    J^\mu 
    &=
    L^\mu +
    \sum_i
    \Big(
    p_i\cdot \hat P
    a_i^\mu
    -
    a_i\cdot \hat P
    p_i^\mu
    \Big)
    \ .
\end{align}
The $N$-kicks of $P^\mu$ and $J^\mu$ (which are zero) may be computed either through their dependence on $p_i^\mu$, $a_i^\mu$ and $b^\mu$ and the $N$-kicks of $p_i^\mu$, $a_i^\mu$ and $b^\mu$, or directly through the brackets.
In particular,
\begin{align}
   \{N,  P^\mu\}&=0
    \ ,
    &
   \{N,  J^\mu\}&=0
    \ ,
\end{align}
which is an alternative way to verify \cref{eq:ConservativeLike}.

The observables inherit the same schematic structure as the on-shell action.
In particular, their 1PM and 2PM contributions are fully determined from the probe limit (i.e. 0SF) and 1SF contributions first appear at 3PM.
A simple special case is the aligned spin configuration, where we assume the spins to take the following form:
\begin{align}
    a^\mu_i
    \to
    G m_i \chi_i \ell^\mu
    \ .
\end{align}
The spin dependence of the on-shell action then reduces to dependence on the two scalars $\chi_i$, which for black holes are constrained to $|\chi_i| \le 1$.
The two signed spin lengths $\chi_i$ may be considered as constants
along with the masses $m_i$, and the only non-zero brackets are then the ones of $b^\mu$ and $v_i^\mu$.

For aligned spins, the motion is planar and the conservative-like motion generated by $N$
is encapsulated by a scattering angle:
\begin{align}
    \sin\left(\frac{\theta_{\rm aligned}}2\right)=
    \Gamma\frac{|\Delta p_{1,N}^\mu|}{2\mu\sqrt{\gamma^2-1}}\,.
\end{align}
Notice, though, that the scattering angle we obtain is the full angle containing both
conservative and dissipative effects ---
this is because we included dissipative effects in our calculation of $N$.
In particular, we show in Appendix~\ref{app:alignedSpins} that the scattering angle
$\theta_{\rm aligned}$ is also given simply by
\begin{align}
    \theta_{\text{aligned}}
    =
    -\frac{\d N_{\rm aligned}}{\d |L|}
    \ ,
\end{align}
which is the usual relationship between the on-shell (radial) action and the scattering angle.
The angle has a finite high-energy limit, wherein we send $\gamma\to\infty$:
\begin{align}\label{HELimit}
\begin{aligned}
    \theta&\xrightarrow[\gamma\to\infty]{}4\frac{GE}{|b|}
    \left(1-\frac{a_+}{|b|}+\frac{a_+^2}{|b|^2}-\frac{a_+^3}{|b|^3}+\frac{a_+^4}{|b|^4}+\cdots\right)\\
    &\qquad\quad+\frac{32}{3}\bigg(\frac{GE}{|b|}\bigg)^3
    \bigg(1-\frac{3 a_+}{|b|}+\frac{123a_+^2+3 a_-^2}{20 |b|^2}
    -\frac{85 a_+^3+3 a_+ a_-^2}{8 |b|^3}\\
    &\qquad\qquad\qquad\qquad\qquad+\frac{4647 a_+^4+162 a_+^2 a_-^2+27 a_-^4}{280 |b|^4}
    +\cdots\bigg)+\cdots\,,
\end{aligned}
\end{align}
$a_\pm$ being the signed lengths of $a_\pm^\mu$~\eqref{aPM}.
This result naturally extends the work of Amati, Ciafaloni and Veneziano~\cite{Amati:1990xe}
up to quartic order in spins, quadratic having already been done by some of the current authors~\cite{Jakobsen:2022fcj}.
Interestingly, the linear-in-$G$ terms are entirely generated by a small-spin expansion of $4GE/(|b|+a_+)$.

\subsection{Checks}

Our results pass several non-trivial checks against the literature.
First, turning off spin of one of the black holes, e.g. $a_2^{\mu}=0$, we reproduce the recent results of \rcite{Akpinar:2025bkt}.
In this case, we may compare directly at the level of the on-shell action where we find a complete match.
As previously mentioned, our new results extend the results of ref.~\cite{Akpinar:2025bkt} to include spins on both black holes, up to quartic order in spins.
For aligned spins, besides verifying finiteness of the high-energy limit~\eqref{HELimit},
we have also checked that a low-velocity PN expansion of our scattering angle reproduces corresponding the PN result of \rcite{Bautista:2024agp}.
In this comparison, we were able to determine the unknown constant $h_{22}$ of ref.~\cite{Bautista:2024agp}:
\begin{align}
    h_{22}=-\frac{1561}{5},
\end{align}
which is the 6PN contribution to the 3PM-1SF scattering angle at $\cO(a_{1}^{2}a_{2}^{2})$.
Additionally, we have checked our complete observables up to $G^3 S^2$ against refs.~\cite{Jakobsen:2022fcj,Jakobsen:2022zsx}, were we find a complete match including the radiative pieces,
and to $G^3 S^4$ at 0SF order against \rcite{Hoogeveen:2025tew}.
Finally, as we will cover in the next section,
all radiative parts of our on-shell action have been checked via linear response against the 2PM loss of angular momentum,
which was recently reported to the required spin order in \rcite{Alessio:2025flu}
(with previous similar results appearing in \rcite{Alessio:2022kwv}).

\subsection{Background invariance of $N$ and linear response}\label{linearResponse}
At one-loop order, it was noticed that the on-shell action is invariant under shifts in its background parameters~\cite{Jakobsen:2021zvh}.
We confirm that this is also the case at this order.
More precisely, we have,
\begin{align}
    N(\cO+\Delta\cO_{N})
    =
    N(\cO)
    +
    \cO(G^5)
    \ ,
\end{align}
where $N(\cO+\Delta \cO_N)$ instructs us to evaluate $N$ in terms of variables shifted by their $N$-generated kicks --- i.e. variables defined at infinity up to dissipative effects not captured by $N$, i.e. flux effects.

If we include flux effects, $\Delta J^{\mu}=\cO(G^{2})$ while $\Delta P^\mu=\cO(G^3)$.
We find, then, that $N$ obeys the following linear response formula at $G^3$ order:
\begin{align}
    \nn
    N\Big|_{\rm rad}
    &=
    \bigg(\frac{GM}{|b|}\bigg)^3
    M|b|\nu^2\sqrt{\gamma^2-1}
    \mathcal{I}(\gamma)
    N^{(\rm rad)}_{(3,1)}
    \\
    &=
    \frac14 
    \frac{\pat N}{\pat J^\mu}
    \Delta J^\mu
    +
    \cO(G^4)
    \ .
\end{align}
Contributions from $\Delta P^\mu$ to this formula are subleading.
In addition we have checked the established linear response relations for the impulse and spin kick~\cite{Jakobsen:2022zsx,Jakobsen:2023hig} which at 3PM order may simply be stated as,
\begin{align}
    \Delta \cO^\mu_N
    \Big|_{\rm rad}
    =
    \frac12
    \frac{
        \partial \Delta \cO^\mu_N
    }{
        \partial J^\sigma
    }
    \Delta J^\sigma
    \ ,
\end{align}
with $\cO$ taken as $p^\mu_i$ or $a^\mu_i$.
In this fashion, our conservative-like 3PM observables uniquely fix the 2PM loss of angular momentum to fourth order in spins.
Explicitly, we find that $\Delta J^\mu$ at 2PM takes the form
\begin{align}\label{eq:LossOfJ}
    \Delta J^\mu
    &=
    -J_{\rm rad}^{\mu}
    =
    -\frac{4 G^2 M^3\nu^2(2\gamma^2-1)}{|b|\Gamma}
    \mathcal{I}(\gamma)
    \\
    &\quad\times\Re
    \bigg[
        (\hat L^\mu+\i \hat b^\mu)
        \bigg(
            1
            +
            \xi^2
            +
            \xi^4
            +
            2\frac{v}{1+v^2}
            (\xi+\xi^3)
            +
            \mathcal{O}(\xi^5)
        \bigg)
    \bigg]
    +
    \mathcal{O}(G^3)
    \ ,
    \nn
\end{align}
with $\xi=(\hat L+\i \hat b)_\mu a^\mu_+/|b|$ and $\gamma=1/\sqrt{1-v^2}$.
The 2PM loss of angular momentum was recently computed to 11th order in spins in Ref.~\cite{Alessio:2025flu} and our result Eq.~\eqref{eq:LossOfJ} reproduces theirs to the relevant spin order.

Our formula~\eqref{eq:LossOfJ} suggests a resummation in terms of a geometric series in $\xi$ (with different coefficients at even or odd orders in spin).
This is natural because, being related to the leading-order memory, this quantity is only sensitive to linear-in-curvature couplings, and thus we would expect the resummation to manifest once higher spins are included \cite{Jakobsen:2021lvp}.
Additionally, the structure is that of the tree-level, all-spin-orders scattering angle \cite{Hoogeveen:2025tew}.

\section{Conclusions}

The existence of a scalar generating function, akin to the Hamiltonian,
through which scattering observables such as the momentum impulse and scattering angle
can be derived has been an important problem since WQFT's inception~\cite{Mogull:2020sak}.
In that original paper,
a precise link was derived between the eikonal phase of a $2\to2$
scalar S-matrix and the free energy of WQFT.
The eikonal phase was also explored in the subsequent
$\cN=2$ supersymmetric extension to WQFT~\cite{Jakobsen:2021zvh},
now superseded by the use of bosonic oscillators~\cite{Haddad:2024ebn}.
However, a key issue has been the appearance of infrared divergences starting at 3PM
order in the non-spinning case (two loops) when the eikonal is naively extended to this precision.
These divergences fail to properly cancel when combining diagrams with different causality prescriptions ---
there, arising from a time-symmetric ``averaged'' worldline propagator.
This has hampered the definition and use of the eikonal phase in higher-loop computations.
Given their better-understood causality properties~\cite{Jakobsen:2022psy},
focus has instead shifted to scattering observables such as the momentum impulse $\Delta p^\mu$.

In this paper, we have introduced and motivated the on-shell action
as a generator of scattering observables.
The on-shell action differs from the eikonal phase by the manner
in which we project on incoming vacuum states $|0\rangle$:
\begin{align}
    \i\chi&=\log(\langle0|\hat{S}|0\rangle)\,,&
    \i N&=\langle0|\log(\hat{S})|0\rangle\,.
\end{align}
In short: the eikonal phase is the logarithm of the $\hat{S}$-matrix element,
whereas the on-shell action is the matrix element of the logarithm of the $\hat{S}$-matrix.
This subtle difference is crucial,
as unitarity $\hat{S}^\dagger\hat{S}=1$ implies that the former is complex,
whereas the latter is real.
From a QFT perspective,
the Feynman rules used to produce the on-shell action ---
including retarded/advanced causality prescriptions on propagators ---
have been shown to arise from an application of the Magnus series~\cite{Kim:2024svw}.
This replaces the Dyson series that is typically used to generate $\hat{S}$-matrix elements,
and naturally gives rise to time-symmetric propagators.

Scattering observables are generated from $N$ using Poisson brackets via exponentiation:
$\Delta\mathcal{O}=\exp(\{N,\bullet\})\mathcal{O}-\mathcal{O}$.
The link works in both directions:
from the scattering observables we may infer the causality prescription of $N$
at a diagrammatic level, bypassing explicit use of the Magnus series.
Alternatively, once $N$ has been calculated we may use the brackets to
obtain explicit, integrated expressions for the scattering observables.
In the latter case, a constrained set of brackets allows us to impose additional requirements
including the SSC and ensure its preservation throughout
the time evolution of the system.
The result is a simple, flexible approach for conveniently handling observables and their symmetries.

Given its close similarity to a 4-point QFT scattering amplitude,
a particular advantage of the on-shell action over direct calculation of scattering observables
is that it is amenable to unitarity-based methods. 
Adopting generalised unitarity on the worldline necessitated the introduction of $n$-graviton functions in \cref{eq:nGravFunction}, which play the role of scattering amplitudes for one massive leg emitting $n$ gravitons; the connection between both constructions was established through direct computation at leading order in $G$ in \cref{eq:12GravAmps}.
A simple analysis of the analyticity properties of the integrals involved in the on-shell action justified taking graviton momenta in the numerator of the integrand on shell.
Combined with its diagrammatic representation --- \cref{eq:N1,eq:N2,eq:N3} --- this allowed for the expression of the on-shell action in terms of these $n$-graviton functions --- \cref{eq:NUnitarity,eq:CutIntegrands} --- completing the adaptation of generalised unitarity to the worldline.
Analysing this cut-constructed form, equality between the worldline and scattering amplitude on-shell actions was demonstrated on general grounds, without explicit computation in a particular theory, up to two-loop order.

Tying everything together, we advanced the precision frontier for the PM scattering of two black holes, completing contributions to observables at two-loop order for both black holes spinning up to total spin order four in both spin vectors.
This includes our result for the on-shell action itself,
from which we derived the momentum impulse, spin kick, and, for aligned-spin scattering,
the scattering angle. Our results are collected in an ancillary file at {\zenodo}.

Our paper leaves promising avenues open for further exploration.
The first, naturally, is performing calculations at higher loop orders.
At three-loop order (representing 4PM in the non-spinning case),
so far only linear-in-spin effects have been
considered~\cite{Jakobsen:2023ndj,Jakobsen:2023hig}.
With our newfound emphasis on the on-shell action, we believe that the extension
to higher spins is now much more computationally feasible.
Rather than having to compute the impulse and spin kick separately,
we can now focus on the on-shell action $N$ alone ---
those observables become secondary, derived quantities.
Our ability to use generalised unitarity should also provide a powerful addition to
our existing computational toolkit, allowing the calculation to be broken down into
more manageable pieces.

Another important avenue to explore will be the full inclusion of radiation-reaction effects.
In order to compute observables such as the radiated energy, angular momentum
and gravitational waveform, knowing the on-shell action alone is not enough,
even if one computes it in all regions of integration.
One option is to include a background for the gravitational field~\cite{Kim:2025hpn},
which enables the introduction of brackets on the background graviton variables
akin to those already used on the worldline.
Alternatively, one may consider taking different matrix elements of the $\hat{N}$ operator
besides merely the vacuum element $\langle0|\hat{N}|0\rangle$:
in particular, one may also consider external graviton states.
These have already been shown to generate classical radiative observables by
a similar application of worldline brackets~\cite{Alessio:2025flu}.
The practical application of such an approach to WQFT remains to be explored,
and we leave this promising avenue open for the future.

\bigskip

\textbf{Note added:} We are grateful to Vincent He and Julio Parra-Martinez for sharing a draft of their upcoming and complementary work \cite{He:2025how}, and coordinating {\tt arXiv.org} submission.

\section*{Acknowledgments}

The authors thank
A.~Brandhuber, R.~Gonzo, J.~W.~Kim, P.~Pichini, B.~Sauer, G.~Travaglini and P.~Vives
for valuable discussions.
G.M.~is supported by The Royal Society under grant URF\textbackslash R1\textbackslash 231578,
``Gravitational Waves from Worldline Quantum Field Theory''.
The work of K.H., G.U.J.~and J.P.~was funded by the European Union through the 
European Research Council under grant ERC Advanced Grant 101097219 (GraWFTy).
Views and opinions expressed are however those of the authors only and do not necessarily reflect those of the European Union or European Research Council Executive Agency. Neither the European Union nor the granting authority can be held responsible for them.

\appendix

\section{Aligned-spin scattering angle}\label{app:alignedSpins}

Here we demonstrate that, for spins aligned with the orbital plane,
the on-shell action $N:=\langle0|\hat{N}|0\rangle$ can be identified with the on-shell action
$I_r:=\int_{r_{\rm min}}^\infty\!\d r\,p_r$.
As we focus on conservative scattering, we work in the center-of-mass frame
defined by the total momentum vector
$\hat{P}^\mu=(p_1^\mu+p_2^\mu)/|p_1+p_2|$.
The relative two-body motion is then conveniently described by
\begin{align}
\begin{aligned}
    p^\mu&=\frac1{E^2}\left[m_2(\gamma m_1+m_2)p_1^\mu-m_1(\gamma m_2+m_1)p_2^\mu\right]\\
    &=(0,\mathbf{p})\,.
\end{aligned}
\end{align}
The change in the momentum $\mathbf{p}$ is due to a rotation by the angle $\theta$:
\begin{align}
\begin{aligned}
    \mathbf{p}'=\mathbf{p}+\Delta\mathbf{p}&=
    |\mathbf{p}|(\cos\theta\,\hat{\mathbf{p}}+\sin\theta\,\hat{\mathbf{b}})\\
    &=\exp(\theta\,\hat{\mathbf L}\times\bullet)\mathbf{p}\,,
\end{aligned}
\end{align}
where $\hat{\mathbf{L}}=-\hat{\mathbf{b}}\times\hat{\mathbf{p}}$.
Comparing this with the key property of the on-shell action~\eqref{expObservables},
which implies that
\begin{align}
    \mathbf{p}'=\exp(\{N,\bullet\})\mathbf{p}\,,
\end{align}
we learn that $\{N,\bullet\}=\theta\,\hat{\mathbf{L}}\times\bullet$,
i.e.~we may identify the bracket with $N$ as the action of a rotation generator by the angle $\theta$.
In order to deduce the scattering angle from the on-shell action,
we simply act on the momentum vector $\mathbf{p}$:
\begin{align}
\begin{aligned}
    \{N,\mathbf{p}\}&=-\frac{\partial N}{\partial\mathbf b}=-
    \frac{\partial N}{\partial|\mathbf{b}|}\hat{\mathbf{b}}\,,\\
    \theta\,\hat{\mathbf{L}}\times\mathbf{p}&=\theta|\mathbf{p}|\hat{\mathbf{b}}\,.
\end{aligned}
\end{align}
We have used the fact that, for aligned spins, the on-shell action depends
on the impact parameter only through its modulus $|\mathbf{b}|$.
Thus, we ultimately learn that
\begin{align}
    \theta=-\frac1{|\mathbf{p}|}\frac{\partial N}{\partial|\mathbf{b}|}\,,
\end{align}
which is a defining property of the on-shell action $I_r$.

\section{On-shell factorisation of SSC dependence}\label{app:SSC}

In the main text, we presented that the bosonic-WQFT action in \cref{eq:sWQFT2} produces $n$-graviton functions whose dependence on the SSC vector factorises:
\begin{align}\label{eq:nGravFuncSSC}
    \mathcal{W}^{(n)}(Z)=\left[1+\frac{\i(q\cdot Z)}{m}\right]\mathcal{W}^{(n)}(0)+\cO(Z^{2}),
\end{align}
where $q^\mu=\sum_{i=1}^{n}q_{i}^{\mu}$ is the sum of all graviton momenta emitted from the worldline.
While our action can generically produce $\cO(Z^{2})$ contributions to $\mathcal{W}^{(n)}$, such contributions will not affect observables when the covariant SSC $Z^{\mu}=0$ is chosen as an initial condition; see ref.~\cite{Haddad:2024ebn}.

Here we derive two consequences from \cref{eq:nGravFuncSSC}.
First, we show that this factorisation of the SSC dependence is a necessary and sufficient condition for the covariant SSC to be (weakly) conserved under evolution with the on-shell action.
In the process, we will demonstrate that the on-shell action effectively inherits this SSC dependence simply through a shift of the spin-independent impact parameter.

To generalise the discussion as cleanly as possible, we introduce the notion of \textit{disconnected} $n$-graviton functions.
An $n$-graviton function is said to be disconnected if it can be written in terms of the connected $n$-graviton functions of \cref{eq:nGravConn} as
\begin{align}\label{eq:nGravDisc}
    \mathcal{W}^{(n)}_{\rm disc.}=\prod_{i=1}^{k}\mathcal{W}^{(n_{i})}_{\rm conn.},
\end{align}
for some $n_{i}$ partitioning $n$.
To accommodate the most general scenario, then, the $n$-graviton functions in this section are allowed to be disconnected.
It is easy to see that if the connected functions obey \cref{eq:nGravFuncSSC} then so will the disconnected function which is their product.

Let us begin with weak conservation of the covariant SSC.
As with other quantities, the evolution of the SSC vector is governed by Poisson brackets with the on-shell action, $N(b;Z_{i})$, whose dependence on the SSC vector of worldline $i$ we have highlighted, and which otherwise depends most generally on the impact parameter $\tilde{b}^{\mu}={P^{\mu}}_{\nu}\left(b_{2}^{\nu}-b_{1}^{\nu}\right)$.

Since we are interested specifically in the change in the covariant SSC, weak conservation means we subsequently evaluate these brackets at the initial value $Z^{\mu}_{i}=0$.
Consequently, conservation of the covariant SSC is given by
\begin{align}
    0=\left[\{N,Z^{\mu}_{i}\}+\frac{1}{2}\{N,\{N,Z^{\mu}_{i}\}\}+\frac{1}{3!}\{N,\{N,\{N,Z^{\mu}_{i}\}\}\}+\dots\right]_{Z_{i}=0}.
\end{align}
It is not difficult to show inductively that
\be
\begin{aligned}
    &\{N(\tilde{b};Z_{i}),Z_{i}^{\mu}\}_{Z=0}=0 \\
    \Rightarrow\quad&\{N(\tilde{b};Z_{i}),\{\dots,\{N(\tilde{b};Z_{i}),Z_{i}^{\mu}\}\dots\}\}_{Z_{i}=0}=0,
\end{aligned}
\ee
for any number of iterations of Poisson brackets with the on-shell action.
Conservation of the covariant SSC is hence equivalent to $\{N(\tilde{b};Z_{i}),Z_{i}^{\mu}\}_{Z_{i}=0}=0$ at any PM order.
For reference, this Poisson bracket written explicitly is
\begin{align}
    \{N,Z_{i}^{\nu}\}&=-\frac{1}{m}\left(\frac{\partial N}{\partial\tilde{b}^{\mu}_{i}}\frac{\partial Z^{\nu}_{i}}{\partial v_{i,\mu}}+\i\frac{\partial N}{\partial\alpha^{\mu}_{i}}\frac{\partial Z_{i}^{\nu}}{\partial\bar\alpha_{i,\mu}}-\i\frac{\partial N}{\partial\bar\alpha_{i,\mu}}\frac{\partial Z^{\nu}_{i}}{\partial\alpha_{i}^{\mu}}\right),
\end{align}
having used that the SSC vector does not depend on the impact parameter to drop a term.
With this, we will now prove the following statement:

Suppose that the on-shell action in momentum space at the $(n-1)^{\rm st}$ subleading order is constructible from the $n$-graviton functions in \cref{eq:nGravFunction} via unitarity cuts.
Then the following are equivalent:
\begin{align*}
    {\rm (i)}\quad &\{N(\tilde{b};Z_{i}),Z_{i}^{\mu}\}_{Z_{i}=0}=0, \\
    {\rm (ii)}\quad &\mathcal{W}_{i}^{(n)}(Z_{i})=\left[1+\frac{\i (q\cdot Z_{i})}{m}\right]\mathcal{W}_{i}^{(n)}(0)+\cO(Z_{i}^{2}).
\end{align*}
This recasts the imposition of the conservation of the covariant SSC in terms of a simple dependence on the SSC vector in the $n$-graviton functions.

Without loss of generality, we will label the worldline whose SSC we are analysing with $i=2$.
As detailed in the previous section, the assumption that the on-shell action in momentum space is amenable to unitarity cuts relates it to a (sum of a) product of two $n$-graviton functions, integrated over the momenta of gravitons exchanged between them.
In momentum space,
\begin{align}\label{eq:OSActionUnitarity}
    &\tilde N(q;Z_{2})\sim\sum\int_{\ell_{1}\dots\ell_{n-1}}\sum_{\lambda_{i}}\frac{\mathcal{W}_{1}^{(n)}\mathcal{W}_{2}^{(n)}(Z_{2})}{\ell_{1}^{2}\dots\ell_{n-1}^{2}(q-\ell)^{2}},
\end{align}
for $\ell^{\mu}=\sum_{i=1}^{n-1}\ell^{\mu}_{i}$.
Whether worldline 1 is spinning or not is immaterial to the analysis in this section, so we suppress its SSC vector in what follows.

To begin the proof, let us assume (i) and show that it implies (ii).
We can split the on-shell action  as
\begin{align*}
    N(\tilde{b};Z_{2})=N(\tilde{b};0)+\frac{1}{m_{2}}N_{\rho}(\tilde{b})Z_{2}^{\rho}+\cO(Z_{2}^{2}),
\end{align*}
where $N_{\rho}(\tilde{b})$ does not depend on $Z_{2}^{\mu}$.
Plugging this into the vanishing Poisson bracket,
\begin{align}\label{eq:iToiiPB}
    &\frac{\partial N(\tilde{b};0)}{\partial b^{\lambda}_{2}}S_{2}^{\mu\lambda} \\
    &=-m_{2}v_{2}^{\lambda}\left[\bar{\alpha}^{\mu}_{2}\frac{\partial}{\partial\bar{\alpha}_{2}^{\lambda}}\left(N(\tilde{b};0)+N_{\rho}(\tilde{b})\frac{Z_{2}^{\rho}}{m_{2}}\right)+\alpha^{\mu}_{2}\frac{\partial}{\partial\alpha_{2}^{\lambda}}\left(N(\tilde{b};0)+N_{\rho}(\tilde{b})\frac{Z_{2}^{\rho}}{m_{2}}\right)\right]_{Z_{2}=0}\nn \\
    &+m_{2}\left[(v_{2}\cdot\bar{\alpha}_{2})\frac{\partial}{\partial\bar{\alpha}_{2,\mu}}\left(N(\tilde{b};0)+N_{\rho}(\tilde{b})\frac{Z_{2}^{\rho}}{m_{2}}\right)+(v_{2}\cdot\alpha_{2})\frac{\partial}{\partial\alpha_{2,\mu}}\left(N(\tilde{b};0)+N_{\rho}(\tilde{b})\frac{Z_{2}^{\rho}}{m_{2}}\right)\right]_{Z_{2}=0}.\nn
\end{align}
Since by construction $N(\tilde{b};0)$ is independent of $Z_{2}^{\mu}$, its only dependence on the oscillators is through the spin vector, $a^{\mu}\propto\epsilon^{\mu v\bar\alpha\alpha}$.
Therefore, by antisymmetry of the Levi-Civita symbol, $v_{2}^{\rho}\partial N(\tilde{b};0)/\partial\alpha_{2}^{\rho}=v^{\rho}_{2}\partial N(\tilde{b};0)/\partial\bar\alpha_{2}^{\rho}=0$.
Moreover, since we set $Z^{\mu}_{2}=0$ after differentiation, we can pass $N_{\rho}(\tilde{b})$ through these derivatives, such that they only act on the explicit SSC vector.
Then, we find that the last line is proportional to $Z_{2}^{\mu}$ and is subsequently set to 0, while the remainder simplifies to
\begin{align}
    \frac{\partial N(\tilde{b};0)}{\partial b_{2}^{\lambda}}S_{2}^{\mu\lambda}&=-N_{\lambda}(\tilde{b})S_{2}^{\mu\lambda}.
\end{align}
Using that this must hold for any initial spin tensor $S_{2}^{\mu\alpha}$ and Fourier transforming to momentum space, we find
\be
\begin{aligned}
    &\tilde N_{\mu}(q)=\i q_{\mu}\tilde N(q;0) \\
    \Rightarrow\quad&\tilde N(q;Z)=\left[1+\frac{\i(q\cdot Z_{2})}{m}\right]\tilde N(q;0)+\cO(Z_{2}^{2}).
\end{aligned}
\ee
Finally, substituting the assumption \cref{eq:OSActionUnitarity} in both sides of this equation gives
\begin{align}
    &\sum\int_{\ell_{1}\dots\ell_{n-1}}\sum_{\lambda_{i}}\frac{\mathcal{W}_{1}^{(n)}\mathcal{W}_{2}^{(n)}(Z_{2})}{\ell_{1}^{2}\dots\ell_{n-1}^{2}(q-\ell)^{2}} \\
    &=\left[1+\frac{\i(q\cdot Z_{2})}{m}\right]\sum\int_{\ell_{1}\dots\ell_{n-1}}\sum_{\lambda_{i}}\frac{\mathcal{W}_{1}^{(n)}\mathcal{W}_{2}^{(n)}(0)}{\ell_{1}^{2}\dots\ell_{n-1}^{2}(q-\ell)^{2}}+\cO(Z_{2}^{2}).\nn
\end{align}
The sum of graviton momenta gives the momentum transferred between the worldlines, which is precisely $q^{\mu}$.
Since this must hold regardless of the properties of the second worldline, we conclude that (i) $\Rightarrow$ (ii).

Let us now prove the converse.
Since we are ambivalent about the structure of terms at $\cO(Z_{2}^{n\geq2})$ -- as mentioned above, such terms will vanish upon imposing the initial condition $Z_{2}^{\mu}=0$ after evaluating the Poisson bracket -- we may select these terms for convenience such that\footnote{This is reminiscent of the exponential factor emerging from the Lorentz product of massive polarisation vectors transforming in reducible representations of the Lorentz group \cite{Bern:2023ity,Alaverdian:2024spu}.}
\begin{align}\label{eq:SSCExp}
    \mathcal{W}_{2}^{(n)}(Z_{2})=\exp\left[\frac{\i(q\cdot Z_{2})}{m_{2}}\right]\mathcal{W}_{2}^{(n)}(0)+\cO(Z_{2}^{2}).
\end{align}
Through unitarity, the on-shell action inherits this overall exponential structure, such that upon transitioning to impact-parameter space we can write $N(\tilde{b};Z_{2})=N(b)$ where $b^{\mu}=\tilde{b}^{\mu}+Z_{2}^{\mu}/m_{2}$ is the invariant impact parameter in \cref{eq:GeneralisedIP} for non-spinning worldline 1.
Evaluating the Poisson bracket,
\begin{align}
    &\{N(b),Z_{2}^{\mu}\}_{Z_{2}=0}\nn \\
    &=-\frac{1}{m_{2}}\left[\frac{\partial N(b)}{\partial b_{2}^{\lambda}}S_{2}^{\mu\lambda}+\frac{\partial N(b)}{\partial\bar{\alpha}_{2}^{\lambda}}m_{2}\bar{\alpha}_{2}^{\mu}v_{2}^{\lambda}+\frac{\partial N(b)}{\partial\alpha_{2}^{\lambda}}m_{2}v_{2}^{\lambda}\alpha_{2}^{\mu}\right]_{Z_{2}=0} \\
    &=-\frac{1}{m_{2}}\underbrace{\frac{\partial N(b)}{\partial b^{\lambda}}S_{2}^{\mu\lambda}}_{\rm Ia}-\underbrace{\left(\frac{\partial N(b)}{\partial b^{\rho}}\frac{\partial b^{\rho}}{\partial\bar{\alpha}_{2}^{\lambda}}\right.}_{\rm Ib}+\underbrace{\left.\frac{\partial N(\tilde{b})}{\partial\bar{\alpha}_{2}^{\lambda}}\right)}_{\rm II}\bar{\alpha}_{2}^{\mu}v_{2}^{\lambda}-\underbrace{\left(\frac{\partial N(b)}{\partial b^{\rho}}\frac{\partial b^{\rho}}{\partial\alpha_{2}^{\lambda}}\right.}_{\rm Ib}+\underbrace{\left.\frac{\partial N(\tilde{b})}{\partial\alpha_{2}^{\lambda}}\right)}_{\rm II}\alpha_{2}^{\mu}v_{2}^{\lambda}.\nn
\end{align}
We have suppressed terms proportional to $v_{2}\cdot\alpha_{2}$ and $v_{2}\cdot\bar\alpha_{2}$, analogous to the last line of \cref{eq:iToiiPB}, as they vanish upon setting $Z_{2}=0$ after differentiating.

The terms labelled II arise from the derivative with respect to the bosonic oscillators acting on the spin-vector dependence of the on-shell action, which vanish when contracted with the initial velocity, as explained above.
Terms labelled Ib encode the action of these derivatives on the SSC vector, contained entirely in the shifted impact parameter.
Upon simplifying, these precisely cancel term Ia, such that the Poisson bracket vanishes.
This completes the proof.

A covariant-SSC-conserving action was determined in ref.~\cite{Haddad:2024ebn} through two separate methods, both of which scale poorly to subleading orders.
While both methods involved carrying along unfixed Wilson coefficients, one required evolving the covariant SSC with the Hamiltonian, and the other computed the linear impulse and spin kick to construct the change in the SSC vector.
Imposing conservation of the covariant SSC then fixed the values of the unknown coefficients.
Now we have proven that a simpler, on-shell analysis of the $Z^\mu$ dependence of $n$-graviton functions can replace the cumbersome methods used there.

As an intermediate step in our proof above, we have shown by conveniently adding irrelevant $\cO(Z^{n\geq2})$ terms to $\mathcal{W}^{(n)}$ that the on-shell action only depends on the SSC vector through a shift of the spinless impact parameter.
We can alternatively reach this conclusion by casting the inherited $[1+\i(q\cdot Z_{2})/m_{2}]$ structure as a differential operator in impact-parameter space:
\begin{align}\label{eq:OSACoM}
    N(\tilde{b};Z_{2})=\left[1+\frac{1}{m_{2}}Z_{2}^{\mu}\frac{\partial}{\partial b_{2}^{\mu}}\right]N(\tilde{b};0)+\cO(Z_{2}^{2})=N(\tilde{b}+Z_{2}/m_{2};0)+\cO(Z_{2}^2).
\end{align}
A consequence of this is that, so long as we are considering scattering that conserves the covariant SSC, we may actually work with an on-shell action with vanishing SSC vector if we in the end upgrade the spinless impact parameter to the invariant impact parameter of \cref{eq:GeneralisedIP}.
This simple dependence on the SSC vector meshes well with the interpretation of the SSC as a choice for the representative point of an extended body \cite{Costa:2011zn,Costa:2014nta,Levi:2015msa}.

\bibliographystyle{JHEP}
\bibliography{wlonshellaction.bib}

\end{document}